\documentclass[aps,prd,twocolumn,superscriptaddress,nofootinbib,floatfix]{revtex4-1}

\usepackage{graphicx}
\usepackage{amsmath}
\usepackage{amssymb}
\usepackage{yhmath}
\usepackage{extarrows}
\usepackage[colorlinks,citecolor=blue,linkcolor=blue,urlcolor=blue,anchorcolor=blue]{hyperref}
\allowdisplaybreaks
\makeatletter
\renewcommand\theequation{\arabic{section}.\arabic{equation}}
\@addtoreset{equation}{section}
\makeatother
\newcommand\dx{{\rm d}}
\newcommand\p{\partial}
\newcommand\etal{{\it et~al.}}
\renewcommand\aa{Astron. Astrophys.}

\renewcommand\apj{Astrophys. J.}
\newcommand\apjl{Astrophys. J. Lett.}
\newcommand\apjs{Astrophys. J. Suppl.}
\newcommand\araa{Annu. Rev. Astron. Astrophys.}
\newcommand\ass{Astrophys. Space Sci.}
\newcommand\cac{Comput. Astrophys. Cosmol.}
\newcommand\cqg{Classical Quantum Gravity}
\newcommand\epjc{Eur. Phys. J. C}
\newcommand\gerg{Gen. Relativ. Gravit.}
\newcommand\jcap{J. Cosmol. Astropart. Phys.}
\newcommand\lrr{Living Rev. Relativity}
\newcommand\mnras{Mon. Not. R. Astron. Soc.}
\newcommand\naa{Nat. Astron.}
\newcommand\pasj{Publ. Astron. Soc. Jpn.}
\newcommand\plb{Phys. Lett. B}
\newcommand\prep{Phys. Rep.}
\renewcommand\prd{Phys. Rev. D}
\renewcommand\prl{Phys. Rev. Lett.}

\begin{document}

\title{Testing the Schwarzschild metric in a strong field region with the Event Horizon Telescope}
\author{S. X. Tian}
\email[]{tshuxun@whu.edu.cn}
\affiliation{School of Physics and Technology, Wuhan University, 430072 Wuhan, China}
\author{Zong-Hong Zhu}
\email[]{zhuzh@whu.edu.cn}
\affiliation{School of Physics and Technology, Wuhan University, 430072 Wuhan, China}
\affiliation{Department of Astronomy, Beijing Normal University, 100875 Beijing, China}
\date{\today}
\begin{abstract}
  Testing gravity theory in the strong field region becomes a reality due to the observations of gravitational waves and black hole shadows. In this paper, we discuss how to constrain the possible deviations of the classical general relativity with the image of M87* observed by the Event Horizon Telescope. More precisely, we want to know where is the event horizon for a non-rotating black hole. General relativity predicts the horizon is located at the Schwarzschild radius $r_\textrm{s}$, while other gravity theories may give different predictions. We propose a parameterized Schwarzschild metric (PSM) in which the horizon is located at $r=nr_\textrm{s}$, where $n$ is a real free parameter, and prove general relativity with nonlinear electrodynamics allows $n\neq1$. In the weak field region, the PSM is equivalent to the Schwarzschild metric regardless of the value of $n$. In the strong field region, the difference between the PSM and Schwarzschild metric would leave an imprint on the shadow image. We present detailed calculations and discussions on the black hole shadows with large background light source and accretion disk in the PSM  framework. More importantly, we point out that $n\approx2$ can be used to explain why the black hole mass measured by the shadow is a factor of about two larger than the previous gas dynamics measurements. If this explanation is confirmed to be right, then this phenomenon, together with the late-time cosmological acceleration, will be very important to test gravity theories.
\end{abstract}
\pacs{}
\maketitle

\section{Introduction}\label{sec:01}
The Event Horizon Telescope (EHT) observed the first image of the black hole \cite{Akiyama2019a,Akiyama2019b,Akiyama2019c,Akiyama2019d,Akiyama2019e,Akiyama2019f}, which provides a new way to test gravity theories in the strong field region. In the framework of general relativity, assuming the black hole is in front of a large light source, \cite{Synge1966} and \cite{Bardeen1973} calculated the apparent size and shape of the Schwarzschild and Kerr black hole, respectively. \cite{Cunningham1972,Cunningham1973} analyzed the optical appearance of a star orbiting a Kerr black hole. A more realistic simulation is given by \cite{Luminet1979}, which calculated the image of a Schwarzschild black hole with the standard thin accretion disk. \cite{Falcke2000,Fraga-Encinas2016,Roelofs2019} simulated the image for the Kerr black hole and other accretion disk models. In order to compare observations with theories, the EHT collaboration \cite{Akiyama2019e,Akiyama2019f,Porth2019} simulated the accretion matters based on the general relativistic magnetohydrodynamic (GRMHD, see \cite{Font2008,*Porth2017,*Olivares2019,*White2019} for pedagogical resources) simulations and produced the images by ray tracing \cite{Dexter2009,*Vincent2011,*Yang2013,*Chan2018,*Davelaar2018}.

For the non-Kerr metrics, which mainly appear in general relativity with exotic matters and modified gravities, most of the work (see \cite{Abdujabbarov2013,*Amarilla2013,*Grenzebach2014,*Cunha2015,*Younsi2016,*Vincent2016,*Cunha2017,*Wang2017,
*Ayzenberg2018,*Cunha2018,*Guo2018,*Hou2018,*Ovgun2018,*Shaikh2018,*Tsukamoto2018,*Wang2018a,*Wang2018b,
*Ali2019,*Cunha2019,*Haroon2019,*Held2019,*Jusufi2019,*Konoplya2019c,*Konoplya2019a,*Konoplya2019b,*Long2019,
*Medeiros2019,*Neves2019,*Rahman2019,*Shaikh2019,*Stuchlik2019,*Vagnozzi2019,*Wang2019,*Zhu2019} for examples) is to analyze the shadow for the case that the black hole is in front of a large light source . These works can be regarded as an extension of \cite{Bardeen1973}. There are few works to discuss the image of the non-Kerr black hole with a realistic accretion disk. Here we briefly introduce three works in this direction. \cite{Broderick2014,Johannsen2016} simulated the images of the quasi-Kerr black hole, which introduces a quadrupolar moment to the Kerr metric \cite{Glampedakis2006}, with radiatively inefficient accretion flow, and discussed the observational constraints on the quadrupolar deviation by the upcoming observations of Sgr A*. \cite{Mizuno2018} simulated the image of a spherically symmetric dilaton black hole with Rezzolla-Zhidenko parametrization \cite{Rezzolla2014} based on GRMHD and general relativistic radiative transfer \cite{Bronzwaer2018} simulations, and pointed out that EHT observations alone are not sufficient to distinguish between the dilaton and Kerr black holes with the parameters they set due to the low resolution. However, \cite{Mizuno2018} did not discuss the results of combining EHT observations with other astronomical observations, which may be a powerful tool to test general relativity as we will see in Sec. \ref{sec:0403}. Very recently, \cite{Gyulchev2019} simulated the image of the Janis-Newman-Winicour naked singularity with the standard thin accretion disk. But \cite{Gyulchev2019}, as well as \cite{Luminet1979}, missed the projection effect as we mentioned in Sec. \ref{sec:0402}. Imaging the non-Kerr black hole with a realistic accretion disk plays an important role in testing gravity theory with the EHT observations.

In this paper, we explore the ability to test gravity theory in the strong field region with the image of M87* \cite{Akiyama2019a,Akiyama2019b,Akiyama2019c,Akiyama2019d,Akiyama2019e,Akiyama2019f}. More precisely, we focus on the issue that where is the event horizon for a non-rotating black hole, which has not been directly tested by observations before. A parameterized metric is needed to do this. General relativity or the Schwarzschild metric has been widely tested in the weak field region \cite{Clifton2012,Will2014,Tian2019}. Thus, in order to recover all successes of the Schwarzschild metric, the desired metric should be equivalent to the Schwarzschild metric in the weak field region. We propose the following parameterized Schwarzschild metric (PSM)
\begin{equation}\label{eq:1.01}
  \dx s^2=-c^2A_n(r)\dx t^2+\frac{\dx r^2}{A_n(r)}+r^2\dx\Omega^2,
\end{equation}
where $\dx\Omega^2=\dx\theta^2+\sin^2\theta\dx\varphi^2$, $r_\textrm{s}=2GM/c^2$, $n$ is a real free parameter, and
\begin{equation}\label{eq:1.02}
  A_n(r)=(1-\frac{nr_\textrm{s}}{r})^{1/n}=1-\frac{r_\textrm{s}}{r}+\frac{(1-n)r_\textrm{s}^2}{2r^2}+\cdots.
\end{equation}
This metric satisfies the weak field equivalence requirement regardless of the value of $n$. The Schwarzschild metric corresponds to $n=1$ and the PSM is significantly different with the Schwarzschild metric at the horizon scale if $n\neq1$. One clear feature of the PSM is that the event horizon is located at $r=nr_\textrm{s}$. EHT observations could be used to constrain $n$ if one can simulate the PSM black hole image. In this paper, our calculations mainly follow \cite{Luminet1979,Gyulchev2019}, which assume a standard thin accretion disk around the black hole.

One important thing is worth mentioning here. EHT observations show M87* is rotating \cite{Akiyama2019e,Dokuchaev2019}, which is also confirmed by the jet structure observations \cite{Doeleman2012,Nakamura2018,Takahashi2018}, while the PSM describes a non-rotating black hole. So why can we compare our results with EHT observations? The main reason is that the shadow size is fairly independent of the spin for the Kerr black hole with an accretion disk \cite{Falcke2000}. This is consistent with the result obtained in \cite{Bardeen1973}, which shows the influence of the spin on the shadow for the Kerr black hole with a large background light source is small if the spin is not extremely high. It is reasonable to assume that this property also holds for the parameterized Kerr metric (PKM, an analogy of the PSM). On the other side, our discussions (see Sec. \ref{sec:0403}) depend only on the shadow size. Therefore, choosing a spherical symmetry metric is reasonable for our purpose. Note that the black hole spin controls the north-south asymmetry \cite{Akiyama2019a} and circularity \cite{Bambi2019,Tamburini2019} of the shadow. Comparing various measurements of the spin (see \cite{Akiyama2019e,Dokuchaev2019,Doeleman2012,Nakamura2018,Takahashi2018,Bambi2019,Tamburini2019,
Li2009,Kawashima2019,Nemmen2019,Nokhrina2019} for examples) is useful for exploring strong gravitational field physics (including not only the gravity theory but also the accretion theory). We would like to leave the work of constraining the PKM with EHT observations to the future.

The other important thing worth mentioning is that the PSM is not regular. Especially, the Ricci scalar $R$ is divergent at $r=nr_{\rm s}$ for a general $n$. The regularity of the spacetime is crucial for the general relativity researches \cite{Yunes2011}. However, it is not hard to avoid the singularity at $r=nr_{\rm s}$ and preserve the main properties we are considering of the PSM. To do this, the key is that the region involved in our calculations is mainly range from approximately $r_{\rm ISCO}$ to infinity, where $r_{\rm ISCO}>2nr_{\rm s}$ [see Eq. (\ref{eq:2.08})]. Thus we can further modify the PSM near the horizon to avoid the singularity. For example, for the case of $|n-1|\ll1$, we can keep only the first three terms of the Taylor expansion of $A_n$. The corresponding metric is equivalent to the Reissner-Nordstr\"om metric, which do avoid the singularity at the horizon. Furthermore, the singularity disappears if we keep only finite terms of the Taylor expansion of $A_n$ (we checked this for $R$ and $R_{\mu\nu}R^{\mu\nu}$). The high-order terms may originate from modified gravities as shown in \cite{Yunes2011} or general relativity with nonlinear electrodynamics as shown in Sec. \ref{sec:05}. This discussion indicates that the PSM can be a good approximation of one suitable regular metric outside the horizon (the domain that is not very close to the event horizon) for our purpose. And we choose the PSM as the background metric in the following calculations due to its simplicity.

This paper is organized as follows: Section \ref{sec:02} analyzes the geodesics in the PSM. Section \ref{sec:03} gives the shadow size for the case that the black hole is in front of a large light source. Section \ref{sec:04} simulates the image of the black hole with the standard thin accretion disk and discusses the observational constraints. In this step, we reanalyze the disk structure, light bending, redshift and projection effects in the PSM. Section \ref{sec:05} proves general relativity with nonlinear electrodynamics allows $n\neq1$ in the PSM. Our conclusions will be presented in Sec. \ref{sec:06}. We set $8\pi G=1$ and $c=1$ hereafter.

\section{Geodesics}\label{sec:02}
In this section, we analyze the geodesics in the PSM, which are the basis for the following calculations.

\subsection{Time-like geodesic}
Without loss of generality, we assume a test particle with nonzero rest mass moving in the equatorial plane, i.e., $\theta=\pi/2$ and the $\theta$-component 4-velocity $u^\theta=0$. The PSM is independent of $t$ and $\varphi$ gives
\begin{subequations}\label{eq:2.01}
\begin{align}
  u_t&=g_{tt}u^t=-A_n\frac{\dx t}{\dx\tau}=-E,\\
  u_\varphi&=g_{\varphi\varphi}u^\varphi=r^2\frac{\dx\varphi}{\dx\tau}=L,
\end{align}
\end{subequations}
where $E$ and $L$ are constants, and $\dx\tau^2=-\dx s^2$, which gives $u^\mu u_\mu=-1$, i.e.,
\begin{align}\label{eq:2.02}
  -A_n\cdot(\frac{\dx t}{\dx\tau})^2+\frac{1}{A_n}(\frac{\dx r}{\dx\tau})^2
  +r^2(\frac{\dx\varphi}{\dx\tau})^2=-1.
\end{align}
Substituting Eq. (\ref{eq:2.01}) into Eq. (\ref{eq:2.02}), we obtain
\begin{equation}
  (\frac{\dx r}{\dx\tau})^2=E^2-A_n\cdot(1+\frac{L^2}{r^2}).
\end{equation}
Taking the derivative of the above equation yields
\begin{equation}\label{eq:2.04}
  \frac{\dx^2r}{\dx\tau^2}=A_n\frac{2nL^2r_\textrm{s}+L^2r_\textrm{s}-2L^2r+r_\textrm{s}r^2}{2r^3(nr_\textrm{s}-r)}.
\end{equation}
The circular orbits require $\frac{\dx^2r}{\dx\tau^2}=\frac{\dx r}{\dx\tau}=0$, which gives
\begin{subequations}\label{eq:2.05}
\begin{align}
  L&=\sqrt{\frac{r_\textrm{s}r^2}{2r-(2n+1)r_\textrm{s}}},\\
  E&=\sqrt{A_n\cdot\left[1+\frac{r_\textrm{s}}{2r-(2n+1)r_\textrm{s}}\right]},
\end{align}
\end{subequations}
Note that here $L>0$ means the test particle moving along the positive $\varphi$-direction. Substituting Eq. (\ref{eq:2.05}) into Eq. (\ref{eq:2.01}), we obtain the nonzero 4-velocity components for the circular orbits are
\begin{subequations}\label{eq:2.06}
\begin{equation}
  u^t=\sqrt{\frac{1}{A_n}\left[1+\frac{r_\textrm{s}}{2r-(2n+1)r_\textrm{s}}\right]},
\end{equation}
and
\begin{equation}
  u^\varphi=\sqrt{\frac{r_\textrm{s}}{r^2[2r-(2n+1)r_\textrm{s}]}}.
\end{equation}
\end{subequations}
In order to calculate the radius of the innermost stable circular orbit (ISCO), we can solve $\frac{\dx^2r}{\dx\tau^2}=0$ based on Eq. (\ref{eq:2.04}) to obtain the roots
\begin{equation}
  r_\pm=\frac{L^2\pm\sqrt{L^4-r_\textrm{s}(2nL^2r_\textrm{s}+L^2r_\textrm{s})}}{r_\textrm{s}}.
\end{equation}
The ISCO corresponds to $r_+=r_-$, which gives $L^2=(2n+1)r_\textrm{s}^2$ and the ISCO radius
\begin{equation}\label{eq:2.08}
  r_\textrm{ISCO}=(2n+1)r_\textrm{s}.
\end{equation}
These results will be used to calculate the disk structure in Sec. \ref{sec:0401}.

\subsection{Null geodesic}
In Sec. \ref{sec:0402}, we will discuss the motion of photons from the disk to the observer. In this case, the photon orbit is not on the equatorial plane. However, we can rotate the coordinates to make the orbital plane coincide with the equatorial plane, and denote $\tilde{\varphi}$ as the deflection angle in the orbital plane. Similar to Eq. (\ref{eq:2.01}), we obtain
\begin{subequations}\label{eq:2.09}
\begin{gather}
  A_n\frac{\dx t}{\dx\lambda}=E,\label{eq:2.09a}\\
  r^2\frac{\dx\tilde{\varphi}}{\dx\lambda}=L,\label{eq:2.09b}
\end{gather}
\end{subequations}
where $\lambda$ is an affine parameter. When $r\gg r_\textrm{s}$, we obtain $L/E=r^2\dx\tilde{\varphi}/\dx t=b$, which is the impact parameter. Substituting Eq. (\ref{eq:2.09}) into $u^\mu u_\mu=0$, we obtain
\begin{equation}
  (\frac{\dx r}{\dx\lambda})^2=E^2-A_n\frac{L^2}{r^2}.
\end{equation}
Considering the angular momentum conservation Eq. (\ref{eq:2.09b}), the above equation gives
\begin{equation}
  (\frac{\dx r}{\dx\tilde{\varphi}})^2=\frac{r^4}{b^2}-A_nr^2,
\end{equation}
which is the orbit equation for photons. Letting $u=1/r$, the above equation can be simplified to
\begin{equation}\label{eq:2.12}
  (\frac{\dx u}{\dx\tilde{\varphi}})^2
  =\frac{1}{b^2}-u^2(1-nr_\textrm{s}u)^{1/n}\equiv G(u).
\end{equation}

\section{Shadow: Large background light source}\label{sec:03}
\begin{figure}[b!]
  \centering
  \includegraphics[width=0.99\linewidth]{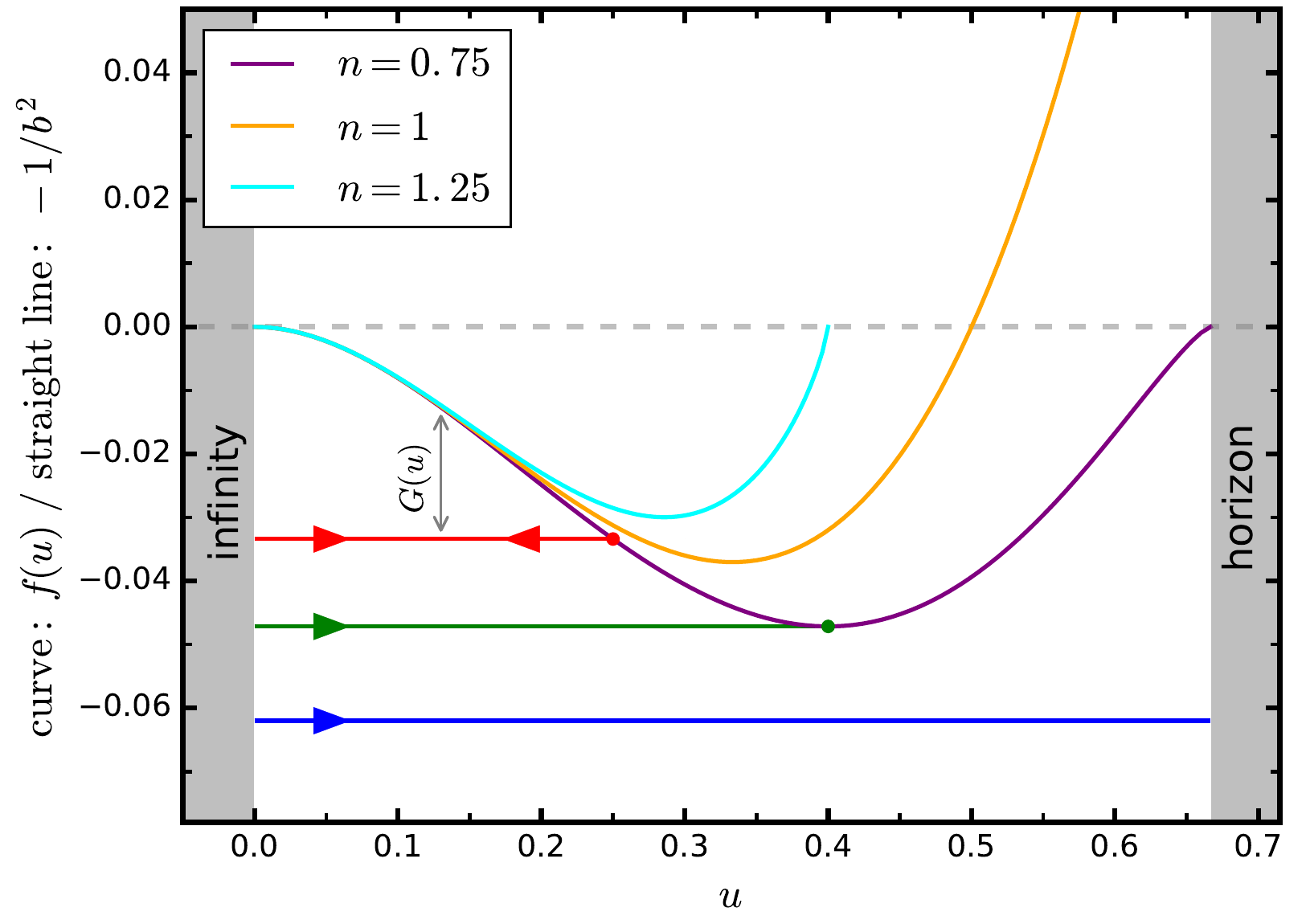}
  \caption{Representative plot shows the critical impact parameter. The value of $G(u)$ equals to the distance between the straight line and the curve line with constant $u$. Here we set $r_\textrm{s}=2$.}
  \label{fig:01}
\end{figure}

If there is a black hole between the observer and an infinity large background light source, then the observer will see a shadow caused by this black hole. For this scenario, \cite{Synge1966} calculated the shadow size of the Schwarzschild black hole. Here we calculate the shadow size for the PSM with the same method. Based on Eq. (\ref{eq:2.12}), we define $f(u)=-u^2(1-nr_\textrm{s}u)^{1/n}$. Fig. \ref{fig:01} plots $f(u)$ for $n=0.75,1,1.25$ and three representative photon orbits for $n=0.75$. Note that the light source is far from the black hole, which means $u=0$ initially. The left side of Eq. (\ref{eq:2.12}) is a square requires $G(u)>0$ along the orbit. In Fig. \ref{fig:01}, the red line, which corresponds to a big $b$, intersects $f(u)$ and the photon will back to $u=0$ after the intersection. Therefore, there will be photons with such large impact parameter observed by the observer. The blue line, which corresponds to a small $b$, do not intersect $f(u)$ and the photon will be absorbed by the black hole. So, no photon with such small impact parameter can be observed by the observer. The green line corresponds to the critical orbit and the photons will be trapped in the so called photon sphere. $f(u)$ takes its minimum value at $u=\frac{2}{(2n+1)r_\textrm{s}}$ and the minimum value $f_\textrm{min}=-\frac{4(2n+1)^{-2-\frac{1}{n}}}{r_\textrm{s}^2}$. The above discussions give the critical impact parameter
\begin{equation}\label{eq:3.01}
  b_\textrm{c}=\frac{1}{\sqrt{-f_\textrm{min}}}=\frac{(2n+1)^{1+\frac{1}{2n}}}{2}r_\textrm{s},
\end{equation}
and the shadow size equals to $\pi b_\textrm{c}^2$.

\section{Shadow: Accretion disk}\label{sec:04}
The scenario described in Sec. \ref{sec:03} is too ideal to compare with the EHT observations. The realistic light source should be an accretion disk orbiting around the black hole. In this paper, we assume the standard thin accretion disk model, i.e., the disk is optically thick and geometrically thin \cite{Shakura1973,Pringle1981,Chakrabarti1996,Abramowicz2013,Kato2008}. We also neglect the influence of the jet \cite{Punsly2019} on the black hole image. For the Schwarzschild metric, \cite{Novikov1973,Page1974} calculated the structure of an optically thick and geometrically thin accretion disk, and \cite{Luminet1979} simulated a photograph of such black hole-accretion disk system. In this section, we firstly analyze the disk radial structure in the PSM, which gives the intrinsic radiation flux from the disk surface. Then we discuss how to transfer the flux from the disk surface to the photographic plate, and plot the images for several examples. Especially, we take into account a projection effect that missed in \cite{Luminet1979}.

\subsection{Accretion disk}\label{sec:0401}
Our method mainly follows \cite{Novikov1973}, while there are minor differences in the details of dealing with the conservation of angular momentum. As the disk is assumed to be geometrically thin and the disk can be put on the equatorial plane, the metric can be written as
\begin{equation}
  \dx s^2=-A_n\dx t^2+A_n^{-1}\dx r^2+\dx z^2+r^2\dx\varphi^2.
\end{equation}
The energy-momentum tensor for the fluid with shear viscosity is \cite{Misner1973}
\begin{equation}\label{eq:4.02}
  T^{\mu\nu}=\rho u^\mu u^\nu-2\varepsilon\eta\sigma^{\mu\nu}+q^\mu u^\nu+q^\nu u^\mu,
\end{equation}
where the density $\rho=\rho(r,z)$, the coefficient of dynamic viscosity $\eta=\eta(r,z)$, the shear tensor
\begin{equation}
  \sigma^{\mu\nu}=\frac{1}{2}(u^\mu_{\ ;\alpha}P^{\alpha\nu}+u^\nu_{\ ;\alpha}P^{\alpha\mu})-\frac{1}{3}\theta P^{\mu\nu},
\end{equation}
the projection tensor $P^{\mu\nu}=g^{\mu\nu}+u^\mu u^\nu$, the isotropic expansion coefficient $\theta=u^\mu_{\ ;\mu}$, and $q^\mu$ is the heat-flux 4-vector. Here, we neglect the internal energy, which is valid as discussed in \cite{Novikov1973}. $\varepsilon$ is a mathematical infinitesimals, and one can set $\varepsilon=1$ after calculations. $\varepsilon$ is mathematical infinitesimal indicates the viscosity is weak, the inflow velocity is small, and the radiation is weak. Thus the disk is a nearly relativistic Kepler disk and the 4-velocity of the fluid element can be written as
\begin{align}\label{eq:4.04}
  u^\mu=\{u^t,\varepsilon u^r,0,u^\varphi\},
\end{align}
where $u^t$ and $u^\varphi$ are given by Eq. (\ref{eq:2.06}), and the inflow velocity $u^r=u^r(r,z)$. The thinness of the disk indicates $q^\mu$ is in the vertical direction \cite{Novikov1973}, which gives
\begin{align}\label{eq:4.05}
  q^\mu=\{0,0,\varepsilon q^z,0\},
\end{align}
where $q^z=q^z(r,z)$. The boundary condition is $q^z(r,h)=-q^z(r,-h)=F_\textrm{s}(r)$, where $h$ is the height of the disk and $F_\textrm{s}(r)$ is the surface radiation flux. In order to describe the radial structure of the disk, we can define the surface density
\begin{equation}
  \Sigma(r)\equiv\int_{-h}^{+h}\rho(r,z)\dx z,
\end{equation}
the mass-averaged inflow velocity
\begin{equation}
  \bar{u}^r(r)\equiv\frac{1}{\Sigma}\int_{-h}^{+h}\rho(r,z)u^r(r,z)\dx z,
\end{equation}
and the integral viscosity coefficient
\begin{equation}
  \bar{\eta}(r)\equiv\int_{-h}^{+h}\eta(r,z)\dx z.
\end{equation}
$\varepsilon$ appears in Eqs. (\ref{eq:4.02}), (\ref{eq:4.04}) and (\ref{eq:4.05}), and all the following calculations only need to preserve the $\mathcal{O}(\varepsilon)$ terms.

\begin{figure}[t]
  \centering
  \includegraphics[width=0.99\linewidth]{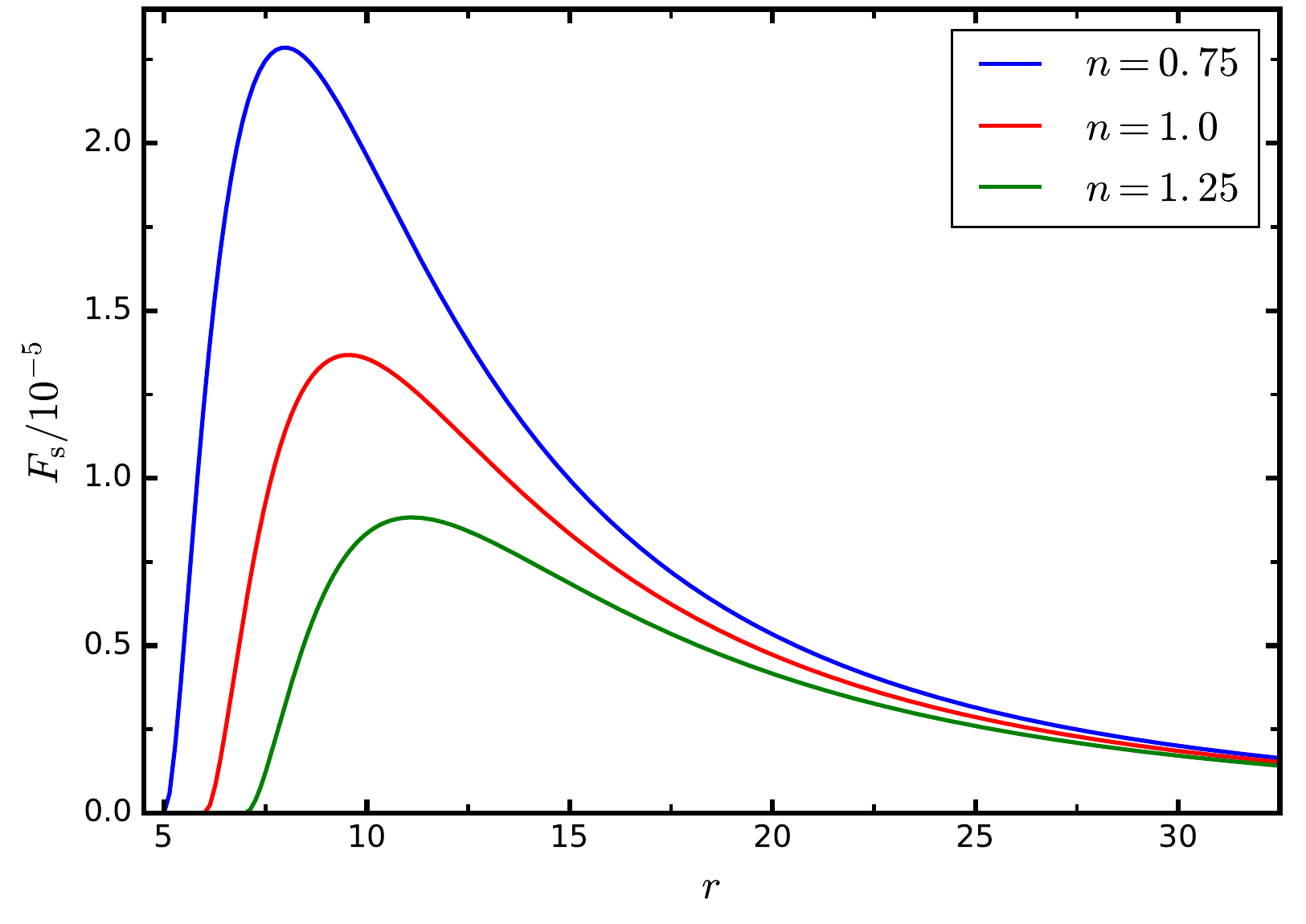}
  \caption{The surface radiation flux with $n=0.75,1,1.25$. Here we set $r_\textrm{s}=2$ and $\dot{M}=1$.}
  \label{fig:02}
\end{figure}

The conservation of rest mass $(\rho u^\mu)_{;\mu}=0$ gives
\begin{equation}
  \frac{\p(\rho u^rr)}{\p r}=0.
\end{equation}
Integrating the above equation in the $r$ and $z$-directions gives
\begin{equation}\label{eq:4.10}
  \dot{M}=-2\pi r\Sigma\bar{u}^r,
\end{equation}
where the integral constant $\dot{M}$ is the mass accretion rate.
\begin{widetext}\noindent
Energy-momentum conservation gives $J^\mu\equiv T^{\mu\nu}_{\ \ ;\nu}=0$. One can directly verify $J^r=\mathcal{O}(\varepsilon^2)$ and $J^z=\mathcal{O}(\varepsilon^2)$. The angular momentum conservation $J^\varphi=0$ gives
\begin{align}
  &r(nr_\textrm{s}-r)(2r-r_\textrm{ISCO})^2\frac{\p(\rho u^rr)}{\p r}
  +r(nr_\textrm{s}-r)(2r-r_\textrm{ISCO})(r-r_\textrm{ISCO})\rho u^r
  +r^2(nr_\textrm{s}-r)(2r-r_\textrm{ISCO})^2\frac{\p q^z}{\p z}\nonumber\\
  &+(4n^3r_\textrm{s}^3-14n^2r_\textrm{s}^2r+13nr_\textrm{s}r^2+nr_\textrm{s}^2r-3nr_\textrm{s}^3-3r^3-r_\textrm{s}r^2+4r_\textrm{s}^2r-r_\textrm{s}^3)A_n\eta\nonumber\\
  &+(nr_\textrm{s}-r)(3r-r_\textrm{ISCO})(2r-r_\textrm{ISCO})rA_n\frac{\p\eta}{\p r}=0,
\end{align}
where $r_\textrm{ISCO}$ is given by Eq. (\ref{eq:2.08}). Integrating the above equation in the $z$-direction, and substituting Eq. (\ref{eq:4.10}) into the result, we obtain
\begin{align}\label{eq:4.12}
  &\frac{\p\bar{\eta}}{\p r}+\frac{(4n^3r_\textrm{s}^3-14n^2r_\textrm{s}^2r+13nr_\textrm{s}r^2+nr_\textrm{s}^2r-3nr_\textrm{s}^3-3r^3-r_\textrm{s}r^2+4r_\textrm{s}^2r-r_\textrm{s}^3)}
  {r(nr_\textrm{s}-r)(3r-r_\textrm{ISCO})(2r-r_\textrm{ISCO})}\bar{\eta}\nonumber\\
  &\quad-\frac{\dot{M}(r-r_\textrm{ISCO})}{2\pi r(3r-r_\textrm{ISCO})A_n}
  +\frac{2r(2r-r_\textrm{ISCO})}{(3r-r_\textrm{ISCO})A_n}F_\textrm{s}=0.
\end{align}
In principle, we can regard $J^t=0$ as another constraint equation. However, as calculated in \cite{Novikov1973}, energy conservation $u_\mu J^\mu=0$ is a simpler approach, which gives
\begin{align}
  2r(nr_\textrm{s}-r)(2r-r_\textrm{ISCO})^2\frac{\p(\rho u^rr)}{\p r}
  +2r^2(nr_\textrm{s}-r)(2r-r_\textrm{ISCO})^2\frac{\p q^z}{\p z}
  +(3r-r_\textrm{ISCO})^2r_\textrm{s}A_n\eta=0.
\end{align}
Integrating the above equation in the $z$-direction, and substituting Eq. (\ref{eq:4.10}) into the result, we obtain
\begin{equation}\label{eq:4.14}
  F_\textrm{s}=\frac{(3r-r_\textrm{ISCO})^2r_\textrm{s}A_n\bar{\eta}}{4r^2(r-nr_\textrm{s})(2r-r_\textrm{ISCO})^2}.
\end{equation}
Substituting Eq. (\ref{eq:4.14}) into Eq. (\ref{eq:4.12}), we obtain
\begin{align}
  &\frac{\p\bar{\eta}}{\p r}+\frac{8n^3r_\textrm{s}^3-28n^2r_\textrm{s}^2r-4n^2r_\textrm{s}^3+26nr_\textrm{s}r^2+14nr_\textrm{s}^2r-10nr_\textrm{s}^3-6r^3-11r_\textrm{s}r^2+14r_\textrm{s}^2r-3r_\textrm{s}^3}
  {2r(nr_\textrm{s}-r)(2r-r_\textrm{ISCO})(3r-r_\textrm{ISCO})}\bar{\eta}\nonumber\\
  &\quad-\frac{\dot{M}(r-r_\textrm{ISCO})}{2\pi r(3r-r_\textrm{ISCO})A_n}=0,
\end{align}
which is a first-order ordinary differential equation for $\bar{\eta}(r)$. Solving the above equation with the boundary condition $\bar{\eta}(r_\textrm{ISCO})=0$ \cite{Novikov1973}, we obtain
\begin{equation}
  \bar{\eta}(r)=\frac{\dot{M}}{2\pi}\frac{(2r-r_\textrm{ISCO})\sqrt{r-nr_\textrm{s}}}{r(3r-r_\textrm{ISCO})A_n^{3/2}(r)}\cdot
  \int_{r_\textrm{ISCO}}^r\frac{(\tilde{r}-r_\textrm{ISCO})A_n^{1/2}(\tilde{r})}{(2\tilde{r}-r_\textrm{ISCO})\sqrt{\tilde{r}-nr_\textrm{s}}}\dx\tilde{r}.
\end{equation}
Substituting the above equation into Eq. (\ref{eq:4.14}), we obtain the surface radiation flux
\begin{equation}\label{eq:4.17}
  F_\textrm{s}(r)=\frac{\dot{M}}{8\pi}\frac{r_\textrm{s}(3r-r_\textrm{ISCO})}{r^3(2r-r_\textrm{ISCO})\sqrt{r-nr_\textrm{s}}A_n^{1/2}(r)}\cdot
  \int_{r_\textrm{ISCO}}^r\frac{(\tilde{r}-r_\textrm{ISCO})A_n^{1/2}(\tilde{r})}{(2\tilde{r}-r_\textrm{ISCO})\sqrt{\tilde{r}-nr_\textrm{s}}}\dx\tilde{r}.
\end{equation}
\end{widetext}
It is not hard to verify Eq. (\ref{eq:4.17}) is consistent with the result obtained in \cite{Novikov1973,Page1974} for the Schwarzschild metric. The integral part of Eq. (\ref{eq:4.17}) approximately equals to $\sqrt{r}$ when $r\gg r_\textrm{ISCO}$. Thus $\lim_{r\gg r_\textrm{ISCO}}F_\textrm{s}=3M\dot{M}/(8\pi r^3)$, which is consistent with the result in the Newtonian case \cite{Shakura1973,Kato2008}. Actually, the result obtained in \cite{Novikov1973,Page1974} (see Eq. (9) in \cite{Cardenas-Avendano2019} for a clearer expression) also applies to the PSM, and could directly give Eq. (\ref{eq:4.17}). We give the above detailed derivation because our calculations are more straightforward than the details presented in \cite{Novikov1973,Page1974}. Fig. \ref{fig:02} plots $F_\textrm{s}(r)$ for $n=0.75,1,1.25$, and shows the smaller horizon corresponds to the brighter disk.

\subsection{From disk to observer}\label{sec:0402}
In order to obtain the observed flux, we need to find the position and flux correspondences between the disk and the photographic plate as did in \cite{Luminet1979}. Solving the null geodesic gives the position correspondence. There are three factors that affect the observed flux for one specific point. One is the distance from the observer to the black hole. However, the leading terms of the distances from different disk elements to the observer are same, which means the distance does not affect the relative brightness at different points in the photographic plate. So, we can neglect this factor hereafter. The second one is the redshift including Doppler redshift and gravitational redshift. The third one is the projection effect that missed in the previous simulations, e.g., \cite{Luminet1979,Fukue1988,Kato2008,Gyulchev2019}. This effect is related to the fact that the photon trajectory is not perpendicular to the disk surface. For clarity, one can see the factor $\cos i$ appeared in Eq. (23) of \cite{Horne1986} or Eq. (3.80) of \cite{Kato2008} for the Newtonian case.

\begin{figure}[t]
  \centering
  \includegraphics[width=0.99\linewidth]{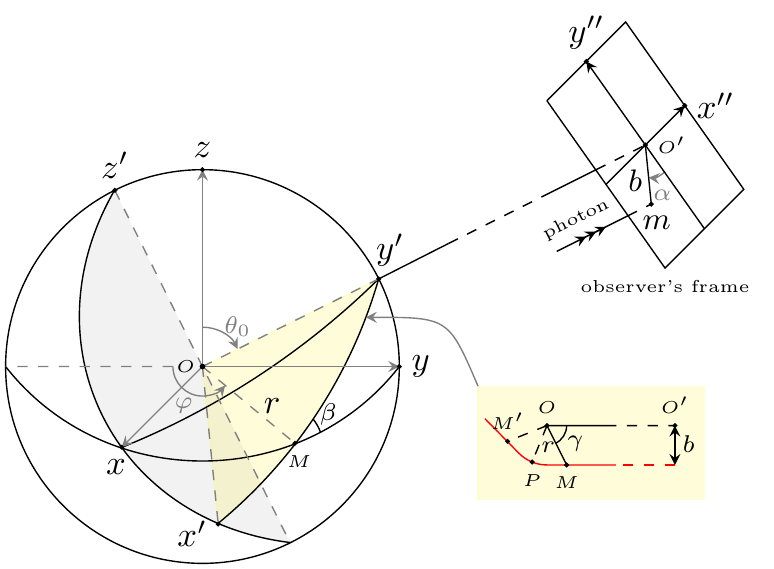}
  \caption{The coordinate system (see main text). The design of this figure is mainly taken from \cite{Luminet1979}.}
  \label{fig:03}
\end{figure}

The coordinate system is plotted in Fig. \ref{fig:03}. The design is mainly taken from \cite{Luminet1979}, and we make minor modifications. The radius of the sphere equals to $r$. The black hole is placed at point $o$ and the disk is placed on the $\overline{xoy}$ plane. Point $M$ means a disk element, and the light it emits reaches point $m$ in the photographic plate. The corresponding photon trajectory is in the yellow plane, and the photon trajectory is parallel to $\overline{oo'}$ when the photon is far from the black hole. The coordinates of $M$ are marked by $(r,\varphi)$, and the coordinates of $m$ are marked by $(b,\alpha)$. In the bottom right subgraph, the red line is the photon trajectory, point $P$ is the perihelion of the photon trajectory with the impact parameter $b$, and point $M'$ satisfies $\overline{oM'}=\overline{oM}=r$ but is on the other side of the perihelion. Note that the light emitted from $M'$ does not pass through $M$ (Do not be disturbed by this subgraph). In the figure, we also point out the start and positive directions of the angle $\varphi$, $\theta$ (i.e., $\theta_0$) and $\alpha$. $\beta$ and $\gamma$ means the value of the corresponding angle with no positive direction. It is clear that $\theta_0\in[0,\pi/2]$, $\alpha\in[0,2\pi]$, $\varphi\in[0,2\pi]$ and $\gamma=[\pi/2-\theta_0,\pi/2+\theta_0]$. Several useful parallel and perpendicular relations: $\overline{xoz'}\parallel\overline{x''o'y''}$, $\overline{o'm}\parallel\overline{ox'}$, $\overline{oo'}\perp\overline{xoz'}$ and $\wideparen{My}\perp\wideparen{yy'}$. Here $\overline{AB}$ means a line, $\overline{ABC}$ means a plane and $\wideparen{AB}$ means an arc. We denote $\widehat{ABC}$ as an angle hereafter.

\subsubsection{Point to point}\label{sec:040201}
At first, we assume $\alpha\in(0,\pi/2)$. Using the sine theorem of spherical triangle in $\Delta Myy'$, we obtain $\sin\widehat{Myy'}/\sin(\wideparen{My'}/r)=\sin\widehat{y'My}/\sin(\wideparen{yy'}/r)$. Substituting $\widehat{Myy'}=\pi/2$, $\wideparen{My'}/r=\gamma$, $\widehat{y'My}=\beta$ and $\wideparen{yy'}/r=\pi/2-\theta_0$ into this result, we obtain $\sin\beta=\cos\theta_0/\sin\gamma$. The parallel and perpendicular relations in Fig. \ref{fig:03} gives $\widehat{x'oM}=\pi/2-\gamma>0$, $\widehat{xox'}=\pi/2-\alpha>0$, $\widehat{Mxx'}=\theta_0$ and $\widehat{xMx'}=\beta$. Using the sine theorem of spherical triangle in $\Delta Mxx'$, we obtain $\sin\widehat{Mxx'}/\sin\widehat{x'oM}=\sin\widehat{xMx'}/\sin\widehat{xox'}$, i.e., $\sin\theta_0/\cos\gamma=\sin\beta/\cos\alpha$. Eliminating $\sin\beta$ and using $\sin\gamma>0$, we obtain
\begin{equation}\label{eq:4.18}
  \cos\gamma=\cos\alpha(\cos^2\alpha+\cot^2\theta_0)^{-1/2}.
\end{equation}
One can easily verify Eq. (\ref{eq:4.18}) holds for $\alpha\in[0,2\pi]$ following the above proof. \cite{Luminet1979} showed this result directly without proof. So we present detailed proof here.

Eq. (\ref{eq:2.12}) gives the deflection angle of photons from $M$ to the observer
\begin{equation}\label{eq:4.19}
  \gamma=\int_0^{1/r}\frac{\dx u}{\sqrt{G(u)}},
\end{equation}
in which $b$ is a parameter. We can obtain $\alpha=\alpha(\gamma)$ and $b=b(\gamma)$ from Eq. (\ref{eq:4.18}) and Eq. (\ref{eq:4.19}), respectively. Then we obtain $b=b(\alpha)$ for a given $r$, which is the iso-radial curve (see the definition in \cite{Luminet1979}) as seen by the observer. Numerically we can do this by interpolation. One thing worth mentioning here is that one $b$ corresponding to two $\gamma$ as shown in Fig. \ref{fig:03}. Actually, Eq. (\ref{eq:4.19}) gives $\gamma(M)$ directly because $M$ is on the right side of the perihelion $P$. For $\gamma(M')$, we can calculate its value based on symmetry. At first, we should find the distance from $P$ to $o$, which is given by $G(1/r)=0$. $P$ corresponds to the intersection of the red line and the purple curve in Fig. \ref{fig:01}. Then we can calculate the deflection angle from $P$ to the observer using $\gamma(P)=\int_0^{1/r(P)}\dx u/G(u)$. The PSM is spherically symmetric and static, and the photon orbit should be symmetrical about the line $\overline{oP}$. So $\gamma(M')=2[\gamma(P)-\gamma(M)]+\gamma(M)$. One $b$ corresponding to two $\gamma$ does not influence the numerical interpolation because one $\gamma$ only corresponding to one $b$. In the numerical calculation, we use the above method to obtain $b(\alpha)$ for $\alpha\in[0,\pi]$, and then use the symmetry to calculate $b(\alpha)$ for $\alpha\in[\pi,2\pi]$.

\subsubsection{Redshift}
As discussed in \cite{Luminet1979,Ellis1971,*Ellis2009}, the redshift will contribute a factor $(1+z)^{-4}$ to the observed flux. Here we denote $p^\mu$ as the 4-momentum of photons. The PSM is independent of $t$ and $\varphi$ gives $p_t$ and $p_\varphi$ are conserved. However, the photon orbit is not in the equatorial plane, and we cannot set $\theta=\pi/2$ anymore. In the rest frame of the emitting disk element, the photon energy is \cite{Luminet1979}
\begin{equation}
  E_\textrm{em}=p_\mu u^\mu=p_tu^t+p_\varphi u^\varphi,
\end{equation}
where $u^t$ and $u^\varphi$ are given by Eq. (\ref{eq:2.06}). The only nonzero component of the observer's 4-velocity is $u^t_\textrm{ob}=1$, and the observed energy of the same photon is $E_\textrm{ob}=p_\mu u^\mu_\textrm{ob}=p_t$. Thus the redshift of photon is given by
\begin{align}\label{eq:4.21}
  1+z&=\frac{E_\textrm{em}}{E_\textrm{ob}}=u^t+u^\varphi\frac{p_\varphi}{p_t}
  =u^t+u^\varphi\frac{g_{\varphi\varphi}}{g_{tt}}\frac{\dx\varphi}{\dx t}\nonumber\\
  &\xlongequal{r\gg r_\textrm{s}}u^t-u^\varphi r^2\sin^2\theta_0\frac{\dx\varphi}{\dx t}.
\end{align}
In the second line of the above equation, we estimate the value of $p_\varphi/p_t$ at $r\gg r_\textrm{s}$. This is valid because $p_\varphi/p_t$ is conserved along the photon orbit. Note that, in Eq. (\ref{eq:4.21}), $u^{t,\varphi}=u^{t,\varphi}(r_{\rm disk-element})$, and the notation $r_{\rm disk-element}$ is different with $r$. Projecting the photon trajectory onto the equatorial plane, one can prove
\begin{equation}\label{eq:4.22}
  \left.\frac{\dx\varphi}{\dx t}\right|_{r\gg r_\textrm{s}}=\frac{b\sin\alpha}{(r^2-b^2)\sin\theta_0}.
\end{equation}
Substituting Eq. (\ref{eq:4.22}) into Eq. (\ref{eq:4.21}), we obtain
\begin{align}\label{eq:4.13}
  1+z=u^t-u^\varphi b\sin\theta_0\sin\alpha.
\end{align}
For a given $r$ in the disk and any given point $(b,\alpha)$ in the photographic plate, Eq. (\ref{eq:4.13}) gives the corresponding redshit. Note that there is a difference of one sign between Eq. (\ref{eq:4.13}) and Eq. (19) in \cite{Luminet1979}, which is due to the different definitions of the positive directions.

\subsubsection{Projection}
\begin{figure*}[t]
  \centering
  \includegraphics[width=0.329\linewidth]{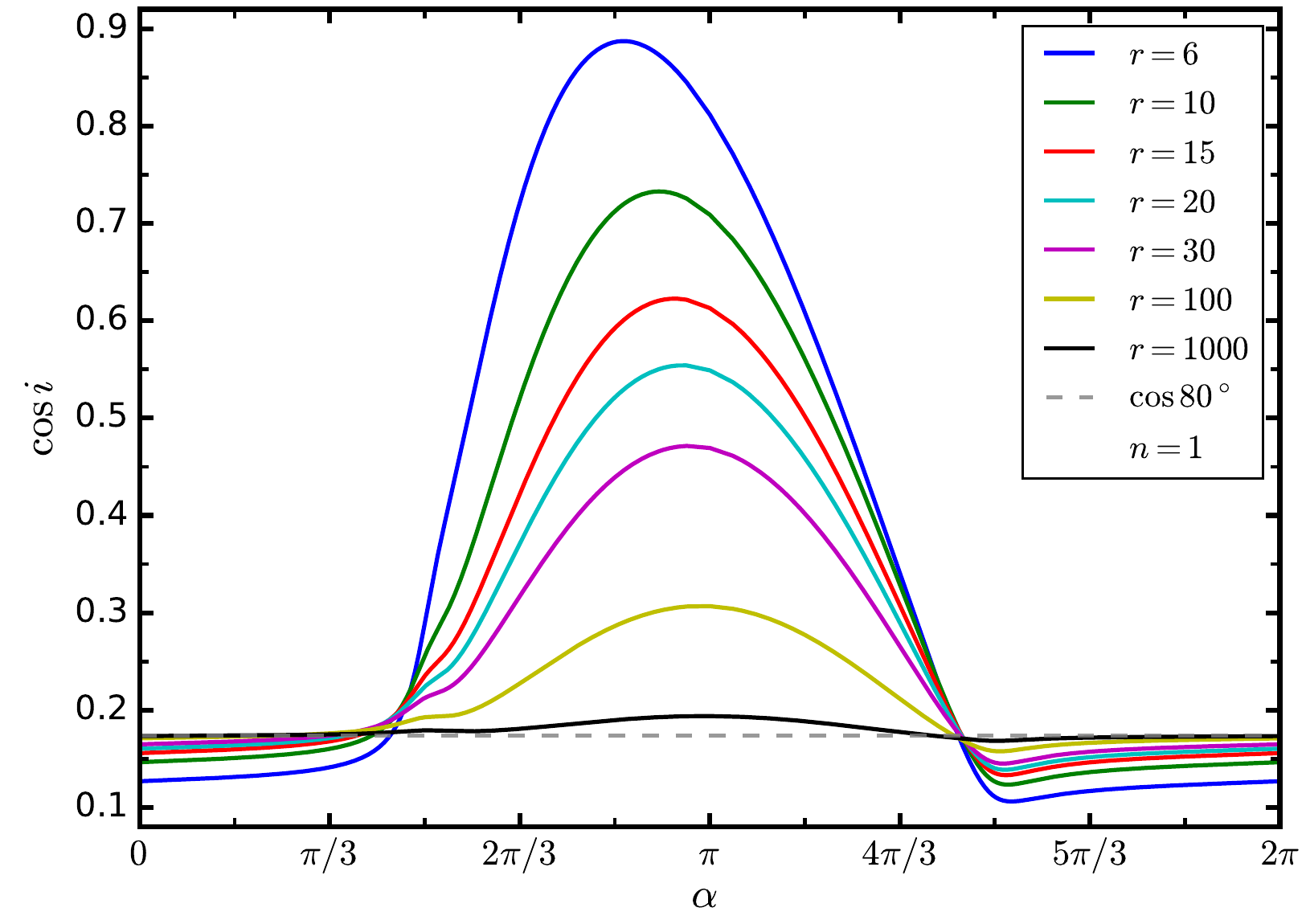}
  \includegraphics[width=0.329\linewidth]{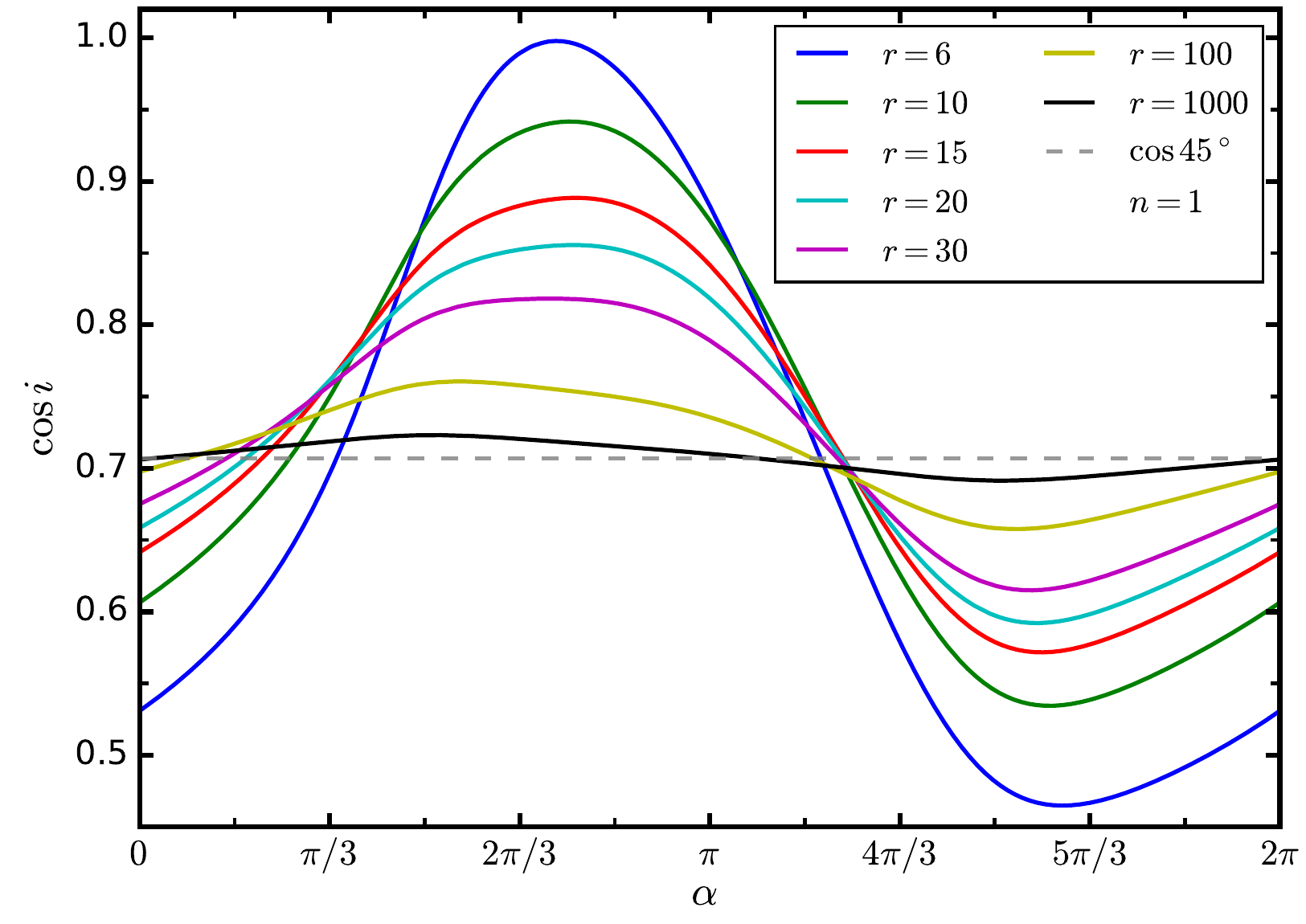}
  \includegraphics[width=0.329\linewidth]{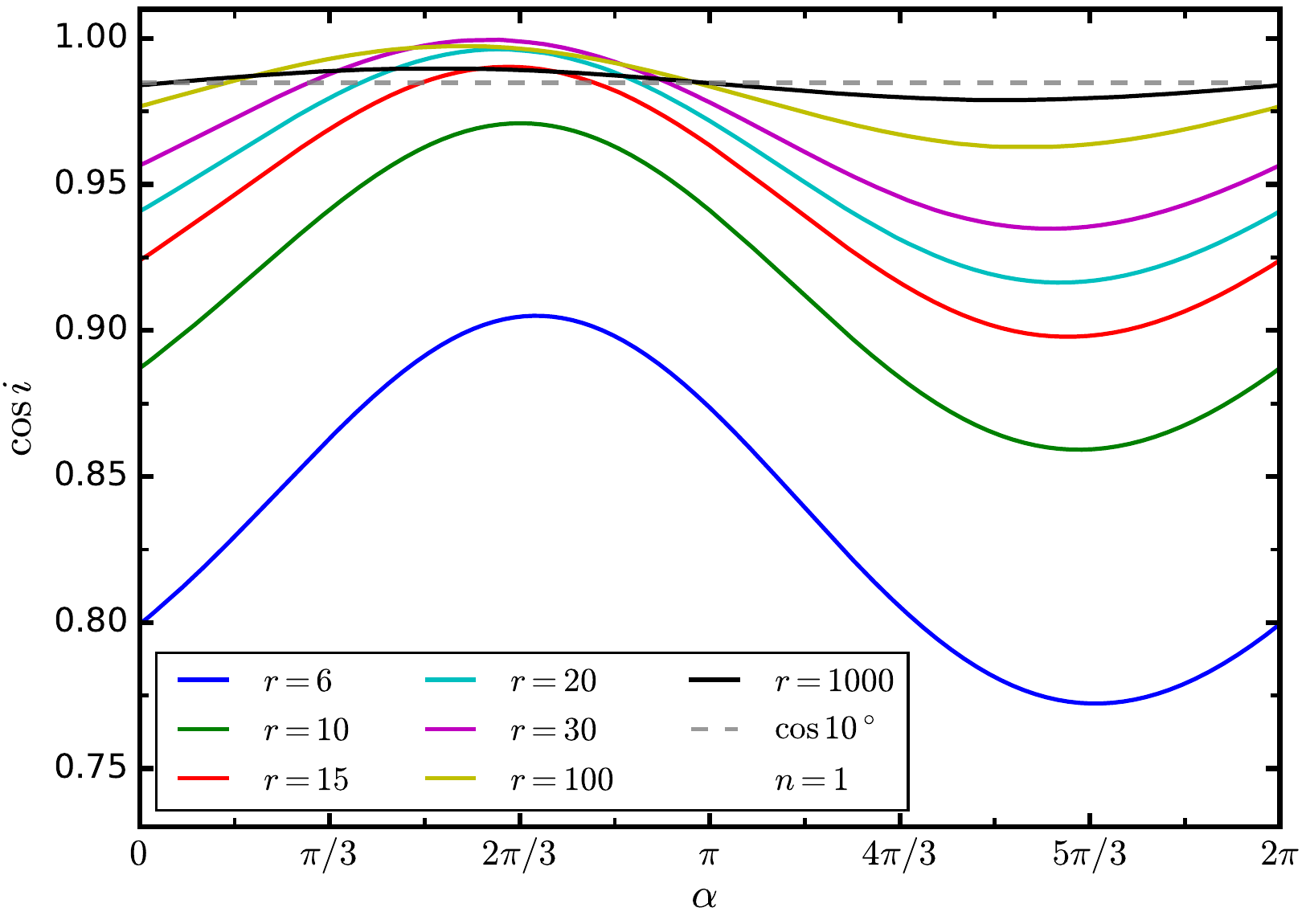}
  \caption{The projection factor $\cos i$ for $n=1$ and $\theta_0=80^\circ\textrm{(left)},\,45^\circ\textrm{(middle)},\,10^\circ\textrm{(right)}$. Here we set $r_\textrm{s}=2$.}
  \label{fig:04}
\end{figure*}

We can calculate the projection factor $\cos i$ in the rest frame of the disk element, where $i$ is the angle between the normal line of the disk surface and the photon trajectory in the three-dimensional space. The 4-velocity of disk element is given by Eq. (\ref{eq:4.04}) with $u^r=0$. Then, we should find the 4-momentum of photon when it just leaves the disk surface. $p_t$ and $p_\varphi$ are conserved gives
\begin{subequations}\label{eq:4.24}
\begin{align}
  p^t&=\frac{\dx t}{\dx\lambda}=\frac{E}{A_n},\\
  p^\varphi&=\frac{\dx\varphi}{\dx\lambda}=\frac{L}{r^2\sin^2\theta}.
\end{align}
\end{subequations}
Note that $\theta$ varies from $\pi/2$ to $\theta_0$ when the photon moves from the disk to the observer, and hence $p^\theta\neq0$. Here the constants $E$ and $L$ are different with the same symbols appeared in Eq. (\ref{eq:2.09}). Combined Eq. (\ref{eq:4.24}) with Eq. (\ref{eq:4.22}), we obtain $L/E=b\sin\theta_0\sin\alpha$. The disk is assumed to be geometrically thin. When the photon just leaves the disk surface, $p_\mu p^\mu=0$ gives
\begin{equation}
  A_n^{-1}(p^r)^2+r^2(p^\theta)^2+r^2(p^\varphi)^2=E^2A_n^{-1}.
\end{equation}
In addition, based on Fig. \ref{fig:03}, we know $\tan\beta=|\dx\theta/\dx\varphi|$, where $\beta$ is given by $\sin\beta=\cos\theta_0/\sin\gamma$ (see Sec. \ref{sec:040201}). This result gives $(p^\theta)^2=(p^\varphi)^2\tan^2\beta$. The above results are sufficient to calculate $\cos i$. The coordinate transformation ($x^\mu\rightarrow x'^\mu$: $t\rightarrow t'=t/u^t$, $r\rightarrow r'=r$, $\theta\rightarrow\theta'=\theta$, $\varphi\rightarrow\varphi'=\varphi-t\cdot u^\varphi/u^t$) gives the rest frame of the disk element, in which $u'^\mu=\frac{\p x'^\mu}{\p x^\alpha}u^\alpha=\{1,0,0,0\}$. In the $x'^\mu$ coordinate system, the space components of the photon momentum are $p'^r=p^r$, $p'^\theta=p^\theta$, and $p'^\varphi=p^\varphi-p^tu^\varphi/u^t$. Thus the projection factor
\begin{align}\label{eq:4.26}
  &\cos i=\frac{|rp'^\theta|}{\sqrt{r^2|p'^\theta|^2+r^2|p'^\varphi|^2+A_n^{-1}|p'^r|^2}}\nonumber\\
  &=\frac{|rp^\varphi\tan\beta|}{\sqrt{E^2A_n^{-1}-2r^2p^\varphi p^tu^\varphi/u^t
  +(rp^tu^\varphi/u^t)^2}},
\end{align}
where $p^t$ and $p^\varphi$ are given by Eq. (\ref{eq:4.24}) with $\theta=\pi/2$, and $u^t$ and $u^\varphi$ are given by Eq. (\ref{eq:2.06}). Note that the value of $\cos i$ is independent of the choice of $E$ as we expected. And we set $E=1$ in the numerical calculations. As shown in Sec. \ref{sec:040201}, one $\alpha$ corresponds to one $\gamma$ and one $\beta$. So far, for a given $r$ and $(b,\alpha)$, we can calculate the value of $\cos i$. Fig. \ref{fig:04} plots $\cos i-\alpha$ for $n=1$ and $\theta_0=80^\circ,45^\circ,10^\circ$, which shows 1) $\lim_{r\gg r_\textrm{s}}\cos i=\cos\theta_0$, i.e., Eq. (\ref{eq:4.26}) is consistent with the result obtained in the flat spacetime as we expected; 2) large $\theta_0$ causes large relative change of $\cos i$, which means the influence of the projection effect will be enhanced if $\theta_0$ is increased. The curve in Fig. \ref{fig:04} is not symmetric about $\alpha=\pi$ because of the Kepler motion of the disk.

Finally, the observed flux is given by
\begin{equation}\label{eq:4.27}
  F_\textrm{ob}=\frac{F_\textrm{s}\cos i}{(1+z)^4}.
\end{equation}
In the simulated figures, we use $\textrm{max}(F_\textrm{ob})$ to rescale the observed flux.

\subsection{Results and discussions}\label{sec:0403}
\begin{figure}[t]
  \centering
  \includegraphics[width=0.99\linewidth]{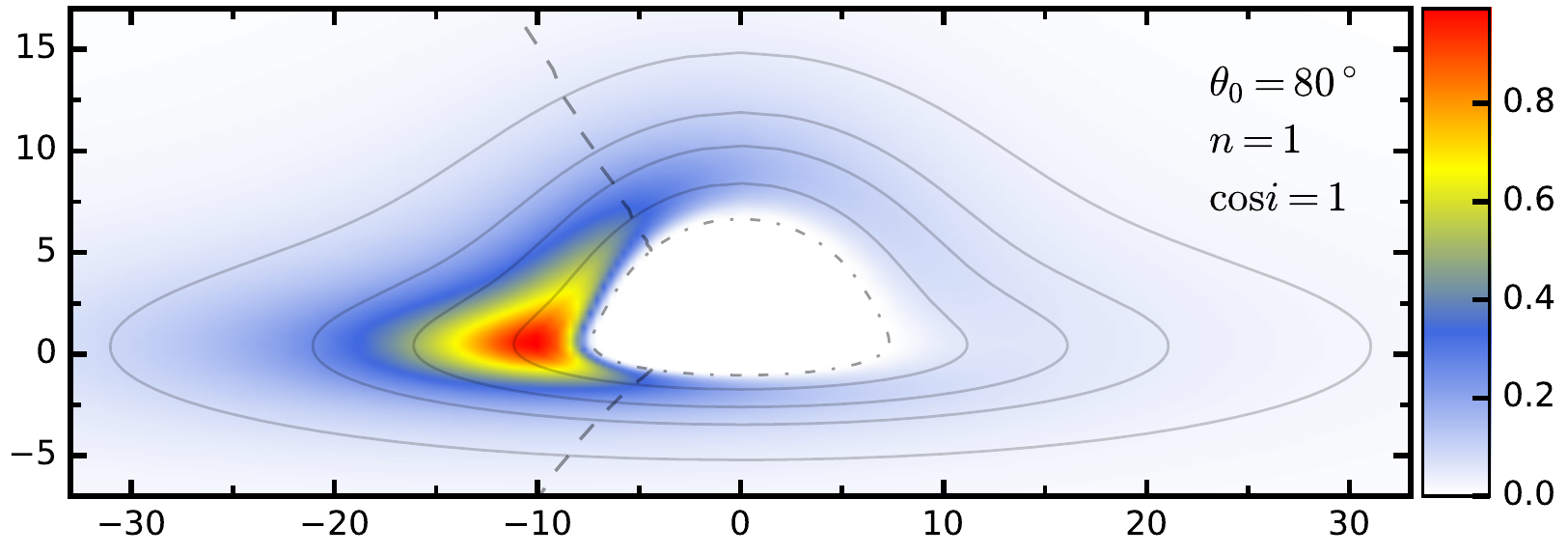}
  \caption{The simulated image without considering the projection effect. The innermost dash-dotted loop corresponds to $r=r_\textrm{ISCO}$, and the outer solid loops correspond to $r=10,15,20,30$, respectively. The dashed line corresponds to $z=0$. The left side of this dashed line has $z<0$ (blueshift), and the right side has $z>0$ (redshift). $\cos i=1$ indicates this figure do not take into account the projection effect. Here, the horizontal and vertical directions correspond to the $x''$ and $y''$ axes in Fig. \ref{fig:03}, respectively. We set $r_\textrm{s}=2$.}
  \label{fig:05}
\end{figure}

\begin{figure}[t]
  \centering
  \includegraphics[width=0.99\linewidth]{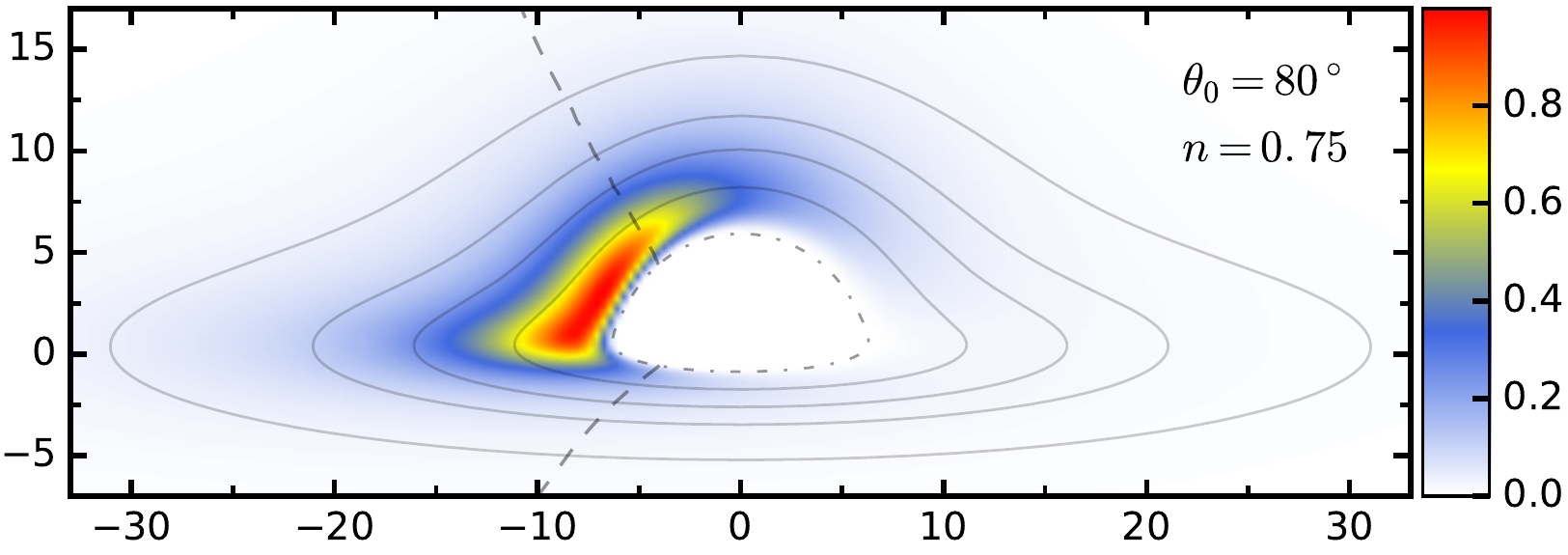}\\
  \includegraphics[width=0.99\linewidth]{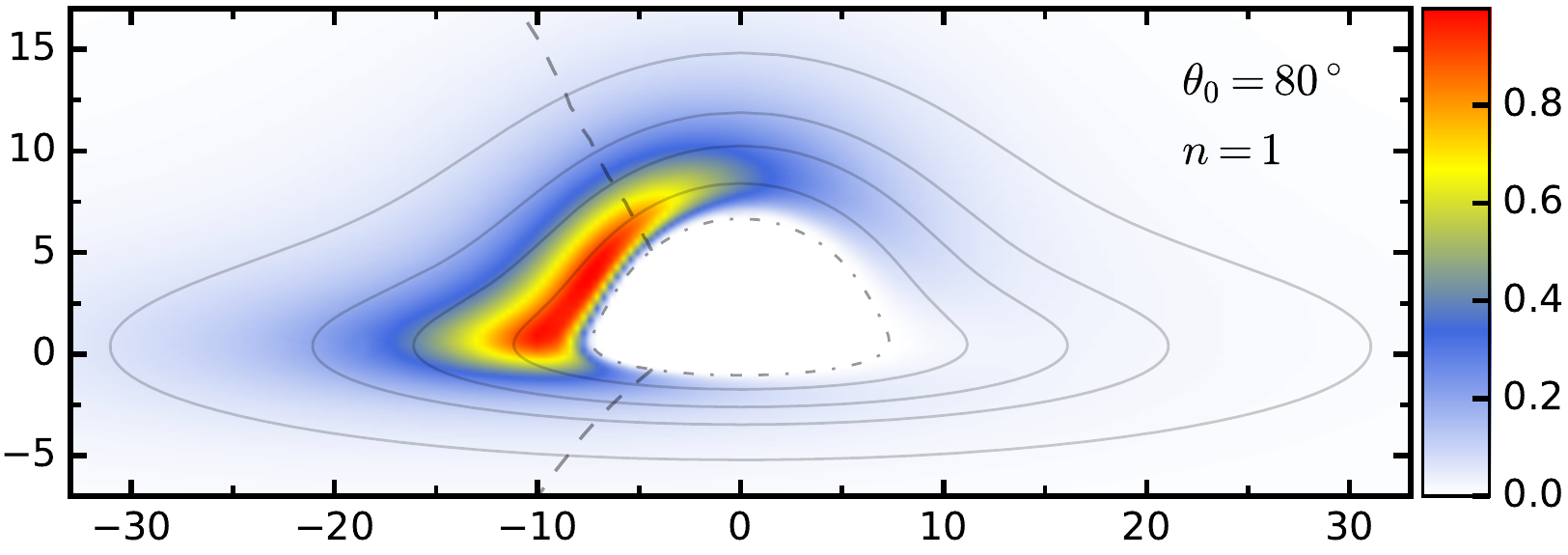}\\
  \includegraphics[width=0.99\linewidth]{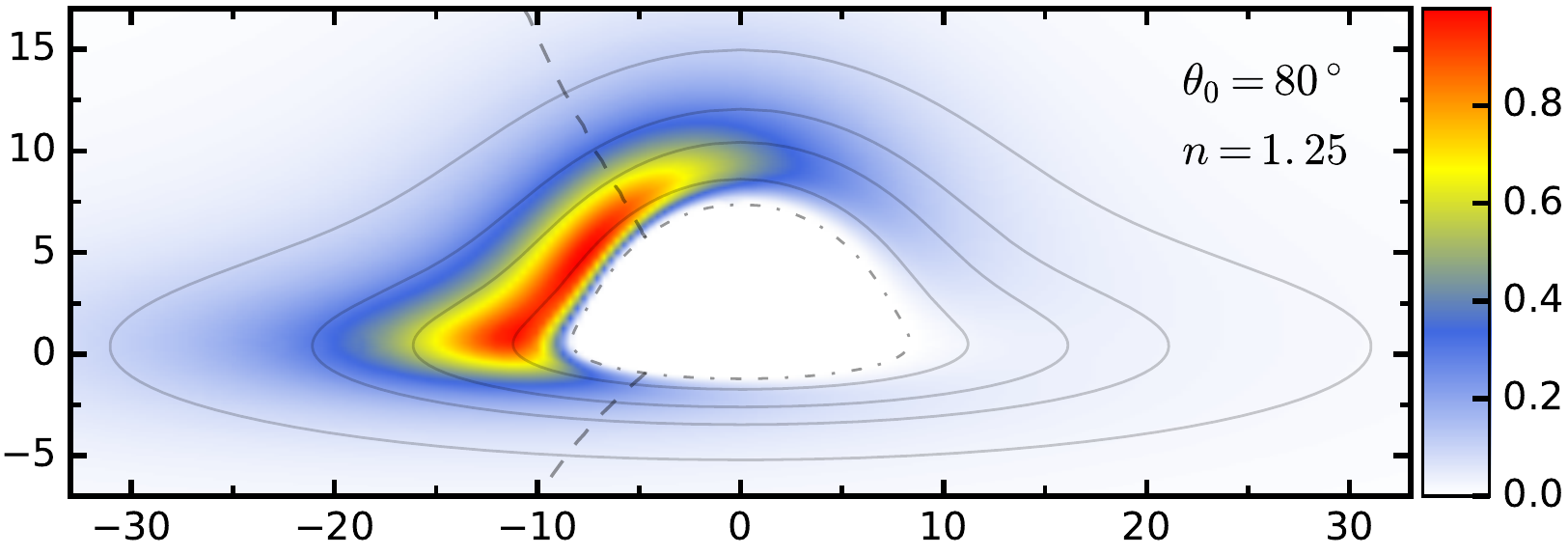}
  \caption{The simulated image with considering the projection effect for $\theta_0=80^\circ$ and $n=0.75\,\textrm{(top)},\,1\,\textrm{(middle)},\,1.25\,\textrm{(bottom)}$. The other marks are the same as in Fig. \ref{fig:05}.}
  \label{fig:06}
\end{figure}

\begin{figure*}[t]
  \centering
  \includegraphics[width=0.32\linewidth]{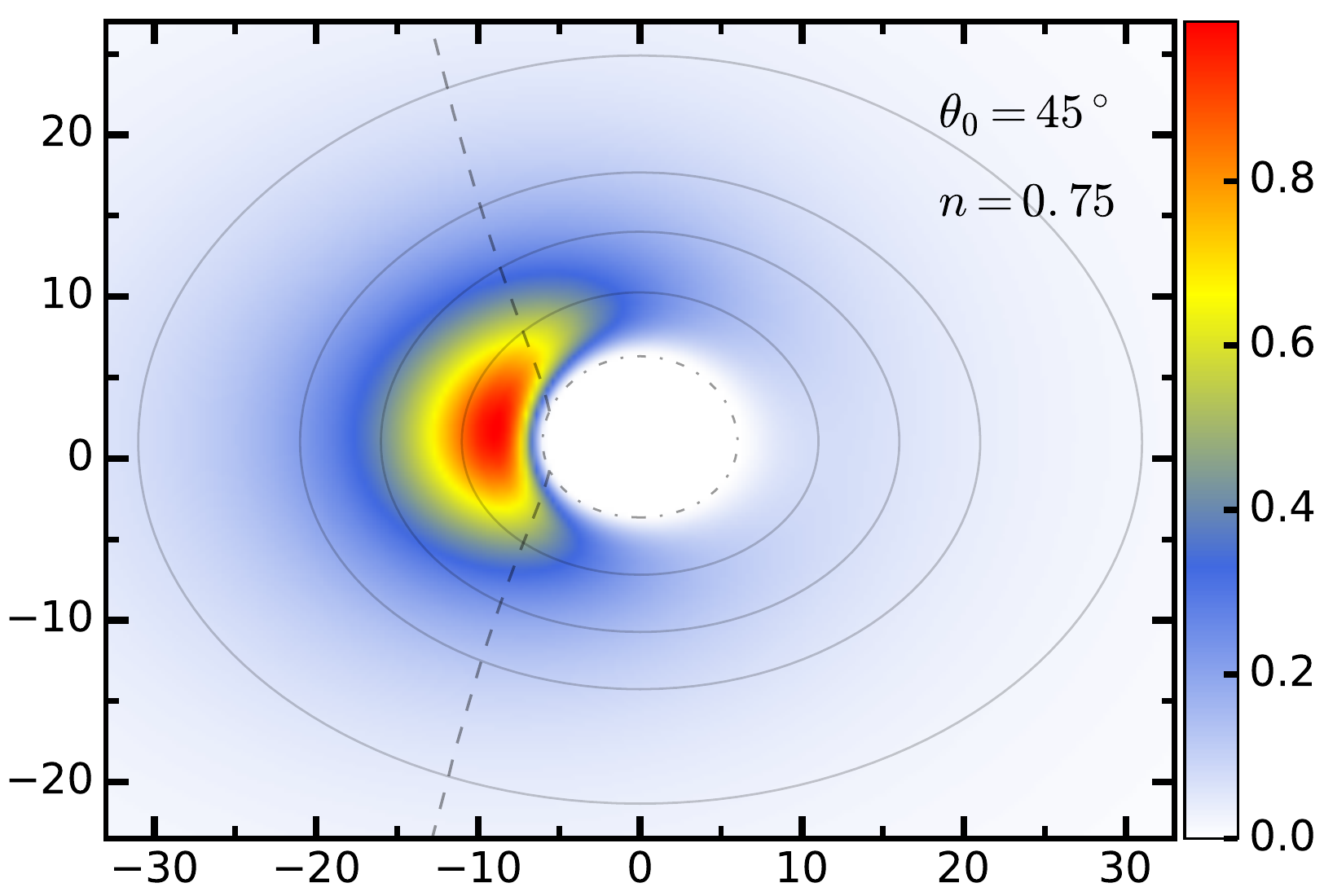}
  \includegraphics[width=0.32\linewidth]{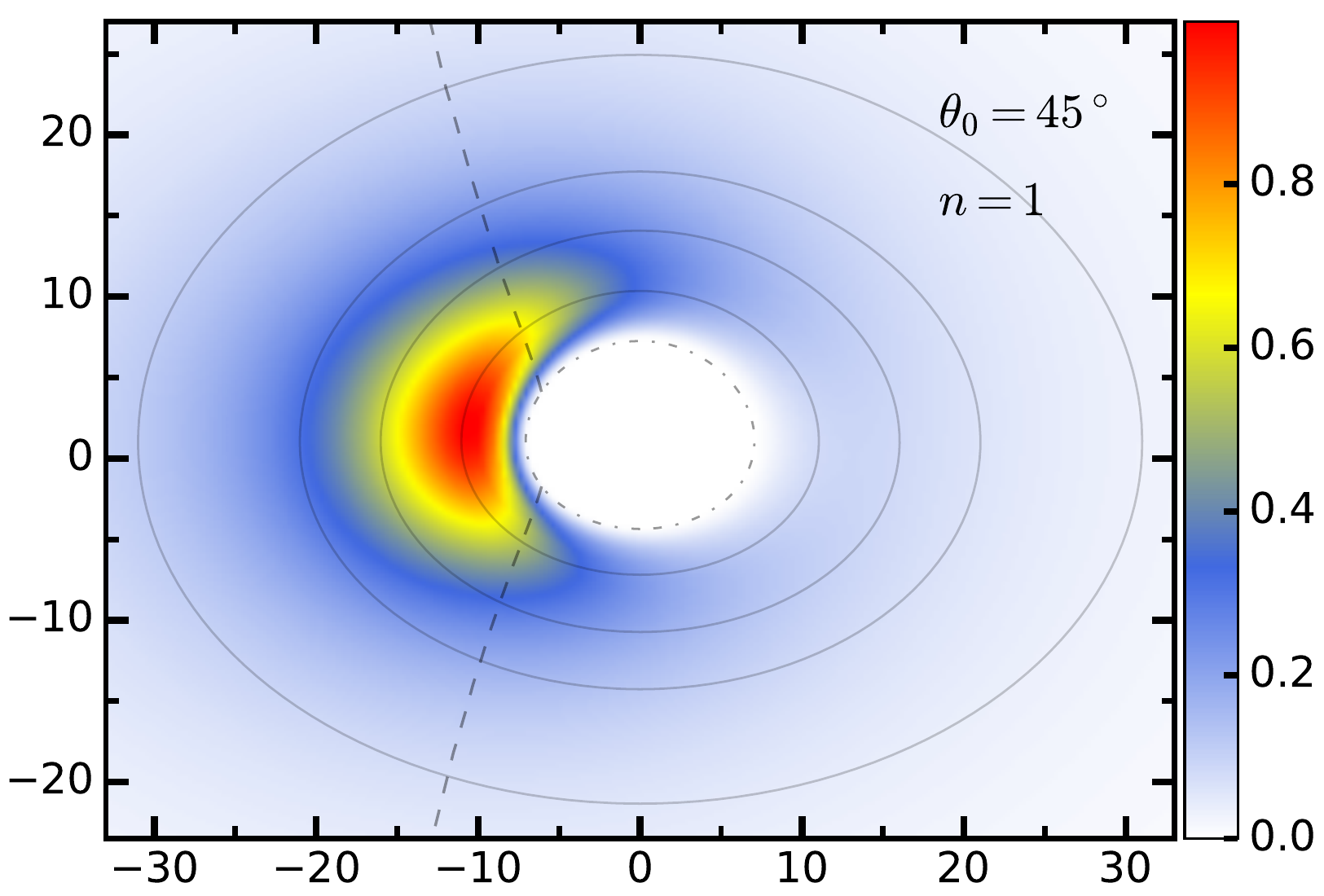}
  \includegraphics[width=0.32\linewidth]{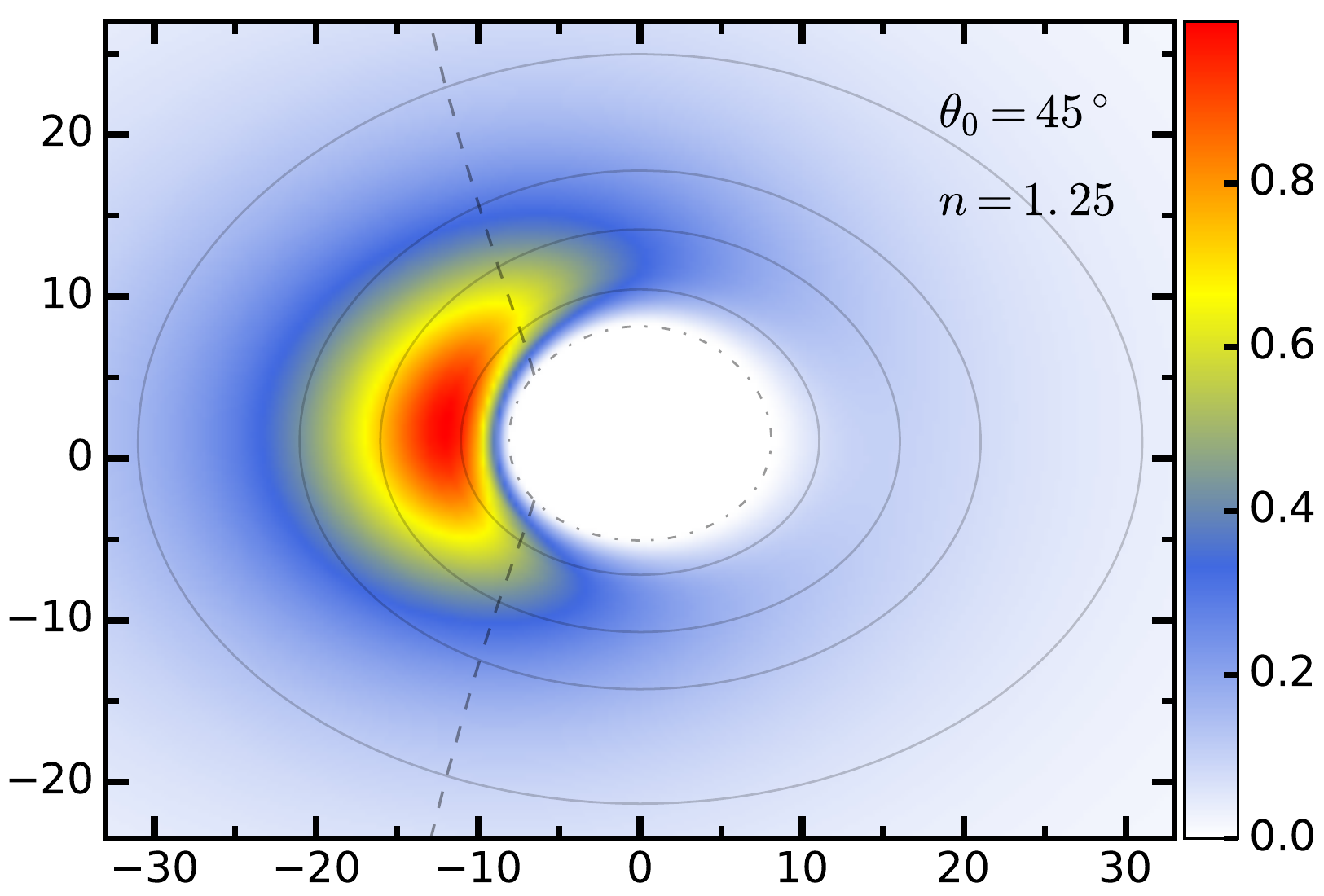}
  \caption{The simulated image with considering the projection effect for $\theta_0=45^\circ$ and $n=0.75\,\textrm{(left)},\,1\,\textrm{(middle)},\,1.25\,\textrm{(right)}$. The other marks are the same as in Fig. \ref{fig:05}.}
  \label{fig:07}
\end{figure*}

\begin{figure*}[t]
  \centering
  \includegraphics[width=0.32\linewidth]{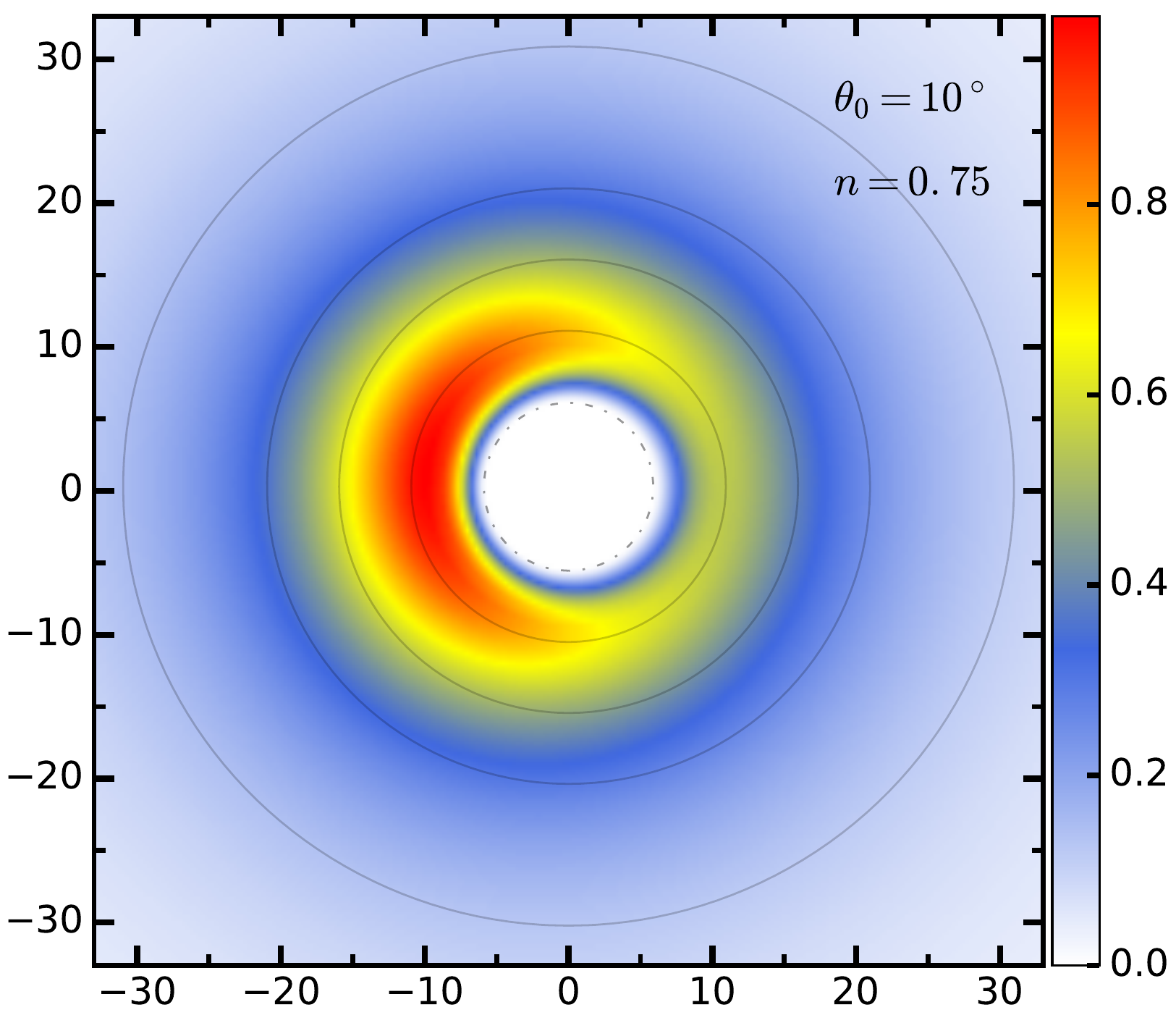}
  \includegraphics[width=0.32\linewidth]{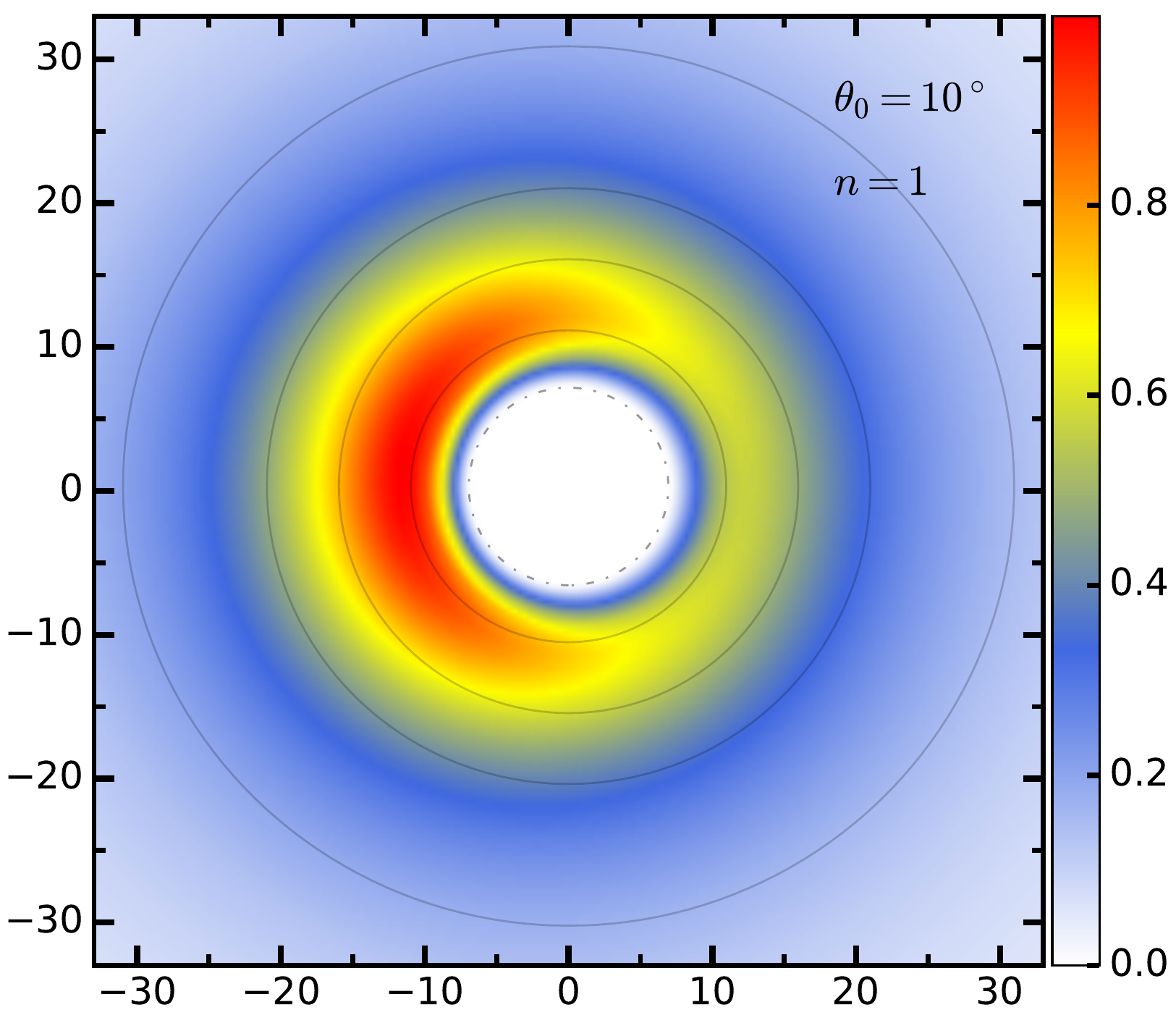}
  \includegraphics[width=0.32\linewidth]{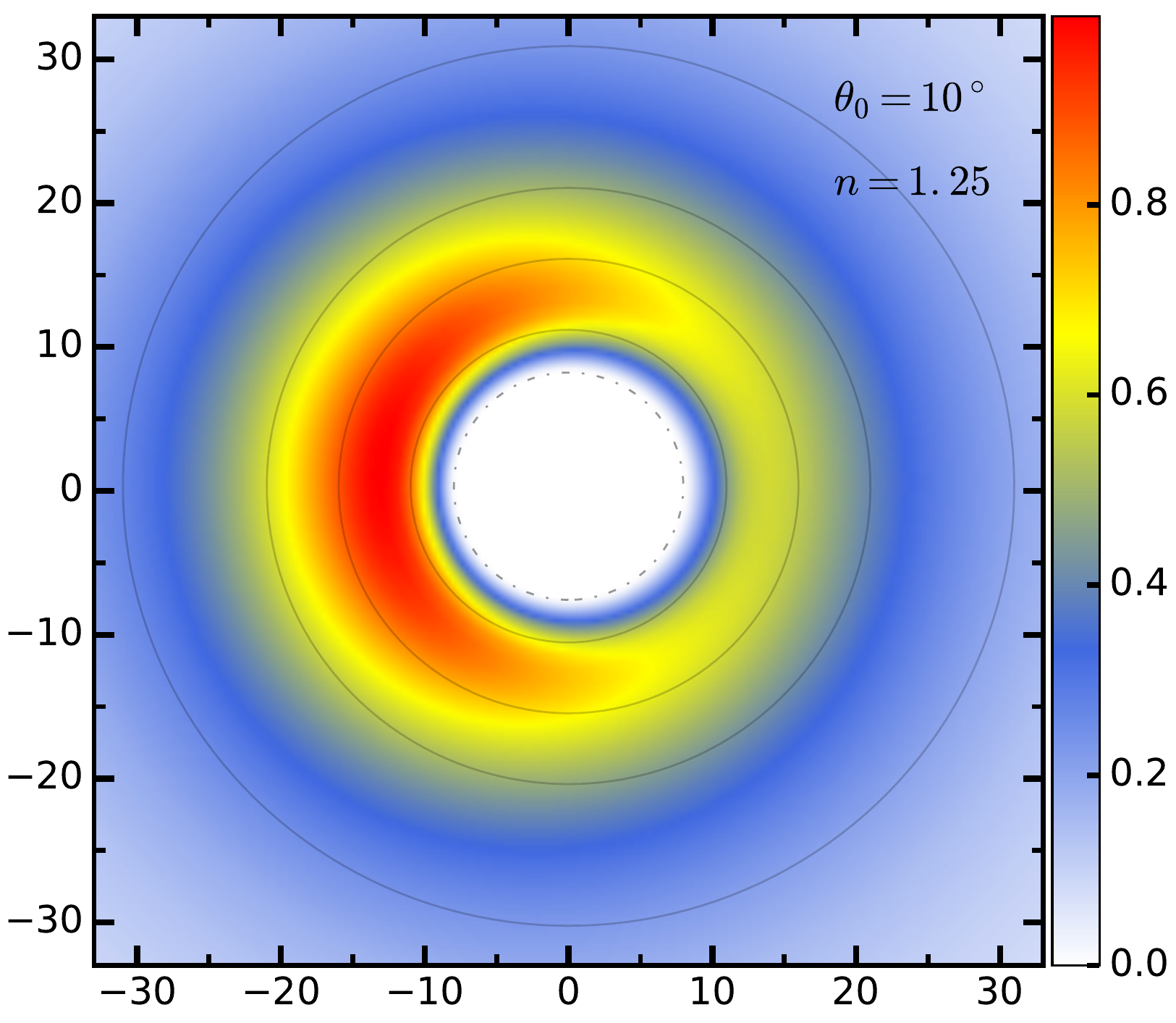}
  \caption{The simulated image with considering the projection effect for $\theta_0=10^\circ$ and $n=0.75\,\textrm{(left)},\,1\,\textrm{(middle)},\,1.25\,\textrm{(right)}$. The other marks are the same as in Fig. \ref{fig:05}. Here no point corresponds to $z=0$ and all the region we are considering has $z>0$.}
  \label{fig:08}
\end{figure*}

We only consider the direct image of the disk because the disk is assumed to be optically thick. Fig. \ref{fig:05} plots the simulated image for $\theta_0=80^\circ$ and $n=1$ without considering the projection effect, which is exactly Figure 11 in \cite{Luminet1979}. Comparing Fig. \ref{fig:05} with the middle one in Fig. \ref{fig:06}, we know the projection effect extends the bright area from the negative $x''$-axis to the positive $y''$-axis. This is consistent with the left panel of Fig. \ref{fig:04}, which shows the positive $y''$-axis corresponds to the bigger $\cos i$. Fig. \ref{fig:06} also plots the simulated image for $\theta_0=80^\circ$ and $n=0.75,1.25$, and shows larger $n$ corresponds to larger shadow size as we expected. However, the shape of the black hole shadow is approximately independent of the value of $n$. This conclusion is also confirmed in Fig. \ref{fig:07} and Fig. \ref{fig:08}, which plot the simulated image for $\theta_0=45^\circ$ and $\theta_0=10^\circ$, respectively.

The mass of the black hole M87* given by the EHT is $M_\textrm{shadow-method}=(6.5\pm0.7)\times10^9M_\odot$ \cite{Akiyama2019a}. There are two other ways to measure the mass of M87*: stellar dynamics measurements yield $M_\textrm{stellar-method}=6.2^{+1.1}_{-0.6}\times10^9M_\odot$ \cite{Gebhardt2009,Gebhardt2011,Akiyama2019f} and gas dynamics observations give $M_\textrm{gas-method}=3.5^{+0.9}_{-0.3}\times10^9M_\odot$ \cite{Harms1994,Macchetto1997,Walsh2013,Akiyama2019f}. The shadow method is performed in the horizon scale and mainly depends on the shadow size (note that the shadow size is nearly independent of the spin for the Kerr black hole \cite{Falcke2000}), while the stellar and gas methods are performed in the scale that much larger than the black hole horizon and mainly depend on the weak field limit of the metric. Thus we obtain $nM\approx M_\textrm{shadow-method}$ and $M=M_\textrm{stellar,gas-method}$ for the PSM. If the results about $M_\textrm{shadow-method}$ and $M_\textrm{stellar-method}$ are correct, then we obtain $|n-1|\lesssim0.1$ based on these observations. If the results about $M_\textrm{shadow-method}$ and $M_\textrm{gas-method}$ are correct, then we can use $n\approx2$ in the PSM to explain why the black hole mass measured by the shadow is a factor of about two larger than the previous gas dynamics measurements. The latter scenario is very important because it is likely to be the second phenomenon that violates the classical general relativity (Einstein equation with the normal matter). Note that the first phenomenon that violates the classical general relativity is the late-time cosmological acceleration \cite{Clifton2012,Ferreira2019}. Finding more observational phonomenon against general relativity will help us to establish the correct theory of gravity.

\section{Existence of the PSM in one specific gravity theory}\label{sec:05}
In Sec. \ref{sec:01}, we directly parameterize the Schwarzschild metric. In this section, we discuss the possible physical origins of the PSM. Here we try three theories: $f(R)$ gravity, general relativity with scalar field, and general relativity with nonlinear electrodynamics. The first two give negative results, while the last one gives a positive result.

\subsection{Non-existence of the PSM in the $f(R)$ gravity}
The action of the $f(R)$ gravity is (see \cite{Clifton2012,DeFelice2010,*Sotiriou2010,*Nojiri2011,*Joyce2015,*Nojiri2017a} for reviews)
\begin{equation}
  S=\frac{1}{2\kappa}\int\dx^4x\sqrt{-g}f(R),
\end{equation}
where $f(R)$ is an analytic function and $\kappa=8\pi G/c^4$, which can be given by the Newtonian approximation \cite{Capozziello2007}. Variation of the action with respect to the metric in the metric formalism gives the field equation
\begin{equation}\label{eq:5.02}
  E_{\mu\nu}\equiv f_RR_{\mu\nu}-\frac{g_{\mu\nu}}{2}f-\nabla_\mu\nabla_\nu f_R+g_{\mu\nu}\Box f_R=0,
\end{equation}
where $f_R\equiv\dx f/\dx R$, $\nabla_\mu$ is the covariant derivative and $\Box\equiv\nabla^\alpha\nabla_\alpha$. $f(R)=R$ gives the Einstein field equation, whose solution corresponds to $n=1$ in the PSM. Contraction of Eq. (\ref{eq:5.02}) gives
\begin{equation}\label{eq:5.03}
  3\Box f_R-2f+Rf_R=0.
\end{equation}
We can substitute the PSM into Eq. (\ref{eq:5.03}) to obtain the solution of $f(R)$, and then verify whether all $E_{\mu\nu}$ are equal to zero for this solution. For simplicity, and as a first step, we assume $n=1+\delta$ and $|\delta|\ll1$. All the following calculations only preserve the $\mathcal{O}(\delta)$ terms.

Direct calculation gives the Ricci scalar
\begin{align}\label{eq:5.04}
  R&=-A_n''-\frac{4A_n'}{r}-\frac{2A_n}{r^2}+\frac{2}{r^2}\nonumber\\
  &=\delta\left[\frac{1}{r^2}\ln\frac{(r-r_\textrm{s})^2}{r^2}+\frac{2r_\textrm{s}r-r_\textrm{s}^2}{r^3(r-r_\textrm{s})}\right]+\mathcal{O}(\delta^2),
\end{align}
where $'\equiv\dx/\dx r$. We can assume $f(R)=R+\mathcal{O}(\delta)$ as we expect $f(R)=R$ if $\delta=0$. Then the Taylor expansion of $f(R)$ at $R=0$ is
\begin{align}
  f(R)=f_0+f_R(0)R+\mathcal{O}(R^2)
  =f_0+R+\mathcal{O}(\delta^2),
\end{align}
where $f_0=f(0)=\mathcal{O}(\delta)$ and $f_R(0)=f_R|_{R=0}=1+\mathcal{O}(\delta)$. Here we also use $R=\mathcal{O}(\delta)$ given by Eq. (\ref{eq:5.04}). We assume $f_R=1+g(r)+\mathcal{O}(\delta^2)$, where $g(r)=\mathcal{O}(\delta)$. $f_R$ is a scalar gives
\begin{align}
  \Box f_R&=\frac{1}{\sqrt{-g}}\frac{\p}{\p x^\mu}(\sqrt{-g}g^{\mu\nu}\frac{\p f_R}{\p x^\nu})\nonumber\\
  &=(1-\frac{r_\textrm{s}}{r})g''+(\frac{2}{r}-\frac{r_\textrm{s}}{r^2})g'+\mathcal{O}(\delta^2).
\end{align}
Substituting the above results into Eq. (\ref{eq:5.03}), we obtain
\begin{align}
  &3(1-\frac{r_\textrm{s}}{r})g''+3(\frac{2}{r}-\frac{r_\textrm{s}}{r^2})g'-2f_0\nonumber\\
  &\quad-\delta\left(\frac{1}{r^2}\ln\frac{(r-r_\textrm{s})^2}{r^2}+\frac{2r_\textrm{s}r-r_\textrm{s}^2}{r^3(r-r_\textrm{s})}\right)=0.
\end{align}
Solving the above equation yields
\begin{align}
  g(r)&=\frac{f_0r_\textrm{s}^2}{9}(B_3+\ln B_2)+C_1\ln B_1+C_2\nonumber\\
  &+\frac{\delta}{6}\left[\ln B_2\ln B_1-4\,\textrm{dilog}(\frac{r_\textrm{s}}{r})-\frac{3}{4}\ln^2B_1\right],
\end{align}
where $C_i$ is integral constant, $B_1=(1-r_\textrm{s}/r)^2$, $B_2=(1-r/r_\textrm{s})^2$, $B_3=(1+r/r_\textrm{s})^2$, and the dilogarithm function
\begin{equation}
  \textrm{dilog}(x)\equiv\int_1^x\frac{\ln\tilde{x}}{1-\tilde{x}}\dx\tilde{x},
\end{equation}
which is defined in $x\in[0,+\infty)$.

So far, we obtain the expressions for $f=f(r)$, $f_R=f_R(r)$. Then we can directly calculate $E_{\mu\nu}$ based on Eq. (\ref{eq:5.02}). Unfortunately, there are nonzero components of $E_{\mu\nu}$, e.g.,
\begin{align}
  E_{00}&=\frac{1}{36r^4}\cdot[-6f_0r^4+10f_0r_\textrm{s}r^3+36C_1r_\textrm{s}^2\nonumber\\
  &\quad+12\delta r_\textrm{s}r-24\delta r_\textrm{s}^2+3\delta(2r^2-r_\textrm{s}^2)\ln B_1].
\end{align}
This result indicates the PSM with $n\neq1$ cannot exist in the $f(R)$ gravity\footnote{Note that the above calculations are performed in the metric formalism, and the Palatini or metric-affine formalism \cite{DeFelice2010,*Sotiriou2010,*Nojiri2011,*Joyce2015,*Nojiri2017a} may give different results. One important thing is that if different formalism gives different black hole solution for the same action, then EHT observations would be useful to distinguish between these formalisms.}. Actually, this conclusion is consistent with the weak field limit of the $f(R)$ gravity as the PSM gives the post-Newtonian parameter $\gamma=1$ and the weak filed limit of the general $f(R)$ gravity gives $\gamma=1/2$ \cite{Chiba2007,*Faraoni2008}. Inspired by this, we think a necessary condition for a gravity theory has the PSM solution is the weak field limit of this gravity theory gives $\gamma=1$. Some modified gravities can indeed give exactly $\gamma=1$ \cite{Tian2018a,Tian2018b}. Finding a modified gravity theory that has the PSM solution is meaningful, and we leave this work to the future. In the next two subsections, we try general relativity with exotic matters.

\subsection{Non-existence of the PSM in general relativity with scalar field}
The action for general relativity with scalar field can be written as
\begin{equation}
  S=\int\dx^4x\sqrt{-g}\left[\frac{R}{2\kappa}-\frac{\mathcal{L}_\phi}{\kappa}\right],
\end{equation}
where $\mathcal{L}_\phi=\p^\mu\phi\p_\mu\phi/2+V(\phi)$. The field equation is
\begin{equation}
  G^\mu_{\ \nu}=T^\mu_{\ \nu},
\end{equation}
where $T^\mu_{\ \nu}=\partial^\mu\phi\partial_\nu\phi-g^\mu_{\ \nu}\mathcal{L}_\phi$. For the PSM, the nonzero components of the Einstein tensor are
\begin{subequations}\label{eq:5.13}
\begin{align}
  G^0_{\ 0}&=G^1_{\ 1}=\frac{A_n'}{r}+\frac{A_n}{r^2}-\frac{1}{r^2},\label{eq:5.13a}\\
  G^2_{\ 2}&=G^3_{\ 3}=\frac{A_n''}{2}+\frac{A_n'}{r},
\end{align}
\end{subequations}
while the nonzero components of the energy-momentum tensor are
\begin{subequations}
\begin{align}
  T^0_{\ 0}&=T^2_{\ 2}=T^3_{\ 3}=-\mathcal{L}_\phi,\\
  T^1_{\ 1}&=A_n\cdot(\phi')^2-\mathcal{L}_\phi.
\end{align}
\end{subequations}
Generally we have $G^0_{\ 0}\neq G^2_{\ 2}$, which is contrary to $T^0_{\ 0}=T^2_{\ 2}$. Thus general relativity with scalar field do not have the PSM solution.

\subsection{Existence of the PSM in general relativity with nonlinear electrodynamics}
In the framework of general relativity, what kind of field is consistent with the PSM? We recognize that $T^0_{\ 0}=T^1_{\ 1}$ for the possible field should be necessary because of Eq. (\ref{eq:5.13a}). The electromagnetic fields satisfy this condition. However, the PSM is not the Reissner-Nordstr\"om metric. To deal with this problem, our method is to use the nonlinear electrodynamics \cite{Ayon-Beato2000,Rodrigues2016,Chinaglia2017,Nojiri2017b,Rodrigues2019}. The action for general relativity with nonlinear electrodynamics can be written as
\begin{equation}\label{eq:5.15}
  S=\int\dx^4x\sqrt{-g}\left[\frac{R}{2\kappa}+\mathcal{L}(I)\right],
\end{equation}
where $I=F^{\mu\nu}F_{\mu\nu}/4$ and $F_{\mu\nu}=\p_\mu A_\nu-\p_\nu A_\mu$. The filed equation is
\begin{equation}\label{eq:5.16}
  G^\mu_{\ \nu}=\kappa\left(\delta^\mu_{\ \nu}\mathcal{L}-F^{\mu\alpha}F_{\nu\alpha}\p_I\mathcal{L}\right)
  \equiv\kappa T^\mu_{\ \nu},
\end{equation}
where $\p_I\mathcal{L}=\p\mathcal{L}/\p I$. The energy-momentum tensor appeared in Eq. (\ref{eq:5.16}) is consistent with \cite{Ayon-Beato2000,Misner1973,Rodrigues2016}, but contrary to \cite{Chinaglia2017,Nojiri2017b,Rodrigues2019}. So we present the details of the derivation of Eq. (\ref{eq:5.16}) in Appendix \ref{sec:A.A}. We set $\kappa=1$ hereafter. The equation of motion for the electromagnetic field is
\begin{equation}\label{eq:5.17}
  0=\nabla_\mu(F^{\mu\nu}\p_I\mathcal{L})=\p_\mu(\sqrt{-g}F^{\mu\nu}\p_I\mathcal{L}),
\end{equation}
where the second equality uses $F_{\mu\nu}$ is an anti-symmetric tensor.

\begin{figure}[t]
  \centering
  \includegraphics[width=0.99\linewidth]{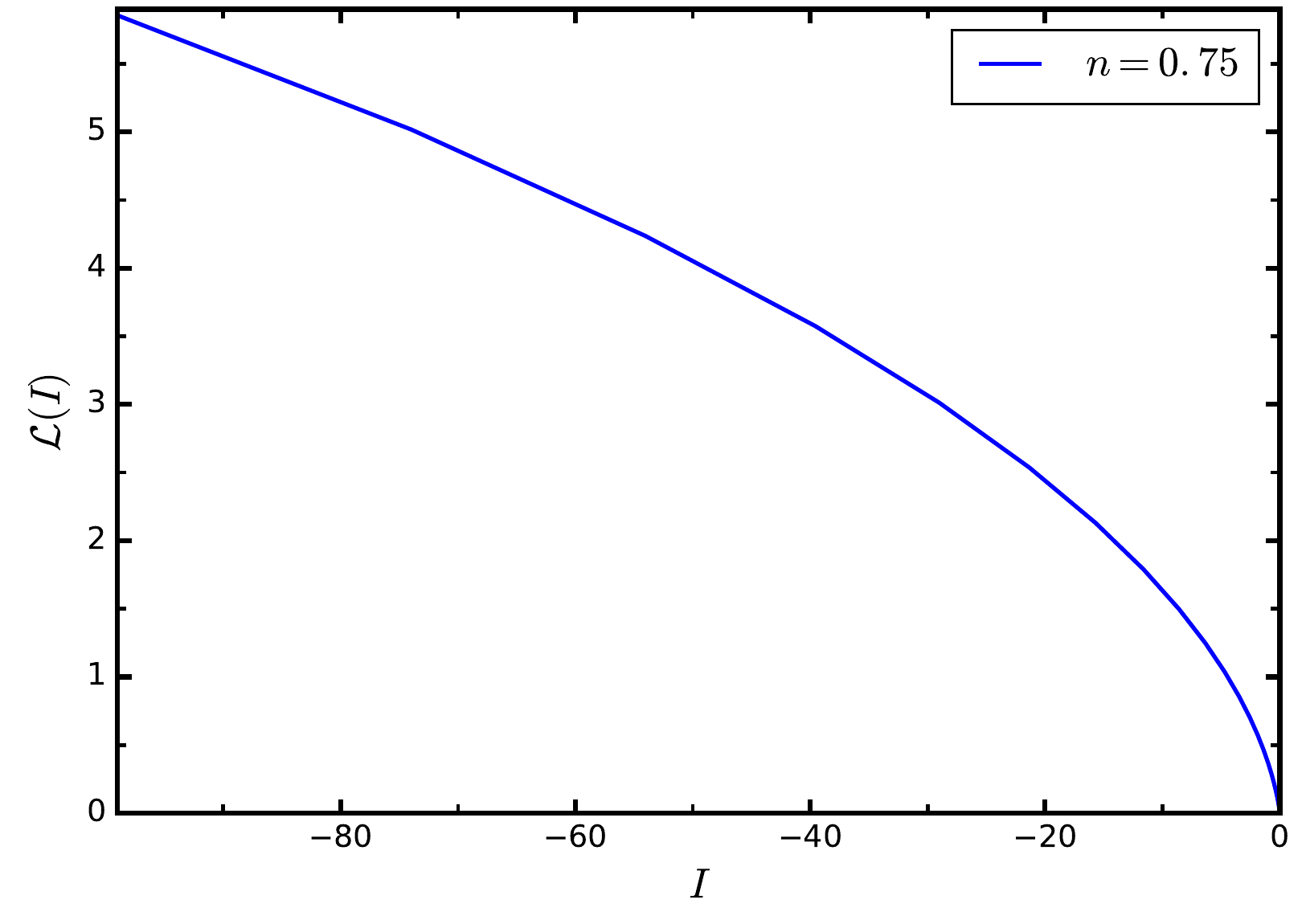}
  \caption{The Lagrangian density $\mathcal{L}(I)$ of nonlinear electrodynamics for $n=0.75$. Here we set $r_\textrm{s}=2$.}
  \label{fig:09}
\end{figure}

As did in \cite{Nojiri2017b}, we assume $A_\mu=\{\mathcal{A},0,0,0\}$, where $\mathcal{A}=\mathcal{A}(r)$. Thus the nonzero components of $F^{\mu\nu}$ are $F^{01}=-F_{01}=\mathcal{A}'$ and $F^{10}=-F_{10}=-\mathcal{A}'$. The scalar $I=-(\mathcal{A}')^2/2$. The nonzero components of $T^\mu_{\ \nu}$ are
\begin{subequations}\label{eq:5.18}
\begin{align}
  T^0_{\ 0}&=T^1_{\ 1}=\mathcal{L}+(\mathcal{A}')^2\p_I\mathcal{L},\\
  T^2_{\ 2}&=T^3_{\ 3}=\mathcal{L}.
\end{align}
\end{subequations}
Eq. (\ref{eq:5.17}) gives $\mathcal{A}'\p_I\mathcal{L}=q/r^2$, where the integral constant $q$ is the charge. Comparing the PSM with the Reissner-Nordstr\"om metric, which replace $A_n(r)$ with $1-r_\textrm{s}/r+q^2/2r^2$ in Eq. (\ref{eq:1.01}), and considering the Taylor expansion Eq. (\ref{eq:1.02}), we obtain
\begin{equation}
  (1-n)r_\textrm{s}^2=q^2,
\end{equation}
which requires $n\leq1$. Then solving Eq. (\ref{eq:5.16}) yields
\begin{align}
  q\mathcal{A}'&=r^2\left(\frac{A_n}{r^2}-\frac{1}{r^2}-\frac{A_n''}{2}\right)
  =\frac{(n-1)r_\textrm{s}^2}{r^2}+\mathcal{O}(r_\textrm{s}^3/r^3),\\
  \mathcal{L}&=\frac{A_n''}{2}+\frac{A_n'}{r}=\frac{(1-n)r_\textrm{s}^2}{2r^4}+\mathcal{O}(r_\textrm{s}^3/r^5)\label{eq:5.21},
\end{align}
and the scalar
\begin{equation}\label{eq:5.22}
  I=-\frac{1}{2}(\mathcal{A}')^2\xlongequal{r\rightarrow+\infty}
  \frac{(n-1)r_\textrm{s}^2}{2r^4}.
\end{equation}
Therefore general relativity with nonlinear electrodynamics allows the existence of the PSM. The Taylor expansions of above results give
\begin{equation}\label{eq:5.23}
  \lim_{I\rightarrow0}\mathcal{L}(I)=-I+\mathcal{O}(I^2),
\end{equation}
which means the theory return to the linear electrodynamics if the field is weak. Comparing with the Fig. 1 in \cite{Rodrigues2019}, which shows $\lim_{I\rightarrow0}\mathcal{L}(I)\neq0$ in their case, we know Eq. (\ref{eq:5.23}) is nontrivial. Fig. \ref{fig:09} plots $\mathcal{L}(I)$ for $n=0.75$ based on Eqs. (\ref{eq:5.21}) and (\ref{eq:5.22}).

\section{Conclusions}\label{sec:06}
In this paper, we discuss the issue that where is the event horizon of a spherically symmetric black hole. More precisely, if the mass of the black hole is defined in the weak field limit, which is called the Newtonian gravitational mass, then we want to know whether the radius of the horizon is equal to the Schwarzschild radius. We first parameterize the Schwarzschild metric in the form of Eq. (\ref{eq:1.01}), which introduces a real free parameter $n$. We call this metric PSM in the main text. The Schwarzschild metric corresponds to $n=1$. In the weak field limit, the PSM is equivalent to the Schwarzschild metric regardless of the value of $n$. The geodesics in the PSM are analyzed and Eq. (\ref{eq:3.01}) gives the black hole shadow size for a large background light source. In order to get closer to the real scenario, we analyze the image of the black hole with an accretion disk. Eq. (\ref{eq:4.17}) gives the brightness of an optically thick and geometrically thin accretion disk. The observed flux is given by Eq. (\ref{eq:4.27}), which takes into account the light bending, redshift and projection effects. Figures \ref{fig:06}, \ref{fig:07} and \ref{fig:08} plot several simulated images. The observational constraints on $n$ are discussed. More importantly, we point out that general relativity may be violated in the horizon region if the results of the EHT measurements \cite{Akiyama2019a} and the gas dynamics measurements \cite{Harms1994,Macchetto1997,Walsh2013,Akiyama2019f} of the mass of M87* are correct. Especially, $n\approx2$ could eliminate the inconsistency between these two kinds of observations. The upcoming EHT observations about Sgr A* are useful to prove or disprove this conjecture. If this conjecture is confirmed, then this phenomenon will be very important to test gravity theories. Finally, we prove that the PSM cannot exist in the $f(R)$ gravity (metric formalism) or general relativity with scalar field. However, general relativity with nonlinear electrodynamics has the PSM solution with $n<1$.

Before 2015 (GW150914), tests of gravity theory are mainly performed in the weak field region (e.g. the post-Newtonian analysis), the dynamical region (mainly the cosmological constraints with the Friedmann-Lema\^{i}tre-Robertson-Walker metric), and their combination (e.g. the cosmological large-scale structure formation). Now, observations of gravitational waves and black hole shadows open a new window to test gravity theory and explore new physics. For example, the simultaneous detection of GW170817 and GRB 170817A rules out many of the parameter space for modified gravities \cite{Baker2017,*Bettoni2017,*Creminelli2017,*Ezquiaga2017,*Sakstein2017,
*Akrami2018,*Amendola2018,*Arai2018,*Boran2018,*Cai2018,*Copeland2019,*Crisostomi2018,
*Gumrukcuoglu2018,*Kreisch2018,*Nojiri2018-170817,*Oost2018,*Pardo2018,*Visinelli2018,
*Casalino2018,*Ganz2019,*Jana2019,*Ramos2019}. Personally, we believe the most important thing is that these observations, for the first time, directly provide physical information about the extremely strong gravitational field (see \cite{Moffat2019,*Davoudiasl2019,*Bar2019,
Abbott2016_GW150914testGR,*Callister2017,*Shao2017,*Abbott2019_GW170817testGR,*Bustillo2019,*Giddings2019,
*Hughes2019,*Isi2019a,*Isi2019b,*Silva2019} for applications). We look forward to seeing a phenomenon that is inconsistent with the prediction of general relativity in the strong field region.

\section*{Acknowledgements}
This work was supported by the National Natural Science Foundation of China under Grant No. 11633001 and the Strategic Priority Research Program of the Chinese Academy of Sciences, Grant No. XDB23000000.

\appendix
\allowdisplaybreaks
\makeatletter
\renewcommand\theequation{\Alph{section}.\arabic{equation}}
\@addtoreset{equation}{section}
\makeatother
\section{Variation of the action Eq. (\ref{eq:5.15})}\label{sec:A.A}
We can rewrite Eq. (\ref{eq:5.15}) as $S=S_\textrm{EH}+S_I$, where $S_\textrm{EH}$ is the Einstein-Hilbert action and its variation is
\begin{equation}
  \delta S_\textrm{EH}=\frac{1}{2\kappa}\int\dx^4x\sqrt{-g}G_{\mu\nu}\delta g^{\mu\nu}.
\end{equation}
$A_\mu$ and $F_{\mu\nu}$ are independent of the metric, and $F^{\mu\nu}$ depends on the metric as $F^{\mu\nu}=g^{\mu\alpha}g^{\nu\beta}F_{\alpha\beta}$. Based on the definition of $I$, we obtain
\begin{align}
  \delta I&=\frac{1}{4}F_{\mu\nu}\delta F^{\mu\nu}\nonumber\\
  &=\frac{1}{4}(F_{\mu\nu}g^{\nu\beta}F_{\alpha\beta}\delta g^{\mu\alpha}
  +F_{\mu\nu}g^{\mu\alpha}F_{\alpha\beta}\delta g^{\nu\beta})\nonumber\\
  &=\frac{1}{2}F_{\mu\alpha}g^{\alpha\beta}F_{\nu\beta}\delta g^{\mu\nu}.
\end{align}
Considering $\delta\sqrt{-g}=-\sqrt{-g}g_{\mu\nu}\delta g^{\mu\nu}/2$, we obtain
\begin{align}
  \delta S_I&=\int\dx^4x\frac{-\sqrt{-g}}{2}\left[g_{\mu\nu}\mathcal{L}
  -F_{\mu\alpha}g^{\alpha\beta}F_{\nu\beta}\p_I\mathcal{L}\right]\delta g^{\mu\nu}.
\end{align}
The above results give Eq. (\ref{eq:5.16}).


\begin{thebibliography}{149}%
\makeatletter
\providecommand \@ifxundefined [1]{%
 \@ifx{#1\undefined}
}%
\providecommand \@ifnum [1]{%
 \ifnum #1\expandafter \@firstoftwo
 \else \expandafter \@secondoftwo
 \fi
}%
\providecommand \@ifx [1]{%
 \ifx #1\expandafter \@firstoftwo
 \else \expandafter \@secondoftwo
 \fi
}%
\providecommand \natexlab [1]{#1}%
\providecommand \enquote  [1]{``#1''}%
\providecommand \bibnamefont  [1]{#1}%
\providecommand \bibfnamefont [1]{#1}%
\providecommand \citenamefont [1]{#1}%
\providecommand \href@noop [0]{\@secondoftwo}%
\providecommand \href [0]{\begingroup \@sanitize@url \@href}%
\providecommand \@href[1]{\@@startlink{#1}\@@href}%
\providecommand \@@href[1]{\endgroup#1\@@endlink}%
\providecommand \@sanitize@url [0]{\catcode `\\12\catcode `\$12\catcode
  `\&12\catcode `\#12\catcode `\^12\catcode `\_12\catcode `\%12\relax}%
\providecommand \@@startlink[1]{}%
\providecommand \@@endlink[0]{}%
\providecommand \url  [0]{\begingroup\@sanitize@url \@url }%
\providecommand \@url [1]{\endgroup\@href {#1}{\urlprefix }}%
\providecommand \urlprefix  [0]{URL }%
\providecommand \Eprint [0]{\href }%
\providecommand \doibase [0]{http://dx.doi.org/}%
\providecommand \selectlanguage [0]{\@gobble}%
\providecommand \bibinfo  [0]{\@secondoftwo}%
\providecommand \bibfield  [0]{\@secondoftwo}%
\providecommand \translation [1]{[#1]}%
\providecommand \BibitemOpen [0]{}%
\providecommand \bibitemStop [0]{}%
\providecommand \bibitemNoStop [0]{.\EOS\space}%
\providecommand \EOS [0]{\spacefactor3000\relax}%
\providecommand \BibitemShut  [1]{\csname bibitem#1\endcsname}%
\let\auto@bib@innerbib\@empty
\bibitem [{\citenamefont {{Akiyama \etal}}(2019{\natexlab{a}})}]{Akiyama2019a}%
  \BibitemOpen
  \bibfield  {author} {\bibinfo {author} {\bibfnamefont {K.}~\bibnamefont
  {{Akiyama \etal}}},\ }\href {\doibase 10.3847/2041-8213/ab0ec7} {\bibfield
  {journal} {\bibinfo  {journal} {\apjl}\ }\textbf {\bibinfo {volume} {875}},\
  \bibinfo {eid} {L1} (\bibinfo {year} {2019}{\natexlab{a}})}\BibitemShut
  {NoStop}%
\bibitem [{\citenamefont {{Akiyama \etal}}(2019{\natexlab{b}})}]{Akiyama2019b}%
  \BibitemOpen
  \bibfield  {author} {\bibinfo {author} {\bibfnamefont {K.}~\bibnamefont
  {{Akiyama \etal}}},\ }\href {\doibase 10.3847/2041-8213/ab0c96} {\bibfield
  {journal} {\bibinfo  {journal} {\apjl}\ }\textbf {\bibinfo {volume} {875}},\
  \bibinfo {eid} {L2} (\bibinfo {year} {2019}{\natexlab{b}})}\BibitemShut
  {NoStop}%
\bibitem [{\citenamefont {{Akiyama \etal}}(2019{\natexlab{c}})}]{Akiyama2019c}%
  \BibitemOpen
  \bibfield  {author} {\bibinfo {author} {\bibfnamefont {K.}~\bibnamefont
  {{Akiyama \etal}}},\ }\href {\doibase 10.3847/2041-8213/ab0c57} {\bibfield
  {journal} {\bibinfo  {journal} {\apjl}\ }\textbf {\bibinfo {volume} {875}},\
  \bibinfo {eid} {L3} (\bibinfo {year} {2019}{\natexlab{c}})}\BibitemShut
  {NoStop}%
\bibitem [{\citenamefont {{Akiyama \etal}}(2019{\natexlab{d}})}]{Akiyama2019d}%
  \BibitemOpen
  \bibfield  {author} {\bibinfo {author} {\bibfnamefont {K.}~\bibnamefont
  {{Akiyama \etal}}},\ }\href {\doibase 10.3847/2041-8213/ab0e85} {\bibfield
  {journal} {\bibinfo  {journal} {\apjl}\ }\textbf {\bibinfo {volume} {875}},\
  \bibinfo {eid} {L4} (\bibinfo {year} {2019}{\natexlab{d}})}\BibitemShut
  {NoStop}%
\bibitem [{\citenamefont {{Akiyama \etal}}(2019{\natexlab{e}})}]{Akiyama2019e}%
  \BibitemOpen
  \bibfield  {author} {\bibinfo {author} {\bibfnamefont {K.}~\bibnamefont
  {{Akiyama \etal}}},\ }\href {\doibase 10.3847/2041-8213/ab0f43} {\bibfield
  {journal} {\bibinfo  {journal} {\apjl}\ }\textbf {\bibinfo {volume} {875}},\
  \bibinfo {eid} {L5} (\bibinfo {year} {2019}{\natexlab{e}})}\BibitemShut
  {NoStop}%
\bibitem [{\citenamefont {{Akiyama \etal}}(2019{\natexlab{f}})}]{Akiyama2019f}%
  \BibitemOpen
  \bibfield  {author} {\bibinfo {author} {\bibfnamefont {K.}~\bibnamefont
  {{Akiyama \etal}}},\ }\href {\doibase 10.3847/2041-8213/ab1141} {\bibfield
  {journal} {\bibinfo  {journal} {\apjl}\ }\textbf {\bibinfo {volume} {875}},\
  \bibinfo {eid} {L6} (\bibinfo {year} {2019}{\natexlab{f}})}\BibitemShut
  {NoStop}%
\bibitem [{\citenamefont {Synge}(1966)}]{Synge1966}%
  \BibitemOpen
  \bibfield  {author} {\bibinfo {author} {\bibfnamefont {J.~L.}\ \bibnamefont
  {Synge}},\ }\href {\doibase 10.1093/mnras/131.3.463} {\bibfield  {journal}
  {\bibinfo  {journal} {\mnras}\ }\textbf {\bibinfo {volume} {131}},\ \bibinfo
  {pages} {463} (\bibinfo {year} {1966})}\BibitemShut {NoStop}%
\bibitem [{\citenamefont {Bardeen}(1973)}]{Bardeen1973}%
  \BibitemOpen
  \bibfield  {author} {\bibinfo {author} {\bibfnamefont {J.~M.}\ \bibnamefont
  {Bardeen}},\ }in\ \href {https://www.twirpx.com/file/1416192/} {\emph
  {\bibinfo {booktitle} {{Black Holes (Les Astres Occlus)}}}},\ \bibinfo
  {editor} {edited by\ \bibinfo {editor} {\bibfnamefont {C.}~\bibnamefont
  {Dewitt}}\ and\ \bibinfo {editor} {\bibfnamefont {B.~S.}\ \bibnamefont
  {Dewitt}}}\ (\bibinfo  {publisher} {Gordon and Breach},\ \bibinfo {address}
  {New York},\ \bibinfo {year} {1973})\ pp.\ \bibinfo {pages}
  {215--239}\BibitemShut {NoStop}%
\bibitem [{\citenamefont {{Cunningham}}\ and\ \citenamefont
  {{Bardeen}}(1972)}]{Cunningham1972}%
  \BibitemOpen
  \bibfield  {author} {\bibinfo {author} {\bibfnamefont {C.~T.}\ \bibnamefont
  {{Cunningham}}}\ and\ \bibinfo {author} {\bibfnamefont {J.~M.}\ \bibnamefont
  {{Bardeen}}},\ }\href {\doibase 10.1086/180933} {\bibfield  {journal}
  {\bibinfo  {journal} {\apjl}\ }\textbf {\bibinfo {volume} {173}},\ \bibinfo
  {pages} {L137} (\bibinfo {year} {1972})}\BibitemShut {NoStop}%
\bibitem [{\citenamefont {{Cunningham}}\ and\ \citenamefont
  {{Bardeen}}(1973)}]{Cunningham1973}%
  \BibitemOpen
  \bibfield  {author} {\bibinfo {author} {\bibfnamefont {C.~T.}\ \bibnamefont
  {{Cunningham}}}\ and\ \bibinfo {author} {\bibfnamefont {J.~M.}\ \bibnamefont
  {{Bardeen}}},\ }\href {\doibase 10.1086/152223} {\bibfield  {journal}
  {\bibinfo  {journal} {\apj}\ }\textbf {\bibinfo {volume} {183}},\ \bibinfo
  {pages} {237} (\bibinfo {year} {1973})}\BibitemShut {NoStop}%
\bibitem [{\citenamefont {{Luminet}}(1979)}]{Luminet1979}%
  \BibitemOpen
  \bibfield  {author} {\bibinfo {author} {\bibfnamefont {J.-P.}\ \bibnamefont
  {{Luminet}}},\ }\href {http://adsabs.harvard.edu/abs/1979A\%26A....75..228L}
  {\bibfield  {journal} {\bibinfo  {journal} {\aa}\ }\textbf {\bibinfo {volume}
  {75}},\ \bibinfo {pages} {228} (\bibinfo {year} {1979})}\BibitemShut
  {NoStop}%
\bibitem [{\citenamefont {Falcke}\ \emph {et~al.}(2000)\citenamefont {Falcke},
  \citenamefont {Melia},\ and\ \citenamefont {Agol}}]{Falcke2000}%
  \BibitemOpen
  \bibfield  {author} {\bibinfo {author} {\bibfnamefont {H.}~\bibnamefont
  {Falcke}}, \bibinfo {author} {\bibfnamefont {F.}~\bibnamefont {Melia}}, \
  and\ \bibinfo {author} {\bibfnamefont {E.}~\bibnamefont {Agol}},\ }\href
  {\doibase 10.1086/312423} {\bibfield  {journal} {\bibinfo  {journal} {\apjl}\
  }\textbf {\bibinfo {volume} {528}},\ \bibinfo {pages} {L13} (\bibinfo {year}
  {2000})}\BibitemShut {NoStop}%
\bibitem [{\citenamefont {{Fraga-Encinas, R.}}\ \emph
  {et~al.}(2016)\citenamefont {{Fraga-Encinas, R.}}, \citenamefont
  {{Mo\'scibrodzka, M.}}, \citenamefont {{Brinkerink, C.}},\ and\ \citenamefont
  {{Falcke, H.}}}]{Fraga-Encinas2016}%
  \BibitemOpen
  \bibfield  {author} {\bibinfo {author} {\bibnamefont {{Fraga-Encinas, R.}}},
  \bibinfo {author} {\bibnamefont {{Mo\'scibrodzka, M.}}}, \bibinfo {author}
  {\bibnamefont {{Brinkerink, C.}}}, \ and\ \bibinfo {author} {\bibnamefont
  {{Falcke, H.}}},\ }\href {\doibase 10.1051/0004-6361/201527599} {\bibfield
  {journal} {\bibinfo  {journal} {\aa}\ }\textbf {\bibinfo {volume} {588}},\
  \bibinfo {pages} {A57} (\bibinfo {year} {2016})}\BibitemShut {NoStop}%
\bibitem [{\citenamefont {Roelofs~\etal}(2019)}]{Roelofs2019}%
  \BibitemOpen
  \bibfield  {author} {\bibinfo {author} {\bibfnamefont {F.}~\bibnamefont
  {Roelofs~\etal}},\ }\href {\doibase 10.1051/0004-6361/201732423} {\bibfield
  {journal} {\bibinfo  {journal} {\aa}\ }\textbf {\bibinfo {volume} {625}},\
  \bibinfo {pages} {A124} (\bibinfo {year} {2019})}\BibitemShut {NoStop}%
\bibitem [{\citenamefont {Porth~\etal}(2019)}]{Porth2019}%
  \BibitemOpen
  \bibfield  {author} {\bibinfo {author} {\bibfnamefont {O.}~\bibnamefont
  {Porth~\etal}},\ }\href {\doibase 10.3847/1538-4365/ab29fd} {\bibfield
  {journal} {\bibinfo  {journal} {\apjs}\ }\textbf {\bibinfo {volume} {243}},\
  \bibinfo {eid} {26} (\bibinfo {year} {2019})}\BibitemShut {NoStop}%
\bibitem [{\citenamefont {Font}(2008)}]{Font2008}%
  \BibitemOpen
  \bibfield  {author} {\bibinfo {author} {\bibfnamefont {J.~A.}\ \bibnamefont
  {Font}},\ }\href {\doibase 10.12942/lrr-2008-7} {\bibfield  {journal}
  {\bibinfo  {journal} {\lrr}\ }\textbf {\bibinfo {volume} {11}},\ \bibinfo
  {pages} {7} (\bibinfo {year} {2008})}\BibitemShut {NoStop}%
\bibitem [{\citenamefont {Porth}\ \emph {et~al.}(2017)\citenamefont {Porth},
  \citenamefont {Olivares}, \citenamefont {Mizuno}, \citenamefont {Younsi},
  \citenamefont {Rezzolla}, \citenamefont {Moscibrodzka}, \citenamefont
  {Falcke},\ and\ \citenamefont {Kramer}}]{Porth2017}%
  \BibitemOpen
  \bibfield  {author} {\bibinfo {author} {\bibfnamefont {O.}~\bibnamefont
  {Porth}}, \bibinfo {author} {\bibfnamefont {H.}~\bibnamefont {Olivares}},
  \bibinfo {author} {\bibfnamefont {Y.}~\bibnamefont {Mizuno}}, \bibinfo
  {author} {\bibfnamefont {Z.}~\bibnamefont {Younsi}}, \bibinfo {author}
  {\bibfnamefont {L.}~\bibnamefont {Rezzolla}}, \bibinfo {author}
  {\bibfnamefont {M.}~\bibnamefont {Moscibrodzka}}, \bibinfo {author}
  {\bibfnamefont {H.}~\bibnamefont {Falcke}}, \ and\ \bibinfo {author}
  {\bibfnamefont {M.}~\bibnamefont {Kramer}},\ }\href {\doibase
  10.1186/s40668-017-0020-2} {\bibfield  {journal} {\bibinfo  {journal} {\cac}\
  }\textbf {\bibinfo {volume} {4}},\ \bibinfo {pages} {1} (\bibinfo {year}
  {2017})}\BibitemShut {NoStop}%
\bibitem [{\citenamefont {Olivares}\ \emph {et~al.}()\citenamefont {Olivares},
  \citenamefont {Porth}, \citenamefont {Davelaar}, \citenamefont {Most},
  \citenamefont {Fromm}, \citenamefont {Mizuno}, \citenamefont {Younsi},\ and\
  \citenamefont {Rezzolla}}]{Olivares2019}%
  \BibitemOpen
  \bibfield  {author} {\bibinfo {author} {\bibfnamefont {H.}~\bibnamefont
  {Olivares}}, \bibinfo {author} {\bibfnamefont {O.}~\bibnamefont {Porth}},
  \bibinfo {author} {\bibfnamefont {J.}~\bibnamefont {Davelaar}}, \bibinfo
  {author} {\bibfnamefont {E.~R.}\ \bibnamefont {Most}}, \bibinfo {author}
  {\bibfnamefont {C.~M.}\ \bibnamefont {Fromm}}, \bibinfo {author}
  {\bibfnamefont {Y.}~\bibnamefont {Mizuno}}, \bibinfo {author} {\bibfnamefont
  {Z.}~\bibnamefont {Younsi}}, \ and\ \bibinfo {author} {\bibfnamefont
  {L.}~\bibnamefont {Rezzolla}},\ }\href@noop {} {}\Eprint
  {http://arxiv.org/abs/1906.10795} {arXiv:1906.10795} \BibitemShut {NoStop}%
\bibitem [{\citenamefont {White}()}]{White2019}%
  \BibitemOpen
  \bibfield  {author} {\bibinfo {author} {\bibfnamefont {C.~J.}\ \bibnamefont
  {White}},\ }\href@noop {} {}\Eprint {http://arxiv.org/abs/1906.09708}
  {arXiv:1906.09708} \BibitemShut {NoStop}%
\bibitem [{\citenamefont {Dexter}\ and\ \citenamefont
  {Agol}(2009)}]{Dexter2009}%
  \BibitemOpen
  \bibfield  {author} {\bibinfo {author} {\bibfnamefont {J.}~\bibnamefont
  {Dexter}}\ and\ \bibinfo {author} {\bibfnamefont {E.}~\bibnamefont {Agol}},\
  }\href {\doibase 10.1088/0004-637x/696/2/1616} {\bibfield  {journal}
  {\bibinfo  {journal} {\apj}\ }\textbf {\bibinfo {volume} {696}},\ \bibinfo
  {pages} {1616} (\bibinfo {year} {2009})}\BibitemShut {NoStop}%
\bibitem [{\citenamefont {Vincent}\ \emph {et~al.}(2011)\citenamefont
  {Vincent}, \citenamefont {Paumard}, \citenamefont {Gourgoulhon},\ and\
  \citenamefont {Perrin}}]{Vincent2011}%
  \BibitemOpen
  \bibfield  {author} {\bibinfo {author} {\bibfnamefont {F.~H.}\ \bibnamefont
  {Vincent}}, \bibinfo {author} {\bibfnamefont {T.}~\bibnamefont {Paumard}},
  \bibinfo {author} {\bibfnamefont {E.}~\bibnamefont {Gourgoulhon}}, \ and\
  \bibinfo {author} {\bibfnamefont {G.}~\bibnamefont {Perrin}},\ }\href
  {\doibase 10.1088/0264-9381/28/22/225011} {\bibfield  {journal} {\bibinfo
  {journal} {\cqg}\ }\textbf {\bibinfo {volume} {28}},\ \bibinfo {pages}
  {225011} (\bibinfo {year} {2011})}\BibitemShut {NoStop}%
\bibitem [{\citenamefont {Yang}\ and\ \citenamefont {Wang}(2013)}]{Yang2013}%
  \BibitemOpen
  \bibfield  {author} {\bibinfo {author} {\bibfnamefont {X.}~\bibnamefont
  {Yang}}\ and\ \bibinfo {author} {\bibfnamefont {J.}~\bibnamefont {Wang}},\
  }\href {\doibase 10.1088/0067-0049/207/1/6} {\bibfield  {journal} {\bibinfo
  {journal} {\apjs}\ }\textbf {\bibinfo {volume} {207}},\ \bibinfo {pages} {6}
  (\bibinfo {year} {2013})}\BibitemShut {NoStop}%
\bibitem [{\citenamefont {k.~Chan}\ \emph {et~al.}(2018)\citenamefont
  {k.~Chan}, \citenamefont {Medeiros}, \citenamefont {\"Ozel},\ and\
  \citenamefont {Psaltis}}]{Chan2018}%
  \BibitemOpen
  \bibfield  {author} {\bibinfo {author} {\bibfnamefont {C.}~\bibnamefont
  {k.~Chan}}, \bibinfo {author} {\bibfnamefont {L.}~\bibnamefont {Medeiros}},
  \bibinfo {author} {\bibfnamefont {F.}~\bibnamefont {\"Ozel}}, \ and\ \bibinfo
  {author} {\bibfnamefont {D.}~\bibnamefont {Psaltis}},\ }\href {\doibase
  10.3847/1538-4357/aadfe5} {\bibfield  {journal} {\bibinfo  {journal} {\apj}\
  }\textbf {\bibinfo {volume} {867}},\ \bibinfo {pages} {59} (\bibinfo {year}
  {2018})}\BibitemShut {NoStop}%
\bibitem [{\citenamefont {Davelaar}\ \emph {et~al.}(2018)\citenamefont
  {Davelaar}, \citenamefont {Bronzwaer}, \citenamefont {Kok}, \citenamefont
  {Younsi}, \citenamefont {Mo{\'{s}}cibrodzka},\ and\ \citenamefont
  {Falcke}}]{Davelaar2018}%
  \BibitemOpen
  \bibfield  {author} {\bibinfo {author} {\bibfnamefont {J.}~\bibnamefont
  {Davelaar}}, \bibinfo {author} {\bibfnamefont {T.}~\bibnamefont {Bronzwaer}},
  \bibinfo {author} {\bibfnamefont {D.}~\bibnamefont {Kok}}, \bibinfo {author}
  {\bibfnamefont {Z.}~\bibnamefont {Younsi}}, \bibinfo {author} {\bibfnamefont
  {M.}~\bibnamefont {Mo{\'{s}}cibrodzka}}, \ and\ \bibinfo {author}
  {\bibfnamefont {H.}~\bibnamefont {Falcke}},\ }\href {\doibase
  10.1186/s40668-018-0023-7} {\bibfield  {journal} {\bibinfo  {journal} {\cac}\
  }\textbf {\bibinfo {volume} {5}},\ \bibinfo {pages} {1} (\bibinfo {year}
  {2018})}\BibitemShut {NoStop}%
\bibitem [{\citenamefont {Abdujabbarov}\ \emph {et~al.}(2013)\citenamefont
  {Abdujabbarov}, \citenamefont {Atamurotov}, \citenamefont {Kucukakca},
  \citenamefont {Ahmedov},\ and\ \citenamefont {Camci}}]{Abdujabbarov2013}%
  \BibitemOpen
  \bibfield  {author} {\bibinfo {author} {\bibfnamefont {A.}~\bibnamefont
  {Abdujabbarov}}, \bibinfo {author} {\bibfnamefont {F.}~\bibnamefont
  {Atamurotov}}, \bibinfo {author} {\bibfnamefont {Y.}~\bibnamefont
  {Kucukakca}}, \bibinfo {author} {\bibfnamefont {B.}~\bibnamefont {Ahmedov}},
  \ and\ \bibinfo {author} {\bibfnamefont {U.}~\bibnamefont {Camci}},\ }\href
  {\doibase 10.1007/s10509-012-1337-6} {\bibfield  {journal} {\bibinfo
  {journal} {\ass}\ }\textbf {\bibinfo {volume} {344}},\ \bibinfo {pages} {429}
  (\bibinfo {year} {2013})}\BibitemShut {NoStop}%
\bibitem [{\citenamefont {Amarilla}\ and\ \citenamefont
  {Eiroa}(2013)}]{Amarilla2013}%
  \BibitemOpen
  \bibfield  {author} {\bibinfo {author} {\bibfnamefont {L.}~\bibnamefont
  {Amarilla}}\ and\ \bibinfo {author} {\bibfnamefont {E.~F.}\ \bibnamefont
  {Eiroa}},\ }\href {\doibase 10.1103/PhysRevD.87.044057} {\bibfield  {journal}
  {\bibinfo  {journal} {Phys. Rev. D}\ }\textbf {\bibinfo {volume} {87}},\
  \bibinfo {pages} {044057} (\bibinfo {year} {2013})}\BibitemShut {NoStop}%
\bibitem [{\citenamefont {Grenzebach}\ \emph {et~al.}(2014)\citenamefont
  {Grenzebach}, \citenamefont {Perlick},\ and\ \citenamefont
  {L\"ammerzahl}}]{Grenzebach2014}%
  \BibitemOpen
  \bibfield  {author} {\bibinfo {author} {\bibfnamefont {A.}~\bibnamefont
  {Grenzebach}}, \bibinfo {author} {\bibfnamefont {V.}~\bibnamefont {Perlick}},
  \ and\ \bibinfo {author} {\bibfnamefont {C.}~\bibnamefont {L\"ammerzahl}},\
  }\href {\doibase 10.1103/PhysRevD.89.124004} {\bibfield  {journal} {\bibinfo
  {journal} {\prd}\ }\textbf {\bibinfo {volume} {89}},\ \bibinfo {pages}
  {124004} (\bibinfo {year} {2014})}\BibitemShut {NoStop}%
\bibitem [{\citenamefont {Cunha}\ \emph {et~al.}(2015)\citenamefont {Cunha},
  \citenamefont {Herdeiro}, \citenamefont {Radu},\ and\ \citenamefont
  {R\'unarsson}}]{Cunha2015}%
  \BibitemOpen
  \bibfield  {author} {\bibinfo {author} {\bibfnamefont {P.~V.~P.}\
  \bibnamefont {Cunha}}, \bibinfo {author} {\bibfnamefont {C.~A.~R.}\
  \bibnamefont {Herdeiro}}, \bibinfo {author} {\bibfnamefont {E.}~\bibnamefont
  {Radu}}, \ and\ \bibinfo {author} {\bibfnamefont {H.~F.}\ \bibnamefont
  {R\'unarsson}},\ }\href {\doibase 10.1103/PhysRevLett.115.211102} {\bibfield
  {journal} {\bibinfo  {journal} {\prl}\ }\textbf {\bibinfo {volume} {115}},\
  \bibinfo {pages} {211102} (\bibinfo {year} {2015})}\BibitemShut {NoStop}%
\bibitem [{\citenamefont {Younsi}\ \emph {et~al.}(2016)\citenamefont {Younsi},
  \citenamefont {Zhidenko}, \citenamefont {Rezzolla}, \citenamefont
  {Konoplya},\ and\ \citenamefont {Mizuno}}]{Younsi2016}%
  \BibitemOpen
  \bibfield  {author} {\bibinfo {author} {\bibfnamefont {Z.}~\bibnamefont
  {Younsi}}, \bibinfo {author} {\bibfnamefont {A.}~\bibnamefont {Zhidenko}},
  \bibinfo {author} {\bibfnamefont {L.}~\bibnamefont {Rezzolla}}, \bibinfo
  {author} {\bibfnamefont {R.}~\bibnamefont {Konoplya}}, \ and\ \bibinfo
  {author} {\bibfnamefont {Y.}~\bibnamefont {Mizuno}},\ }\href {\doibase
  10.1103/PhysRevD.94.084025} {\bibfield  {journal} {\bibinfo  {journal} {Phys.
  Rev. D}\ }\textbf {\bibinfo {volume} {94}},\ \bibinfo {pages} {084025}
  (\bibinfo {year} {2016})}\BibitemShut {NoStop}%
\bibitem [{\citenamefont {Vincent}\ \emph {et~al.}(2016)\citenamefont
  {Vincent}, \citenamefont {Gourgoulhon}, \citenamefont {Herdeiro},\ and\
  \citenamefont {Radu}}]{Vincent2016}%
  \BibitemOpen
  \bibfield  {author} {\bibinfo {author} {\bibfnamefont {F.~H.}\ \bibnamefont
  {Vincent}}, \bibinfo {author} {\bibfnamefont {E.}~\bibnamefont
  {Gourgoulhon}}, \bibinfo {author} {\bibfnamefont {C.}~\bibnamefont
  {Herdeiro}}, \ and\ \bibinfo {author} {\bibfnamefont {E.}~\bibnamefont
  {Radu}},\ }\href {\doibase 10.1103/PhysRevD.94.084045} {\bibfield  {journal}
  {\bibinfo  {journal} {\prd}\ }\textbf {\bibinfo {volume} {94}},\ \bibinfo
  {pages} {084045} (\bibinfo {year} {2016})}\BibitemShut {NoStop}%
\bibitem [{\citenamefont {Cunha}\ \emph {et~al.}(2017)\citenamefont {Cunha},
  \citenamefont {Herdeiro},\ and\ \citenamefont {Radu}}]{Cunha2017}%
  \BibitemOpen
  \bibfield  {author} {\bibinfo {author} {\bibfnamefont {P.~V.~P.}\
  \bibnamefont {Cunha}}, \bibinfo {author} {\bibfnamefont {C.~A.~R.}\
  \bibnamefont {Herdeiro}}, \ and\ \bibinfo {author} {\bibfnamefont
  {E.}~\bibnamefont {Radu}},\ }\href {\doibase 10.1103/PhysRevD.96.024039}
  {\bibfield  {journal} {\bibinfo  {journal} {Phys. Rev. D}\ }\textbf {\bibinfo
  {volume} {96}},\ \bibinfo {pages} {024039} (\bibinfo {year}
  {2017})}\BibitemShut {NoStop}%
\bibitem [{\citenamefont {Wang}\ \emph {et~al.}(2017)\citenamefont {Wang},
  \citenamefont {Chen},\ and\ \citenamefont {Jing}}]{Wang2017}%
  \BibitemOpen
  \bibfield  {author} {\bibinfo {author} {\bibfnamefont {M.}~\bibnamefont
  {Wang}}, \bibinfo {author} {\bibfnamefont {S.}~\bibnamefont {Chen}}, \ and\
  \bibinfo {author} {\bibfnamefont {J.}~\bibnamefont {Jing}},\ }\href {\doibase
  10.1088/1475-7516/2017/10/051} {\bibfield  {journal} {\bibinfo  {journal}
  {\jcap}\ }10 (\bibinfo {year} {2017})\ \bibinfo {pages} {051}}\BibitemShut {NoStop}%
\bibitem [{\citenamefont {Ayzenberg}\ and\ \citenamefont
  {Yunes}(2018)}]{Ayzenberg2018}%
  \BibitemOpen
  \bibfield  {author} {\bibinfo {author} {\bibfnamefont {D.}~\bibnamefont
  {Ayzenberg}}\ and\ \bibinfo {author} {\bibfnamefont {N.}~\bibnamefont
  {Yunes}},\ }\href {\doibase 10.1088/1361-6382/aae87b} {\bibfield  {journal}
  {\bibinfo  {journal} {\cqg}\ }\textbf {\bibinfo {volume} {35}},\ \bibinfo
  {pages} {235002} (\bibinfo {year} {2018})}\BibitemShut {NoStop}%
\bibitem [{\citenamefont {Cunha}\ \emph {et~al.}(2018)\citenamefont {Cunha},
  \citenamefont {Herdeiro},\ and\ \citenamefont {Rodriguez}}]{Cunha2018}%
  \BibitemOpen
  \bibfield  {author} {\bibinfo {author} {\bibfnamefont {P.~V.~P.}\
  \bibnamefont {Cunha}}, \bibinfo {author} {\bibfnamefont {C.~A.~R.}\
  \bibnamefont {Herdeiro}}, \ and\ \bibinfo {author} {\bibfnamefont {M.~J.}\
  \bibnamefont {Rodriguez}},\ }\href {\doibase 10.1103/PhysRevD.97.084020}
  {\bibfield  {journal} {\bibinfo  {journal} {Phys. Rev. D}\ }\textbf {\bibinfo
  {volume} {97}},\ \bibinfo {pages} {084020} (\bibinfo {year}
  {2018})}\BibitemShut {NoStop}%
\bibitem [{\citenamefont {Guo}\ \emph {et~al.}(2018)\citenamefont {Guo},
  \citenamefont {Obers},\ and\ \citenamefont {Yan}}]{Guo2018}%
  \BibitemOpen
  \bibfield  {author} {\bibinfo {author} {\bibfnamefont {M.}~\bibnamefont
  {Guo}}, \bibinfo {author} {\bibfnamefont {N.~A.}\ \bibnamefont {Obers}}, \
  and\ \bibinfo {author} {\bibfnamefont {H.}~\bibnamefont {Yan}},\ }\href
  {\doibase 10.1103/PhysRevD.98.084063} {\bibfield  {journal} {\bibinfo
  {journal} {Phys. Rev. D}\ }\textbf {\bibinfo {volume} {98}},\ \bibinfo
  {pages} {084063} (\bibinfo {year} {2018})}\BibitemShut {NoStop}%
\bibitem [{\citenamefont {Hou}\ \emph {et~al.}(2018)\citenamefont {Hou},
  \citenamefont {Xu},\ and\ \citenamefont {Wang}}]{Hou2018}%
  \BibitemOpen
  \bibfield  {author} {\bibinfo {author} {\bibfnamefont {X.}~\bibnamefont
  {Hou}}, \bibinfo {author} {\bibfnamefont {Z.}~\bibnamefont {Xu}}, \ and\
  \bibinfo {author} {\bibfnamefont {J.}~\bibnamefont {Wang}},\ }\href {\doibase
  10.1088/1475-7516/2018/12/040} {\bibfield  {journal} {\bibinfo  {journal}
  {\jcap}\ }12 (\bibinfo {year} {2018})\ \bibinfo {pages} {040}}\BibitemShut {NoStop}%
\bibitem [{\citenamefont {\"Ovg\"un}\ \emph {et~al.}(2018)\citenamefont
  {\"Ovg\"un}, \citenamefont {Sakall{\i}},\ and\ \citenamefont
  {Saavedra}}]{Ovgun2018}%
  \BibitemOpen
  \bibfield  {author} {\bibinfo {author} {\bibfnamefont {A.}~\bibnamefont
  {\"Ovg\"un}}, \bibinfo {author} {\bibfnamefont {{\.{I}}.}~\bibnamefont
  {Sakall{\i}}}, \ and\ \bibinfo {author} {\bibfnamefont {J.}~\bibnamefont
  {Saavedra}},\ }\href {\doibase 10.1088/1475-7516/2018/10/041} {\bibfield
  {journal} {\bibinfo  {journal} {\jcap}\ }10 (\bibinfo {year} {2018})\
  \bibinfo {pages} {041}}\BibitemShut {NoStop}%
\bibitem [{\citenamefont {Shaikh}(2018)}]{Shaikh2018}%
  \BibitemOpen
  \bibfield  {author} {\bibinfo {author} {\bibfnamefont {R.}~\bibnamefont
  {Shaikh}},\ }\href {\doibase 10.1103/PhysRevD.98.024044} {\bibfield
  {journal} {\bibinfo  {journal} {Phys. Rev. D}\ }\textbf {\bibinfo {volume}
  {98}},\ \bibinfo {pages} {024044} (\bibinfo {year} {2018})}\BibitemShut
  {NoStop}%
\bibitem [{\citenamefont {Tsukamoto}(2018)}]{Tsukamoto2018}%
  \BibitemOpen
  \bibfield  {author} {\bibinfo {author} {\bibfnamefont {N.}~\bibnamefont
  {Tsukamoto}},\ }\href {\doibase 10.1103/PhysRevD.97.064021} {\bibfield
  {journal} {\bibinfo  {journal} {Phys. Rev. D}\ }\textbf {\bibinfo {volume}
  {97}},\ \bibinfo {pages} {064021} (\bibinfo {year} {2018})}\BibitemShut
  {NoStop}%
\bibitem [{\citenamefont {Wang}\ \emph
  {et~al.}(2018{\natexlab{a}})\citenamefont {Wang}, \citenamefont {Chen},\ and\
  \citenamefont {Jing}}]{Wang2018a}%
  \BibitemOpen
  \bibfield  {author} {\bibinfo {author} {\bibfnamefont {M.}~\bibnamefont
  {Wang}}, \bibinfo {author} {\bibfnamefont {S.}~\bibnamefont {Chen}}, \ and\
  \bibinfo {author} {\bibfnamefont {J.}~\bibnamefont {Jing}},\ }\href {\doibase
  10.1103/PhysRevD.97.064029} {\bibfield  {journal} {\bibinfo  {journal} {Phys.
  Rev. D}\ }\textbf {\bibinfo {volume} {97}},\ \bibinfo {pages} {064029}
  (\bibinfo {year} {2018}{\natexlab{a}})}\BibitemShut {NoStop}%
\bibitem [{\citenamefont {Wang}\ \emph
  {et~al.}(2018{\natexlab{b}})\citenamefont {Wang}, \citenamefont {Chen},\ and\
  \citenamefont {Jing}}]{Wang2018b}%
  \BibitemOpen
  \bibfield  {author} {\bibinfo {author} {\bibfnamefont {M.}~\bibnamefont
  {Wang}}, \bibinfo {author} {\bibfnamefont {S.}~\bibnamefont {Chen}}, \ and\
  \bibinfo {author} {\bibfnamefont {J.}~\bibnamefont {Jing}},\ }\href {\doibase
  10.1103/PhysRevD.98.104040} {\bibfield  {journal} {\bibinfo  {journal} {Phys.
  Rev. D}\ }\textbf {\bibinfo {volume} {98}},\ \bibinfo {pages} {104040}
  (\bibinfo {year} {2018}{\natexlab{b}})}\BibitemShut {NoStop}%
\bibitem [{\citenamefont {Ali}\ and\ \citenamefont {Amir}()}]{Ali2019}%
  \BibitemOpen
  \bibfield  {author} {\bibinfo {author} {\bibfnamefont {M.~S.}\ \bibnamefont
  {Ali}}\ and\ \bibinfo {author} {\bibfnamefont {M.}~\bibnamefont {Amir}},\
  }\href@noop {} {}\Eprint {http://arxiv.org/abs/1906.04146} {arXiv:1906.04146}
  \BibitemShut {NoStop}%
\bibitem [{\citenamefont {Cunha}\ \emph {et~al.}(2019)\citenamefont {Cunha},
  \citenamefont {Herdeiro},\ and\ \citenamefont {Radu}}]{Cunha2019}%
  \BibitemOpen
  \bibfield  {author} {\bibinfo {author} {\bibfnamefont {P.~V.~P.}\
  \bibnamefont {Cunha}}, \bibinfo {author} {\bibfnamefont {C.~A.~R.}\
  \bibnamefont {Herdeiro}}, \ and\ \bibinfo {author} {\bibfnamefont
  {E.}~\bibnamefont {Radu}},\ }\href {\doibase 10.1103/PhysRevLett.123.011101}
  {\bibfield  {journal} {\bibinfo  {journal} {Phys. Rev. Lett.}\ }\textbf
  {\bibinfo {volume} {123}},\ \bibinfo {pages} {011101} (\bibinfo {year}
  {2019})}\BibitemShut {NoStop}%
\bibitem [{\citenamefont {Haroon}\ \emph {et~al.}()\citenamefont {Haroon},
  \citenamefont {Jusufi},\ and\ \citenamefont {Jamil}}]{Haroon2019}%
  \BibitemOpen
  \bibfield  {author} {\bibinfo {author} {\bibfnamefont {S.}~\bibnamefont
  {Haroon}}, \bibinfo {author} {\bibfnamefont {K.}~\bibnamefont {Jusufi}}, \
  and\ \bibinfo {author} {\bibfnamefont {M.}~\bibnamefont {Jamil}},\
  }\href@noop {} {}\Eprint {http://arxiv.org/abs/1904.00711} {arXiv:1904.00711}
  \BibitemShut {NoStop}%
\bibitem [{\citenamefont {Held}\ \emph {et~al.}(2019)\citenamefont {Held},
  \citenamefont {Gold},\ and\ \citenamefont {Eichhorn}}]{Held2019}%
  \BibitemOpen
  \bibfield  {author} {\bibinfo {author} {\bibfnamefont {A.}~\bibnamefont
  {Held}}, \bibinfo {author} {\bibfnamefont {R.}~\bibnamefont {Gold}}, \ and\
  \bibinfo {author} {\bibfnamefont {A.}~\bibnamefont {Eichhorn}},\ }\href
  {\doibase 10.1088/1475-7516/2019/06/029} {\bibfield  {journal} {\bibinfo
  {journal} {\jcap}\ }06 (\bibinfo {year} {2019})\ \bibinfo {pages}
  {029}}\BibitemShut {NoStop}%
\bibitem [{\citenamefont {Jusufi}\ \emph {et~al.}(2019)\citenamefont {Jusufi},
  \citenamefont {Jamil}, \citenamefont {Salucci}, \citenamefont {Zhu},\ and\
  \citenamefont {Haroon}}]{Jusufi2019}%
  \BibitemOpen
  \bibfield  {author} {\bibinfo {author} {\bibfnamefont {K.}~\bibnamefont
  {Jusufi}}, \bibinfo {author} {\bibfnamefont {M.}~\bibnamefont {Jamil}},
  \bibinfo {author} {\bibfnamefont {P.}~\bibnamefont {Salucci}}, \bibinfo
  {author} {\bibfnamefont {T.}~\bibnamefont {Zhu}}, \ and\ \bibinfo {author}
  {\bibfnamefont {S.}~\bibnamefont {Haroon}},\ }\href {\doibase
  10.1103/PhysRevD.100.044012} {\bibfield  {journal} {\bibinfo  {journal}
  {Phys. Rev. D}\ }\textbf {\bibinfo {volume} {100}},\ \bibinfo {pages}
  {044012} (\bibinfo {year} {2019})}\BibitemShut {NoStop}%
\bibitem [{\citenamefont {Konoplya}(2019)}]{Konoplya2019c}%
  \BibitemOpen
  \bibfield  {author} {\bibinfo {author} {\bibfnamefont {R.~A.}\ \bibnamefont
  {Konoplya}},\ }\href {\doibase 10.1016/j.physletb.2019.05.043} {\bibfield
  {journal} {\bibinfo  {journal} {\plb}\ }\textbf {\bibinfo {volume} {795}},\
  \bibinfo {pages} {1 } (\bibinfo {year} {2019})}\BibitemShut {NoStop}%
\bibitem [{\citenamefont {Konoplya}\ and\ \citenamefont
  {Zhidenko}(2019)}]{Konoplya2019a}%
  \BibitemOpen
  \bibfield  {author} {\bibinfo {author} {\bibfnamefont {R.~A.}\ \bibnamefont
  {Konoplya}}\ and\ \bibinfo {author} {\bibfnamefont {A.}~\bibnamefont
  {Zhidenko}},\ }\href {\doibase 10.1103/PhysRevD.100.044015} {\bibfield
  {journal} {\bibinfo  {journal} {Phys. Rev. D}\ }\textbf {\bibinfo {volume}
  {100}},\ \bibinfo {pages} {044015} (\bibinfo {year} {2019})}\BibitemShut
  {NoStop}%
\bibitem [{\citenamefont {Konoplya}\ \emph {et~al.}()\citenamefont {Konoplya},
  \citenamefont {Pappas},\ and\ \citenamefont {Zhidenko}}]{Konoplya2019b}%
  \BibitemOpen
  \bibfield  {author} {\bibinfo {author} {\bibfnamefont {R.~A.}\ \bibnamefont
  {Konoplya}}, \bibinfo {author} {\bibfnamefont {T.}~\bibnamefont {Pappas}}, \
  and\ \bibinfo {author} {\bibfnamefont {A.}~\bibnamefont {Zhidenko}},\
  }\href@noop {} {}\Eprint {http://arxiv.org/abs/1907.10112} {arXiv:1907.10112}
  \BibitemShut {NoStop}%
\bibitem [{\citenamefont {Long}\ \emph {et~al.}()\citenamefont {Long},
  \citenamefont {Chen},\ and\ \citenamefont {Jing}}]{Long2019}%
  \BibitemOpen
  \bibfield  {author} {\bibinfo {author} {\bibfnamefont {F.}~\bibnamefont
  {Long}}, \bibinfo {author} {\bibfnamefont {S.}~\bibnamefont {Chen}}, \ and\
  \bibinfo {author} {\bibfnamefont {J.}~\bibnamefont {Jing}},\ }\href@noop {}
  {}\Eprint {http://arxiv.org/abs/1906.04456} {arXiv:1906.04456} \BibitemShut
  {NoStop}%
\bibitem [{\citenamefont {Medeiros}\ \emph {et~al.}()\citenamefont {Medeiros},
  \citenamefont {Psaltis},\ and\ \citenamefont {\"Ozel}}]{Medeiros2019}%
  \BibitemOpen
  \bibfield  {author} {\bibinfo {author} {\bibfnamefont {L.}~\bibnamefont
  {Medeiros}}, \bibinfo {author} {\bibfnamefont {D.}~\bibnamefont {Psaltis}}, \
  and\ \bibinfo {author} {\bibfnamefont {F.}~\bibnamefont {\"Ozel}},\
  }\href@noop {} {}\Eprint {http://arxiv.org/abs/1907.12575} {arXiv:1907.12575}
  \BibitemShut {NoStop}%
\bibitem [{\citenamefont {Neves}()}]{Neves2019}%
  \BibitemOpen
  \bibfield  {author} {\bibinfo {author} {\bibfnamefont {J.~C.~S.}\
  \bibnamefont {Neves}},\ }\href@noop {} {}\Eprint
  {http://arxiv.org/abs/1906.11735} {arXiv:1906.11735} \BibitemShut {NoStop}%
\bibitem [{\citenamefont {Rahman}\ and\ \citenamefont
  {Sen}(2019)}]{Rahman2019}%
  \BibitemOpen
  \bibfield  {author} {\bibinfo {author} {\bibfnamefont {M.}~\bibnamefont
  {Rahman}}\ and\ \bibinfo {author} {\bibfnamefont {A.~A.}\ \bibnamefont
  {Sen}},\ }\href {\doibase 10.1103/PhysRevD.99.024052} {\bibfield  {journal}
  {\bibinfo  {journal} {\prd}\ }\textbf {\bibinfo {volume} {99}},\ \bibinfo
  {pages} {024052} (\bibinfo {year} {2019})}\BibitemShut {NoStop}%
\bibitem [{\citenamefont {Shaikh}(2019)}]{Shaikh2019}%
  \BibitemOpen
  \bibfield  {author} {\bibinfo {author} {\bibfnamefont {R.}~\bibnamefont
  {Shaikh}},\ }\href {\doibase 10.1103/PhysRevD.100.024028} {\bibfield
  {journal} {\bibinfo  {journal} {Phys. Rev. D}\ }\textbf {\bibinfo {volume}
  {100}},\ \bibinfo {pages} {024028} (\bibinfo {year} {2019})}\BibitemShut
  {NoStop}%
\bibitem [{\citenamefont {Stuchl{\'i}k}\ and\ \citenamefont
  {Schee}(2019)}]{Stuchlik2019}%
  \BibitemOpen
  \bibfield  {author} {\bibinfo {author} {\bibfnamefont {Z.}~\bibnamefont
  {Stuchl{\'i}k}}\ and\ \bibinfo {author} {\bibfnamefont {J.}~\bibnamefont
  {Schee}},\ }\href {\doibase 10.1140/epjc/s10052-019-6543-8} {\bibfield
  {journal} {\bibinfo  {journal} {\epjc}\ }\textbf {\bibinfo {volume} {79}},\
  \bibinfo {pages} {44} (\bibinfo {year} {2019})}\BibitemShut {NoStop}%
\bibitem [{\citenamefont {Vagnozzi}\ and\ \citenamefont
  {Visinelli}(2019)}]{Vagnozzi2019}%
  \BibitemOpen
  \bibfield  {author} {\bibinfo {author} {\bibfnamefont {S.}~\bibnamefont
  {Vagnozzi}}\ and\ \bibinfo {author} {\bibfnamefont {L.}~\bibnamefont
  {Visinelli}},\ }\href {\doibase 10.1103/PhysRevD.100.024020} {\bibfield
  {journal} {\bibinfo  {journal} {Phys. Rev. D}\ }\textbf {\bibinfo {volume}
  {100}},\ \bibinfo {pages} {024020} (\bibinfo {year} {2019})}\BibitemShut
  {NoStop}%
\bibitem [{\citenamefont {Wang}\ \emph {et~al.}(2019)\citenamefont {Wang},
  \citenamefont {Xu},\ and\ \citenamefont {Wei}}]{Wang2019}%
  \BibitemOpen
  \bibfield  {author} {\bibinfo {author} {\bibfnamefont {H.-M.}\ \bibnamefont
  {Wang}}, \bibinfo {author} {\bibfnamefont {Y.-M.}\ \bibnamefont {Xu}}, \ and\
  \bibinfo {author} {\bibfnamefont {S.-W.}\ \bibnamefont {Wei}},\ }\href
  {\doibase 10.1088/1475-7516/2019/03/046} {\bibfield  {journal} {\bibinfo
  {journal} {\jcap}\ }03 (\bibinfo {year} {2019})\ \bibinfo {pages}
  {046}}\BibitemShut {NoStop}%
\bibitem [{\citenamefont {Zhu}\ \emph {et~al.}()\citenamefont {Zhu},
  \citenamefont {Wu}, \citenamefont {Jamil},\ and\ \citenamefont
  {Jusufi}}]{Zhu2019}%
  \BibitemOpen
  \bibfield  {author} {\bibinfo {author} {\bibfnamefont {T.}~\bibnamefont
  {Zhu}}, \bibinfo {author} {\bibfnamefont {Q.}~\bibnamefont {Wu}}, \bibinfo
  {author} {\bibfnamefont {M.}~\bibnamefont {Jamil}}, \ and\ \bibinfo {author}
  {\bibfnamefont {K.}~\bibnamefont {Jusufi}},\ }\href@noop {} {}\Eprint
  {http://arxiv.org/abs/1906.05673} {arXiv:1906.05673} \BibitemShut {NoStop}%
\bibitem [{\citenamefont {Broderick}\ \emph {et~al.}(2014)\citenamefont
  {Broderick}, \citenamefont {Johannsen}, \citenamefont {Loeb},\ and\
  \citenamefont {Psaltis}}]{Broderick2014}%
  \BibitemOpen
  \bibfield  {author} {\bibinfo {author} {\bibfnamefont {A.~E.}\ \bibnamefont
  {Broderick}}, \bibinfo {author} {\bibfnamefont {T.}~\bibnamefont
  {Johannsen}}, \bibinfo {author} {\bibfnamefont {A.}~\bibnamefont {Loeb}}, \
  and\ \bibinfo {author} {\bibfnamefont {D.}~\bibnamefont {Psaltis}},\ }\href
  {\doibase 10.1088/0004-637x/784/1/7} {\bibfield  {journal} {\bibinfo
  {journal} {\apj}\ }\textbf {\bibinfo {volume} {784}},\ \bibinfo {pages} {7}
  (\bibinfo {year} {2014})}\BibitemShut {NoStop}%
\bibitem [{\citenamefont {Johannsen}\ \emph {et~al.}(2016)\citenamefont
  {Johannsen}, \citenamefont {Wang}, \citenamefont {Broderick}, \citenamefont
  {Doeleman}, \citenamefont {Fish}, \citenamefont {Loeb},\ and\ \citenamefont
  {Psaltis}}]{Johannsen2016}%
  \BibitemOpen
  \bibfield  {author} {\bibinfo {author} {\bibfnamefont {T.}~\bibnamefont
  {Johannsen}}, \bibinfo {author} {\bibfnamefont {C.}~\bibnamefont {Wang}},
  \bibinfo {author} {\bibfnamefont {A.~E.}\ \bibnamefont {Broderick}}, \bibinfo
  {author} {\bibfnamefont {S.~S.}\ \bibnamefont {Doeleman}}, \bibinfo {author}
  {\bibfnamefont {V.~L.}\ \bibnamefont {Fish}}, \bibinfo {author}
  {\bibfnamefont {A.}~\bibnamefont {Loeb}}, \ and\ \bibinfo {author}
  {\bibfnamefont {D.}~\bibnamefont {Psaltis}},\ }\href {\doibase
  10.1103/PhysRevLett.117.091101} {\bibfield  {journal} {\bibinfo  {journal}
  {Phys. Rev. Lett.}\ }\textbf {\bibinfo {volume} {117}},\ \bibinfo {pages}
  {091101} (\bibinfo {year} {2016})}\BibitemShut {NoStop}%
\bibitem [{\citenamefont {Glampedakis}\ and\ \citenamefont
  {Babak}(2006)}]{Glampedakis2006}%
  \BibitemOpen
  \bibfield  {author} {\bibinfo {author} {\bibfnamefont {K.}~\bibnamefont
  {Glampedakis}}\ and\ \bibinfo {author} {\bibfnamefont {S.}~\bibnamefont
  {Babak}},\ }\href {\doibase 10.1088/0264-9381/23/12/013} {\bibfield
  {journal} {\bibinfo  {journal} {\cqg}\ }\textbf {\bibinfo {volume} {23}},\
  \bibinfo {pages} {4167} (\bibinfo {year} {2006})}\BibitemShut {NoStop}%
\bibitem [{\citenamefont {Mizuno}\ \emph {et~al.}(2018)\citenamefont {Mizuno},
  \citenamefont {Younsi}, \citenamefont {Fromm}, \citenamefont {Porth},
  \citenamefont {Laurentis}, \citenamefont {Olivares}, \citenamefont {Falcke},
  \citenamefont {Kramer},\ and\ \citenamefont {Rezzolla}}]{Mizuno2018}%
  \BibitemOpen
  \bibfield  {author} {\bibinfo {author} {\bibfnamefont {Y.}~\bibnamefont
  {Mizuno}}, \bibinfo {author} {\bibfnamefont {Z.}~\bibnamefont {Younsi}},
  \bibinfo {author} {\bibfnamefont {C.~M.}\ \bibnamefont {Fromm}}, \bibinfo
  {author} {\bibfnamefont {O.}~\bibnamefont {Porth}}, \bibinfo {author}
  {\bibfnamefont {M.~D.}\ \bibnamefont {Laurentis}}, \bibinfo {author}
  {\bibfnamefont {H.}~\bibnamefont {Olivares}}, \bibinfo {author}
  {\bibfnamefont {H.}~\bibnamefont {Falcke}}, \bibinfo {author} {\bibfnamefont
  {M.}~\bibnamefont {Kramer}}, \ and\ \bibinfo {author} {\bibfnamefont
  {L.}~\bibnamefont {Rezzolla}},\ }\href {\doibase 10.1038/s41550-018-0449-5}
  {\bibfield  {journal} {\bibinfo  {journal} {\naa}\ }\textbf {\bibinfo
  {volume} {2}},\ \bibinfo {pages} {585} (\bibinfo {year} {2018})}\BibitemShut
  {NoStop}%
\bibitem [{\citenamefont {Rezzolla}\ and\ \citenamefont
  {Zhidenko}(2014)}]{Rezzolla2014}%
  \BibitemOpen
  \bibfield  {author} {\bibinfo {author} {\bibfnamefont {L.}~\bibnamefont
  {Rezzolla}}\ and\ \bibinfo {author} {\bibfnamefont {A.}~\bibnamefont
  {Zhidenko}},\ }\href {\doibase 10.1103/PhysRevD.90.084009} {\bibfield
  {journal} {\bibinfo  {journal} {Phys. Rev. D}\ }\textbf {\bibinfo {volume}
  {90}},\ \bibinfo {pages} {084009} (\bibinfo {year} {2014})}\BibitemShut
  {NoStop}%
\bibitem [{\citenamefont {Bronzwaer}\ \emph {et~al.}(2018)\citenamefont
  {Bronzwaer}, \citenamefont {Davelaar}, \citenamefont {Younsi}, \citenamefont
  {Mo\'scibrodzka}, \citenamefont {Falcke}, \citenamefont {Kramer},\ and\
  \citenamefont {Rezzolla}}]{Bronzwaer2018}%
  \BibitemOpen
  \bibfield  {author} {\bibinfo {author} {\bibfnamefont {T.}~\bibnamefont
  {Bronzwaer}}, \bibinfo {author} {\bibfnamefont {J.}~\bibnamefont {Davelaar}},
  \bibinfo {author} {\bibfnamefont {Z.}~\bibnamefont {Younsi}}, \bibinfo
  {author} {\bibfnamefont {M.}~\bibnamefont {Mo\'scibrodzka}}, \bibinfo
  {author} {\bibfnamefont {H.}~\bibnamefont {Falcke}}, \bibinfo {author}
  {\bibfnamefont {M.}~\bibnamefont {Kramer}}, \ and\ \bibinfo {author}
  {\bibfnamefont {L.}~\bibnamefont {Rezzolla}},\ }\href {\doibase
  10.1051/0004-6361/201732149} {\bibfield  {journal} {\bibinfo  {journal}
  {\aa}\ }\textbf {\bibinfo {volume} {613}},\ \bibinfo {pages} {A2} (\bibinfo
  {year} {2018})}\BibitemShut {NoStop}%
\bibitem [{\citenamefont {Gyulchev}\ \emph {et~al.}(2019)\citenamefont
  {Gyulchev}, \citenamefont {Nedkova}, \citenamefont {Vetsov},\ and\
  \citenamefont {Yazadjiev}}]{Gyulchev2019}%
  \BibitemOpen
  \bibfield  {author} {\bibinfo {author} {\bibfnamefont {G.}~\bibnamefont
  {Gyulchev}}, \bibinfo {author} {\bibfnamefont {P.}~\bibnamefont {Nedkova}},
  \bibinfo {author} {\bibfnamefont {T.}~\bibnamefont {Vetsov}}, \ and\ \bibinfo
  {author} {\bibfnamefont {S.}~\bibnamefont {Yazadjiev}},\ }\href {\doibase
  10.1103/PhysRevD.100.024055} {\bibfield  {journal} {\bibinfo  {journal}
  {Phys. Rev. D}\ }\textbf {\bibinfo {volume} {100}},\ \bibinfo {pages}
  {024055} (\bibinfo {year} {2019})}\BibitemShut {NoStop}%
\bibitem [{\citenamefont {Clifton}\ \emph {et~al.}(2012)\citenamefont
  {Clifton}, \citenamefont {Ferreira}, \citenamefont {Padilla},\ and\
  \citenamefont {Skordis}}]{Clifton2012}%
  \BibitemOpen
  \bibfield  {author} {\bibinfo {author} {\bibfnamefont {T.}~\bibnamefont
  {Clifton}}, \bibinfo {author} {\bibfnamefont {P.~G.}\ \bibnamefont
  {Ferreira}}, \bibinfo {author} {\bibfnamefont {A.}~\bibnamefont {Padilla}}, \
  and\ \bibinfo {author} {\bibfnamefont {C.}~\bibnamefont {Skordis}},\ }\href
  {\doibase 10.1016/j.physrep.2012.01.001} {\bibfield  {journal} {\bibinfo
  {journal} {\prep}\ }\textbf {\bibinfo {volume} {513}},\ \bibinfo {pages} {1 }
  (\bibinfo {year} {2012})}\BibitemShut {NoStop}%
\bibitem [{\citenamefont {Will}(2014)}]{Will2014}%
  \BibitemOpen
  \bibfield  {author} {\bibinfo {author} {\bibfnamefont {C.~M.}\ \bibnamefont
  {Will}},\ }\href {\doibase 10.12942/lrr-2014-4} {\bibfield  {journal}
  {\bibinfo  {journal} {\lrr}\ }\textbf {\bibinfo {volume} {17}},\ \bibinfo
  {pages} {4} (\bibinfo {year} {2014})}\BibitemShut {NoStop}%
\bibitem [{\citenamefont {Tian}\ and\ \citenamefont {Zhu}(2019)}]{Tian2019}%
  \BibitemOpen
  \bibfield  {author} {\bibinfo {author} {\bibfnamefont {S.~X.}\ \bibnamefont
  {Tian}}\ and\ \bibinfo {author} {\bibfnamefont {Z.-H.}\ \bibnamefont {Zhu}},\
  }\href {\doibase 10.1103/PhysRevD.99.064044} {\bibfield  {journal} {\bibinfo
  {journal} {Phys. Rev. D}\ }\textbf {\bibinfo {volume} {99}},\ \bibinfo
  {pages} {064044} (\bibinfo {year} {2019})}\BibitemShut {NoStop}%
\bibitem [{\citenamefont {Dokuchaev}\ and\ \citenamefont
  {Nazarova}(2019)}]{Dokuchaev2019}%
  \BibitemOpen
  \bibfield  {author} {\bibinfo {author} {\bibfnamefont {V.~I.}\ \bibnamefont
  {Dokuchaev}}\ and\ \bibinfo {author} {\bibfnamefont {N.~O.}\ \bibnamefont
  {Nazarova}},\ }\href {\doibase 10.3390/universe5080183} {\bibfield  {journal}
  {\bibinfo  {journal} {Universe}\ }\textbf {\bibinfo {volume} {5}},\ \bibinfo
  {pages} {183} (\bibinfo {year} {2019})}\BibitemShut {NoStop}%
\bibitem [{\citenamefont {Doeleman~\etal}(2012)}]{Doeleman2012}%
  \BibitemOpen
  \bibfield  {author} {\bibinfo {author} {\bibfnamefont {S.~S.}\ \bibnamefont
  {Doeleman~\etal}},\ }\href {\doibase 10.1126/science.1224768} {\bibfield
  {journal} {\bibinfo  {journal} {Science}\ }\textbf {\bibinfo {volume}
  {338}},\ \bibinfo {pages} {355} (\bibinfo {year} {2012})}\BibitemShut
  {NoStop}%
\bibitem [{\citenamefont {Masanori~\etal}(2018)}]{Nakamura2018}%
  \BibitemOpen
  \bibfield  {author} {\bibinfo {author} {\bibfnamefont {N.}~\bibnamefont
  {Masanori~\etal}},\ }\href {\doibase 10.3847/1538-4357/aaeb2d} {\bibfield
  {journal} {\bibinfo  {journal} {\apj}\ }\textbf {\bibinfo {volume} {868}},\
  \bibinfo {pages} {146} (\bibinfo {year} {2018})}\BibitemShut {NoStop}%
\bibitem [{\citenamefont {Takahashi}\ \emph {et~al.}(2018)\citenamefont
  {Takahashi}, \citenamefont {Toma}, \citenamefont {Kino}, \citenamefont
  {Nakamura},\ and\ \citenamefont {Hada}}]{Takahashi2018}%
  \BibitemOpen
  \bibfield  {author} {\bibinfo {author} {\bibfnamefont {K.}~\bibnamefont
  {Takahashi}}, \bibinfo {author} {\bibfnamefont {K.}~\bibnamefont {Toma}},
  \bibinfo {author} {\bibfnamefont {M.}~\bibnamefont {Kino}}, \bibinfo {author}
  {\bibfnamefont {M.}~\bibnamefont {Nakamura}}, \ and\ \bibinfo {author}
  {\bibfnamefont {K.}~\bibnamefont {Hada}},\ }\href {\doibase
  10.3847/1538-4357/aae832} {\bibfield  {journal} {\bibinfo  {journal} {\apj}\
  }\textbf {\bibinfo {volume} {868}},\ \bibinfo {pages} {82} (\bibinfo {year}
  {2018})}\BibitemShut {NoStop}%
\bibitem [{\citenamefont {Bambi}\ \emph {et~al.}()\citenamefont {Bambi},
  \citenamefont {Freese}, \citenamefont {Vagnozzi},\ and\ \citenamefont
  {Visinelli}}]{Bambi2019}%
  \BibitemOpen
  \bibfield  {author} {\bibinfo {author} {\bibfnamefont {C.}~\bibnamefont
  {Bambi}}, \bibinfo {author} {\bibfnamefont {K.}~\bibnamefont {Freese}},
  \bibinfo {author} {\bibfnamefont {S.}~\bibnamefont {Vagnozzi}}, \ and\
  \bibinfo {author} {\bibfnamefont {L.}~\bibnamefont {Visinelli}},\ }\href@noop
  {} {}\Eprint {http://arxiv.org/abs/1904.12983} {arXiv:1904.12983}
  \BibitemShut {NoStop}%
\bibitem [{\citenamefont {Tamburini}\ \emph {et~al.}()\citenamefont
  {Tamburini}, \citenamefont {Thid\'e},\ and\ \citenamefont
  {Valle}}]{Tamburini2019}%
  \BibitemOpen
  \bibfield  {author} {\bibinfo {author} {\bibfnamefont {F.}~\bibnamefont
  {Tamburini}}, \bibinfo {author} {\bibfnamefont {B.}~\bibnamefont {Thid\'e}},
  \ and\ \bibinfo {author} {\bibfnamefont {M.~D.}\ \bibnamefont {Valle}},\
  }\href@noop {} {}\Eprint {http://arxiv.org/abs/1904.07923} {arXiv:1904.07923}
  \BibitemShut {NoStop}%
\bibitem [{\citenamefont {Li}\ \emph {et~al.}(2009)\citenamefont {Li},
  \citenamefont {Yuan}, \citenamefont {Wang}, \citenamefont {Wang},\ and\
  \citenamefont {Zhang}}]{Li2009}%
  \BibitemOpen
  \bibfield  {author} {\bibinfo {author} {\bibfnamefont {Y.-R.}\ \bibnamefont
  {Li}}, \bibinfo {author} {\bibfnamefont {Y.-F.}\ \bibnamefont {Yuan}},
  \bibinfo {author} {\bibfnamefont {J.-M.}\ \bibnamefont {Wang}}, \bibinfo
  {author} {\bibfnamefont {J.-C.}\ \bibnamefont {Wang}}, \ and\ \bibinfo
  {author} {\bibfnamefont {S.}~\bibnamefont {Zhang}},\ }\href {\doibase
  10.1088/0004-637x/699/1/513} {\bibfield  {journal} {\bibinfo  {journal}
  {\apj}\ }\textbf {\bibinfo {volume} {699}},\ \bibinfo {pages} {513} (\bibinfo
  {year} {2009})}\BibitemShut {NoStop}%
\bibitem [{\citenamefont {Kawashima}\ \emph {et~al.}(2019)\citenamefont
  {Kawashima}, \citenamefont {Kino},\ and\ \citenamefont
  {Akiyama}}]{Kawashima2019}%
  \BibitemOpen
  \bibfield  {author} {\bibinfo {author} {\bibfnamefont {T.}~\bibnamefont
  {Kawashima}}, \bibinfo {author} {\bibfnamefont {M.}~\bibnamefont {Kino}}, \
  and\ \bibinfo {author} {\bibfnamefont {K.}~\bibnamefont {Akiyama}},\ }\href
  {\doibase 10.3847/1538-4357/ab19c0} {\bibfield  {journal} {\bibinfo
  {journal} {\apj}\ }\textbf {\bibinfo {volume} {878}},\ \bibinfo {pages} {27}
  (\bibinfo {year} {2019})}\BibitemShut {NoStop}%
\bibitem [{\citenamefont {Nemmen}(2019)}]{Nemmen2019}%
  \BibitemOpen
  \bibfield  {author} {\bibinfo {author} {\bibfnamefont {R.}~\bibnamefont
  {Nemmen}},\ }\href {\doibase 10.3847/2041-8213/ab2fd3} {\bibfield  {journal}
  {\bibinfo  {journal} {\apjl}\ }\textbf {\bibinfo {volume} {880}},\ \bibinfo
  {pages} {L26} (\bibinfo {year} {2019})}\BibitemShut {NoStop}%
\bibitem [{\citenamefont {Nokhrina}\ \emph {et~al.}()\citenamefont {Nokhrina},
  \citenamefont {Gurvits}, \citenamefont {Beskin}, \citenamefont {Nakamura},
  \citenamefont {Asada},\ and\ \citenamefont {Hada}}]{Nokhrina2019}%
  \BibitemOpen
  \bibfield  {author} {\bibinfo {author} {\bibfnamefont {E.~E.}\ \bibnamefont
  {Nokhrina}}, \bibinfo {author} {\bibfnamefont {L.~I.}\ \bibnamefont
  {Gurvits}}, \bibinfo {author} {\bibfnamefont {V.~S.}\ \bibnamefont {Beskin}},
  \bibinfo {author} {\bibfnamefont {M.}~\bibnamefont {Nakamura}}, \bibinfo
  {author} {\bibfnamefont {K.}~\bibnamefont {Asada}}, \ and\ \bibinfo {author}
  {\bibfnamefont {K.}~\bibnamefont {Hada}},\ }\href@noop {} {}\Eprint
  {http://arxiv.org/abs/1904.05665} {arXiv:1904.05665} \BibitemShut {NoStop}%
\bibitem [{\citenamefont {Yunes}\ and\ \citenamefont
  {Stein}(2011)}]{Yunes2011}%
  \BibitemOpen
  \bibfield  {author} {\bibinfo {author} {\bibfnamefont {N.}~\bibnamefont
  {Yunes}}\ and\ \bibinfo {author} {\bibfnamefont {L.~C.}\ \bibnamefont
  {Stein}},\ }\href {\doibase 10.1103/PhysRevD.83.104002} {\bibfield  {journal}
  {\bibinfo  {journal} {Phys. Rev. D}\ }\textbf {\bibinfo {volume} {83}},\
  \bibinfo {pages} {104002} (\bibinfo {year} {2011})}\BibitemShut {NoStop}%
\bibitem [{\citenamefont {Shakura}\ and\ \citenamefont
  {Sunyaev}(1973)}]{Shakura1973}%
  \BibitemOpen
  \bibfield  {author} {\bibinfo {author} {\bibfnamefont {N.~I.}\ \bibnamefont
  {Shakura}}\ and\ \bibinfo {author} {\bibfnamefont {R.~A.}\ \bibnamefont
  {Sunyaev}},\ }\href {http://adsabs.harvard.edu/abs/1973A\%26A....24..337S}
  {\bibfield  {journal} {\bibinfo  {journal} {\aa}\ }\textbf {\bibinfo {volume}
  {24}},\ \bibinfo {pages} {337} (\bibinfo {year} {1973})}\BibitemShut
  {NoStop}%
\bibitem [{\citenamefont {Pringle}(1981)}]{Pringle1981}%
  \BibitemOpen
  \bibfield  {author} {\bibinfo {author} {\bibfnamefont {J.~E.}\ \bibnamefont
  {Pringle}},\ }\href {\doibase 10.1146/annurev.aa.19.090181.001033} {\bibfield
   {journal} {\bibinfo  {journal} {\araa}\ }\textbf {\bibinfo {volume} {19}},\
  \bibinfo {pages} {137} (\bibinfo {year} {1981})}\BibitemShut {NoStop}%
\bibitem [{\citenamefont {Chakrabarti}(1996)}]{Chakrabarti1996}%
  \BibitemOpen
  \bibfield  {author} {\bibinfo {author} {\bibfnamefont {S.~K.}\ \bibnamefont
  {Chakrabarti}},\ }\href {\doibase 10.1016/0370-1573(95)00057-7} {\bibfield
  {journal} {\bibinfo  {journal} {\prep}\ }\textbf {\bibinfo {volume} {266}},\
  \bibinfo {pages} {229 } (\bibinfo {year} {1996})}\BibitemShut {NoStop}%
\bibitem [{\citenamefont {Abramowicz}\ and\ \citenamefont
  {Fragile}(2013)}]{Abramowicz2013}%
  \BibitemOpen
  \bibfield  {author} {\bibinfo {author} {\bibfnamefont {M.~A.}\ \bibnamefont
  {Abramowicz}}\ and\ \bibinfo {author} {\bibfnamefont {P.~C.}\ \bibnamefont
  {Fragile}},\ }\href {\doibase 10.12942/lrr-2013-1} {\bibfield  {journal}
  {\bibinfo  {journal} {\lrr}\ }\textbf {\bibinfo {volume} {16}},\ \bibinfo
  {pages} {1} (\bibinfo {year} {2013})}\BibitemShut {NoStop}%
\bibitem [{\citenamefont {Kato}\ \emph {et~al.}(2008)\citenamefont {Kato},
  \citenamefont {Fukue},\ and\ \citenamefont {Mineshige}}]{Kato2008}%
  \BibitemOpen
  \bibfield  {author} {\bibinfo {author} {\bibfnamefont {S.}~\bibnamefont
  {Kato}}, \bibinfo {author} {\bibfnamefont {J.}~\bibnamefont {Fukue}}, \ and\
  \bibinfo {author} {\bibfnamefont {S.}~\bibnamefont {Mineshige}},\ }\href@noop
  {} {\emph {\bibinfo {title} {Black-Hole Accretion Disks: Towards a New
  Paradigm}}}\ (\bibinfo  {publisher} {Kyoto University Press},\ \bibinfo
  {address} {Kyoto},\ \bibinfo {year} {2008})\ pp.\ \bibinfo {pages}
  {118,214}\BibitemShut {NoStop}%
\bibitem [{\citenamefont {Punsly}(2019)}]{Punsly2019}%
  \BibitemOpen
  \bibfield  {author} {\bibinfo {author} {\bibfnamefont {B.}~\bibnamefont
  {Punsly}},\ }\href {\doibase 10.3847/2041-8213/ab2a0e} {\bibfield  {journal}
  {\bibinfo  {journal} {\apjl}\ }\textbf {\bibinfo {volume} {879}},\ \bibinfo
  {pages} {L11} (\bibinfo {year} {2019})}\BibitemShut {NoStop}%
\bibitem [{\citenamefont {{Novikov}}\ and\ \citenamefont
  {{Thorne}}(1973)}]{Novikov1973}%
  \BibitemOpen
  \bibfield  {author} {\bibinfo {author} {\bibfnamefont {I.~D.}\ \bibnamefont
  {{Novikov}}}\ and\ \bibinfo {author} {\bibfnamefont {K.~S.}\ \bibnamefont
  {{Thorne}}},\ }in\ \href
  {https://www.its.caltech.edu/~kip/scripts/PubScans/II-48.pdf} {\emph
  {\bibinfo {booktitle} {Black Holes (Les Astres Occlus)}}},\ \bibinfo {editor}
  {edited by\ \bibinfo {editor} {\bibfnamefont {C.}~\bibnamefont {{Dewitt}}}\
  and\ \bibinfo {editor} {\bibfnamefont {B.~S.}\ \bibnamefont {{Dewitt}}}}\
  (\bibinfo  {publisher} {Gordon and Breach},\ \bibinfo {address} {New York},\
  \bibinfo {year} {1973})\ pp.\ \bibinfo {pages} {343--450}\BibitemShut
  {NoStop}%
\bibitem [{\citenamefont {Page}\ and\ \citenamefont {Thorne}(1974)}]{Page1974}%
  \BibitemOpen
  \bibfield  {author} {\bibinfo {author} {\bibfnamefont {D.~N.}\ \bibnamefont
  {Page}}\ and\ \bibinfo {author} {\bibfnamefont {K.~S.}\ \bibnamefont
  {Thorne}},\ }\href {\doibase 10.1086/152990} {\bibfield  {journal} {\bibinfo
  {journal} {Astrophys. J.}\ }\textbf {\bibinfo {volume} {191}},\ \bibinfo
  {pages} {499} (\bibinfo {year} {1974})}\BibitemShut {NoStop}%
\bibitem [{\citenamefont {Misner}\ \emph {et~al.}(1973)\citenamefont {Misner},
  \citenamefont {Thorne},\ and\ \citenamefont {Wheeler}}]{Misner1973}%
  \BibitemOpen
  \bibfield  {author} {\bibinfo {author} {\bibfnamefont {C.~W.}\ \bibnamefont
  {Misner}}, \bibinfo {author} {\bibfnamefont {K.~S.}\ \bibnamefont {Thorne}},
  \ and\ \bibinfo {author} {\bibfnamefont {J.~A.}\ \bibnamefont {Wheeler}},\
  }\href@noop {} {\emph {\bibinfo {title} {Gravitation}}}\ (\bibinfo
  {publisher} {W. H. Freeman \& Co.},\ \bibinfo {address} {San Francisco},\
  \bibinfo {year} {1973})\ pp.\ \bibinfo {pages} {567,570}\BibitemShut
  {NoStop}%
\bibitem [{\citenamefont {C\'ardenas-Avenda\~no}\ \emph
  {et~al.}(2019)\citenamefont {C\'ardenas-Avenda\~no}, \citenamefont {Godfrey},
  \citenamefont {Yunes},\ and\ \citenamefont
  {Lohfink}}]{Cardenas-Avendano2019}%
  \BibitemOpen
  \bibfield  {author} {\bibinfo {author} {\bibfnamefont {A.}~\bibnamefont
  {C\'ardenas-Avenda\~no}}, \bibinfo {author} {\bibfnamefont {J.}~\bibnamefont
  {Godfrey}}, \bibinfo {author} {\bibfnamefont {N.}~\bibnamefont {Yunes}}, \
  and\ \bibinfo {author} {\bibfnamefont {A.}~\bibnamefont {Lohfink}},\ }\href
  {\doibase 10.1103/PhysRevD.100.024039} {\bibfield  {journal} {\bibinfo
  {journal} {Phys. Rev. D}\ }\textbf {\bibinfo {volume} {100}},\ \bibinfo
  {pages} {024039} (\bibinfo {year} {2019})}\BibitemShut {NoStop}%
\bibitem [{\citenamefont {Fukue}\ and\ \citenamefont
  {Yokoyama}(1988)}]{Fukue1988}%
  \BibitemOpen
  \bibfield  {author} {\bibinfo {author} {\bibfnamefont {J.}~\bibnamefont
  {Fukue}}\ and\ \bibinfo {author} {\bibfnamefont {T.}~\bibnamefont
  {Yokoyama}},\ }\href {http://adsabs.harvard.edu/abs/1988PASJ...40...15F}
  {\bibfield  {journal} {\bibinfo  {journal} {\pasj}\ }\textbf {\bibinfo
  {volume} {40}},\ \bibinfo {pages} {15} (\bibinfo {year} {1988})}\BibitemShut
  {NoStop}%
\bibitem [{\citenamefont {{Horne}}\ and\ \citenamefont
  {{Marsh}}(1986)}]{Horne1986}%
  \BibitemOpen
  \bibfield  {author} {\bibinfo {author} {\bibfnamefont {K.}~\bibnamefont
  {{Horne}}}\ and\ \bibinfo {author} {\bibfnamefont {T.~R.}\ \bibnamefont
  {{Marsh}}},\ }\href {\doibase 10.1093/mnras/218.4.761} {\bibfield  {journal}
  {\bibinfo  {journal} {\mnras}\ }\textbf {\bibinfo {volume} {218}},\ \bibinfo
  {pages} {761} (\bibinfo {year} {1986})}\BibitemShut {NoStop}%
\bibitem [{\citenamefont {{Ellis}}(1971)}]{Ellis1971}%
  \BibitemOpen
  \bibfield  {author} {\bibinfo {author} {\bibfnamefont {G.~F.~R.}\
  \bibnamefont {{Ellis}}},\ }in\ \href
  {https://ui.adsabs.harvard.edu/abs/1971grc..conf..104E} {\emph {\bibinfo
  {booktitle} {General Relativity and Cosmology}}},\ \bibinfo {editor} {edited
  by\ \bibinfo {editor} {\bibfnamefont {R.~K.}\ \bibnamefont {{Sachs}}}}\
  (\bibinfo  {publisher} {Academic Press},\ \bibinfo {address} {New York and
  London},\ \bibinfo {year} {1971})\ pp.\ \bibinfo {pages}
  {104--182}\BibitemShut {NoStop}%
\bibitem [{\citenamefont {Ellis}(2009)}]{Ellis2009}%
  \BibitemOpen
  \bibfield  {author} {\bibinfo {author} {\bibfnamefont {G.~F.~R.}\
  \bibnamefont {Ellis}},\ }\href {\doibase 10.1007/s10714-009-0760-7}
  {\bibfield  {journal} {\bibinfo  {journal} {\gerg}\ }\textbf {\bibinfo
  {volume} {41}},\ \bibinfo {pages} {581} (\bibinfo {year} {2009})}\BibitemShut
  {NoStop}%
\bibitem [{\citenamefont {Gebhardt}\ and\ \citenamefont
  {Thomas}(2009)}]{Gebhardt2009}%
  \BibitemOpen
  \bibfield  {author} {\bibinfo {author} {\bibfnamefont {K.}~\bibnamefont
  {Gebhardt}}\ and\ \bibinfo {author} {\bibfnamefont {J.}~\bibnamefont
  {Thomas}},\ }\href {\doibase 10.1088/0004-637x/700/2/1690} {\bibfield
  {journal} {\bibinfo  {journal} {\apj}\ }\textbf {\bibinfo {volume} {700}},\
  \bibinfo {pages} {1690} (\bibinfo {year} {2009})}\BibitemShut {NoStop}%
\bibitem [{\citenamefont {Gebhardt}\ \emph {et~al.}(2011)\citenamefont
  {Gebhardt}, \citenamefont {Adams}, \citenamefont {Richstone}, \citenamefont
  {Lauer}, \citenamefont {Faber}, \citenamefont {Gültekin}, \citenamefont
  {Murphy},\ and\ \citenamefont {Tremaine}}]{Gebhardt2011}%
  \BibitemOpen
  \bibfield  {author} {\bibinfo {author} {\bibfnamefont {K.}~\bibnamefont
  {Gebhardt}}, \bibinfo {author} {\bibfnamefont {J.}~\bibnamefont {Adams}},
  \bibinfo {author} {\bibfnamefont {D.}~\bibnamefont {Richstone}}, \bibinfo
  {author} {\bibfnamefont {T.~R.}\ \bibnamefont {Lauer}}, \bibinfo {author}
  {\bibfnamefont {S.~M.}\ \bibnamefont {Faber}}, \bibinfo {author}
  {\bibfnamefont {K.}~\bibnamefont {Gültekin}}, \bibinfo {author}
  {\bibfnamefont {J.}~\bibnamefont {Murphy}}, \ and\ \bibinfo {author}
  {\bibfnamefont {S.}~\bibnamefont {Tremaine}},\ }\href {\doibase
  10.1088/0004-637x/729/2/119} {\bibfield  {journal} {\bibinfo  {journal}
  {\apj}\ }\textbf {\bibinfo {volume} {729}},\ \bibinfo {pages} {119} (\bibinfo
  {year} {2011})}\BibitemShut {NoStop}%
\bibitem [{\citenamefont {{Harms}}\ \emph {et~al.}(1994)\citenamefont
  {{Harms}}, \citenamefont {{Ford}}, \citenamefont {{Tsvetanov}}, \citenamefont
  {{Hartig}}, \citenamefont {{Dressel}}, \citenamefont {{Kriss}}, \citenamefont
  {{Bohlin}}, \citenamefont {{Davidsen}}, \citenamefont {{Margon}},\ and\
  \citenamefont {{Kochhar}}}]{Harms1994}%
  \BibitemOpen
  \bibfield  {author} {\bibinfo {author} {\bibfnamefont {R.~J.}\ \bibnamefont
  {{Harms}}}, \bibinfo {author} {\bibfnamefont {H.~C.}\ \bibnamefont {{Ford}}},
  \bibinfo {author} {\bibfnamefont {Z.~I.}\ \bibnamefont {{Tsvetanov}}},
  \bibinfo {author} {\bibfnamefont {G.~F.}\ \bibnamefont {{Hartig}}}, \bibinfo
  {author} {\bibfnamefont {L.~L.}\ \bibnamefont {{Dressel}}}, \bibinfo {author}
  {\bibfnamefont {G.~A.}\ \bibnamefont {{Kriss}}}, \bibinfo {author}
  {\bibfnamefont {R.}~\bibnamefont {{Bohlin}}}, \bibinfo {author}
  {\bibfnamefont {A.~F.}\ \bibnamefont {{Davidsen}}}, \bibinfo {author}
  {\bibfnamefont {B.}~\bibnamefont {{Margon}}}, \ and\ \bibinfo {author}
  {\bibfnamefont {A.~K.}\ \bibnamefont {{Kochhar}}},\ }\href {\doibase
  10.1086/187588} {\bibfield  {journal} {\bibinfo  {journal} {\apjl}\ }\textbf
  {\bibinfo {volume} {435}},\ \bibinfo {pages} {L35} (\bibinfo {year}
  {1994})}\BibitemShut {NoStop}%
\bibitem [{\citenamefont {Macchetto}\ \emph {et~al.}(1997)\citenamefont
  {Macchetto}, \citenamefont {Marconi}, \citenamefont {Axon}, \citenamefont
  {Capetti}, \citenamefont {Sparks},\ and\ \citenamefont
  {Crane}}]{Macchetto1997}%
  \BibitemOpen
  \bibfield  {author} {\bibinfo {author} {\bibfnamefont {F.}~\bibnamefont
  {Macchetto}}, \bibinfo {author} {\bibfnamefont {A.}~\bibnamefont {Marconi}},
  \bibinfo {author} {\bibfnamefont {D.~J.}\ \bibnamefont {Axon}}, \bibinfo
  {author} {\bibfnamefont {A.}~\bibnamefont {Capetti}}, \bibinfo {author}
  {\bibfnamefont {W.}~\bibnamefont {Sparks}}, \ and\ \bibinfo {author}
  {\bibfnamefont {P.}~\bibnamefont {Crane}},\ }\href {\doibase 10.1086/304823}
  {\bibfield  {journal} {\bibinfo  {journal} {\apj}\ }\textbf {\bibinfo
  {volume} {489}},\ \bibinfo {pages} {579} (\bibinfo {year}
  {1997})}\BibitemShut {NoStop}%
\bibitem [{\citenamefont {Walsh}\ \emph {et~al.}(2013)\citenamefont {Walsh},
  \citenamefont {Barth}, \citenamefont {Ho},\ and\ \citenamefont
  {Sarzi}}]{Walsh2013}%
  \BibitemOpen
  \bibfield  {author} {\bibinfo {author} {\bibfnamefont {J.~L.}\ \bibnamefont
  {Walsh}}, \bibinfo {author} {\bibfnamefont {A.~J.}\ \bibnamefont {Barth}},
  \bibinfo {author} {\bibfnamefont {L.~C.}\ \bibnamefont {Ho}}, \ and\ \bibinfo
  {author} {\bibfnamefont {M.}~\bibnamefont {Sarzi}},\ }\href {\doibase
  10.1088/0004-637x/770/2/86} {\bibfield  {journal} {\bibinfo  {journal}
  {\apj}\ }\textbf {\bibinfo {volume} {770}},\ \bibinfo {pages} {86} (\bibinfo
  {year} {2013})}\BibitemShut {NoStop}%
\bibitem [{\citenamefont {Ferreira}()}]{Ferreira2019}%
  \BibitemOpen
  \bibfield  {author} {\bibinfo {author} {\bibfnamefont {P.~G.}\ \bibnamefont
  {Ferreira}},\ }\href@noop {} {}\Eprint {http://arxiv.org/abs/1902.10503}
  {arXiv:1902.10503} \BibitemShut {NoStop}%
\bibitem [{\citenamefont {De~Felice}\ and\ \citenamefont
  {Tsujikawa}(2010)}]{DeFelice2010}%
  \BibitemOpen
  \bibfield  {author} {\bibinfo {author} {\bibfnamefont {A.}~\bibnamefont
  {De~Felice}}\ and\ \bibinfo {author} {\bibfnamefont {S.}~\bibnamefont
  {Tsujikawa}},\ }\href {\doibase 10.12942/lrr-2010-3} {\bibfield  {journal}
  {\bibinfo  {journal} {\lrr}\ }\textbf {\bibinfo {volume} {13}},\ \bibinfo
  {pages} {3} (\bibinfo {year} {2010})}\BibitemShut {NoStop}%
\bibitem [{\citenamefont {Sotiriou}\ and\ \citenamefont
  {Faraoni}(2010)}]{Sotiriou2010}%
  \BibitemOpen
  \bibfield  {author} {\bibinfo {author} {\bibfnamefont {T.~P.}\ \bibnamefont
  {Sotiriou}}\ and\ \bibinfo {author} {\bibfnamefont {V.}~\bibnamefont
  {Faraoni}},\ }\href {\doibase 10.1103/RevModPhys.82.451} {\bibfield
  {journal} {\bibinfo  {journal} {Rev. Mod. Phys.}\ }\textbf {\bibinfo {volume}
  {82}},\ \bibinfo {pages} {451} (\bibinfo {year} {2010})}\BibitemShut
  {NoStop}%
\bibitem [{\citenamefont {Nojiri}\ and\ \citenamefont
  {Odintsov}(2011)}]{Nojiri2011}%
  \BibitemOpen
  \bibfield  {author} {\bibinfo {author} {\bibfnamefont {S.}~\bibnamefont
  {Nojiri}}\ and\ \bibinfo {author} {\bibfnamefont {S.~D.}\ \bibnamefont
  {Odintsov}},\ }\href {\doibase 10.1016/j.physrep.2011.04.001} {\bibfield
  {journal} {\bibinfo  {journal} {\prep}\ }\textbf {\bibinfo {volume} {505}},\
  \bibinfo {pages} {59 } (\bibinfo {year} {2011})}\BibitemShut {NoStop}%
\bibitem [{\citenamefont {Joyce}\ \emph {et~al.}(2015)\citenamefont {Joyce},
  \citenamefont {Jain}, \citenamefont {Khoury},\ and\ \citenamefont
  {Trodden}}]{Joyce2015}%
  \BibitemOpen
  \bibfield  {author} {\bibinfo {author} {\bibfnamefont {A.}~\bibnamefont
  {Joyce}}, \bibinfo {author} {\bibfnamefont {B.}~\bibnamefont {Jain}},
  \bibinfo {author} {\bibfnamefont {J.}~\bibnamefont {Khoury}}, \ and\ \bibinfo
  {author} {\bibfnamefont {M.}~\bibnamefont {Trodden}},\ }\href {\doibase
  10.1016/j.physrep.2014.12.002} {\bibfield  {journal} {\bibinfo  {journal}
  {\prep}\ }\textbf {\bibinfo {volume} {568}},\ \bibinfo {pages} {1 } (\bibinfo
  {year} {2015})}\BibitemShut {NoStop}%
\bibitem [{\citenamefont {Nojiri}\ \emph {et~al.}(2017)\citenamefont {Nojiri},
  \citenamefont {Odintsov},\ and\ \citenamefont {Oikonomou}}]{Nojiri2017a}%
  \BibitemOpen
  \bibfield  {author} {\bibinfo {author} {\bibfnamefont {S.}~\bibnamefont
  {Nojiri}}, \bibinfo {author} {\bibfnamefont {S.}~\bibnamefont {Odintsov}}, \
  and\ \bibinfo {author} {\bibfnamefont {V.}~\bibnamefont {Oikonomou}},\ }\href
  {\doibase 10.1016/j.physrep.2017.06.001} {\bibfield  {journal} {\bibinfo
  {journal} {\prep}\ }\textbf {\bibinfo {volume} {692}},\ \bibinfo {pages} {1 }
  (\bibinfo {year} {2017})}\BibitemShut {NoStop}%
\bibitem [{\citenamefont {Capozziello}\ \emph {et~al.}(2007)\citenamefont
  {Capozziello}, \citenamefont {Stabile},\ and\ \citenamefont
  {Troisi}}]{Capozziello2007}%
  \BibitemOpen
  \bibfield  {author} {\bibinfo {author} {\bibfnamefont {S.}~\bibnamefont
  {Capozziello}}, \bibinfo {author} {\bibfnamefont {A.}~\bibnamefont
  {Stabile}}, \ and\ \bibinfo {author} {\bibfnamefont {A.}~\bibnamefont
  {Troisi}},\ }\href {\doibase 10.1103/PhysRevD.76.104019} {\bibfield
  {journal} {\bibinfo  {journal} {Phys. Rev. D}\ }\textbf {\bibinfo {volume}
  {76}},\ \bibinfo {pages} {104019} (\bibinfo {year} {2007})}\BibitemShut
  {NoStop}%
\bibitem [{\citenamefont {Chiba}\ \emph {et~al.}(2007)\citenamefont {Chiba},
  \citenamefont {Smith},\ and\ \citenamefont {Erickcek}}]{Chiba2007}%
  \BibitemOpen
  \bibfield  {author} {\bibinfo {author} {\bibfnamefont {T.}~\bibnamefont
  {Chiba}}, \bibinfo {author} {\bibfnamefont {T.~L.}\ \bibnamefont {Smith}}, \
  and\ \bibinfo {author} {\bibfnamefont {A.~L.}\ \bibnamefont {Erickcek}},\
  }\href {\doibase 10.1103/PhysRevD.75.124014} {\bibfield  {journal} {\bibinfo
  {journal} {Phys. Rev. D}\ }\textbf {\bibinfo {volume} {75}},\ \bibinfo
  {pages} {124014} (\bibinfo {year} {2007})}\BibitemShut {NoStop}%
\bibitem [{\citenamefont {Faraoni}\ and\ \citenamefont
  {Lanahan-Tremblay}(2008)}]{Faraoni2008}%
  \BibitemOpen
  \bibfield  {author} {\bibinfo {author} {\bibfnamefont {V.}~\bibnamefont
  {Faraoni}}\ and\ \bibinfo {author} {\bibfnamefont {N.}~\bibnamefont
  {Lanahan-Tremblay}},\ }\href {\doibase 10.1103/PhysRevD.77.108501} {\bibfield
   {journal} {\bibinfo  {journal} {Phys. Rev. D}\ }\textbf {\bibinfo {volume}
  {77}},\ \bibinfo {pages} {108501} (\bibinfo {year} {2008})}\BibitemShut
  {NoStop}%
\bibitem [{\citenamefont {Tian}()}]{Tian2018a}%
  \BibitemOpen
  \bibfield  {author} {\bibinfo {author} {\bibfnamefont {S.}~\bibnamefont
  {Tian}},\ }\href@noop {} {}\Eprint {http://arxiv.org/abs/1807.06432}
  {arXiv:1807.06432} \BibitemShut {NoStop}%
\bibitem [{\citenamefont {Tian}(2018)}]{Tian2018b}%
  \BibitemOpen
  \bibfield  {author} {\bibinfo {author} {\bibfnamefont {S.}~\bibnamefont
  {Tian}},\ }\href {\doibase 10.1103/PhysRevD.98.084040} {\bibfield  {journal}
  {\bibinfo  {journal} {Phys. Rev. D}\ }\textbf {\bibinfo {volume} {98}},\
  \bibinfo {pages} {084040} (\bibinfo {year} {2018})}\BibitemShut {NoStop}%
\bibitem [{\citenamefont {Ay\'on-Beato}\ and\ \citenamefont
  {Garcia}(2000)}]{Ayon-Beato2000}%
  \BibitemOpen
  \bibfield  {author} {\bibinfo {author} {\bibfnamefont {E.}~\bibnamefont
  {Ay\'on-Beato}}\ and\ \bibinfo {author} {\bibfnamefont {A.}~\bibnamefont
  {Garcia}},\ }\href {\doibase 10.1016/S0370-2693(00)01125-4} {\bibfield
  {journal} {\bibinfo  {journal} {\plb}\ }\textbf {\bibinfo {volume} {493}},\
  \bibinfo {pages} {149 } (\bibinfo {year} {2000})}\BibitemShut {NoStop}%
\bibitem [{\citenamefont {Rodrigues}\ \emph {et~al.}(2016)\citenamefont
  {Rodrigues}, \citenamefont {Junior}, \citenamefont {Marques},\ and\
  \citenamefont {Zanchin}}]{Rodrigues2016}%
  \BibitemOpen
  \bibfield  {author} {\bibinfo {author} {\bibfnamefont {M.~E.}\ \bibnamefont
  {Rodrigues}}, \bibinfo {author} {\bibfnamefont {E.~L.~B.}\ \bibnamefont
  {Junior}}, \bibinfo {author} {\bibfnamefont {G.~T.}\ \bibnamefont {Marques}},
  \ and\ \bibinfo {author} {\bibfnamefont {V.~T.}\ \bibnamefont {Zanchin}},\
  }\href {\doibase 10.1103/PhysRevD.94.024062} {\bibfield  {journal} {\bibinfo
  {journal} {Phys. Rev. D}\ }\textbf {\bibinfo {volume} {94}},\ \bibinfo
  {pages} {024062} (\bibinfo {year} {2016})}\BibitemShut {NoStop}%
\bibitem [{\citenamefont {Chinaglia}\ and\ \citenamefont
  {Zerbini}(2017)}]{Chinaglia2017}%
  \BibitemOpen
  \bibfield  {author} {\bibinfo {author} {\bibfnamefont {S.}~\bibnamefont
  {Chinaglia}}\ and\ \bibinfo {author} {\bibfnamefont {S.}~\bibnamefont
  {Zerbini}},\ }\href {\doibase 10.1007/s10714-017-2235-6} {\bibfield
  {journal} {\bibinfo  {journal} {\gerg}\ }\textbf {\bibinfo {volume} {49}},\
  \bibinfo {pages} {75} (\bibinfo {year} {2017})}\BibitemShut {NoStop}%
\bibitem [{\citenamefont {Nojiri}\ and\ \citenamefont
  {Odintsov}(2017)}]{Nojiri2017b}%
  \BibitemOpen
  \bibfield  {author} {\bibinfo {author} {\bibfnamefont {S.}~\bibnamefont
  {Nojiri}}\ and\ \bibinfo {author} {\bibfnamefont {S.~D.}\ \bibnamefont
  {Odintsov}},\ }\href {\doibase 10.1103/PhysRevD.96.104008} {\bibfield
  {journal} {\bibinfo  {journal} {Phys. Rev. D}\ }\textbf {\bibinfo {volume}
  {96}},\ \bibinfo {pages} {104008} (\bibinfo {year} {2017})}\BibitemShut
  {NoStop}%
\bibitem [{\citenamefont {Rodrigues}\ and\ \citenamefont
  {Silva}(2019)}]{Rodrigues2019}%
  \BibitemOpen
  \bibfield  {author} {\bibinfo {author} {\bibfnamefont {M.~E.}\ \bibnamefont
  {Rodrigues}}\ and\ \bibinfo {author} {\bibfnamefont {M.~V. d.~S.}\
  \bibnamefont {Silva}},\ }\href {\doibase 10.1103/PhysRevD.99.124010}
  {\bibfield  {journal} {\bibinfo  {journal} {Phys. Rev. D}\ }\textbf {\bibinfo
  {volume} {99}},\ \bibinfo {pages} {124010} (\bibinfo {year}
  {2019})}\BibitemShut {NoStop}%
\bibitem [{\citenamefont {Baker}\ \emph {et~al.}(2017)\citenamefont {Baker},
  \citenamefont {Bellini}, \citenamefont {Ferreira}, \citenamefont {Lagos},
  \citenamefont {Noller},\ and\ \citenamefont {Sawicki}}]{Baker2017}%
  \BibitemOpen
  \bibfield  {author} {\bibinfo {author} {\bibfnamefont {T.}~\bibnamefont
  {Baker}}, \bibinfo {author} {\bibfnamefont {E.}~\bibnamefont {Bellini}},
  \bibinfo {author} {\bibfnamefont {P.~G.}\ \bibnamefont {Ferreira}}, \bibinfo
  {author} {\bibfnamefont {M.}~\bibnamefont {Lagos}}, \bibinfo {author}
  {\bibfnamefont {J.}~\bibnamefont {Noller}}, \ and\ \bibinfo {author}
  {\bibfnamefont {I.}~\bibnamefont {Sawicki}},\ }\href {\doibase
  10.1103/PhysRevLett.119.251301} {\bibfield  {journal} {\bibinfo  {journal}
  {Phys. Rev. Lett.}\ }\textbf {\bibinfo {volume} {119}},\ \bibinfo {pages}
  {251301} (\bibinfo {year} {2017})}\BibitemShut {NoStop}%
\bibitem [{\citenamefont {Bettoni}\ \emph {et~al.}(2017)\citenamefont
  {Bettoni}, \citenamefont {Ezquiaga}, \citenamefont {Hinterbichler},\ and\
  \citenamefont {Zumalac\'arregui}}]{Bettoni2017}%
  \BibitemOpen
  \bibfield  {author} {\bibinfo {author} {\bibfnamefont {D.}~\bibnamefont
  {Bettoni}}, \bibinfo {author} {\bibfnamefont {J.~M.}\ \bibnamefont
  {Ezquiaga}}, \bibinfo {author} {\bibfnamefont {K.}~\bibnamefont
  {Hinterbichler}}, \ and\ \bibinfo {author} {\bibfnamefont {M.}~\bibnamefont
  {Zumalac\'arregui}},\ }\href {\doibase 10.1103/PhysRevD.95.084029} {\bibfield
   {journal} {\bibinfo  {journal} {Phys. Rev. D}\ }\textbf {\bibinfo {volume}
  {95}},\ \bibinfo {pages} {084029} (\bibinfo {year} {2017})}\BibitemShut
  {NoStop}%
\bibitem [{\citenamefont {Creminelli}\ and\ \citenamefont
  {Vernizzi}(2017)}]{Creminelli2017}%
  \BibitemOpen
  \bibfield  {author} {\bibinfo {author} {\bibfnamefont {P.}~\bibnamefont
  {Creminelli}}\ and\ \bibinfo {author} {\bibfnamefont {F.}~\bibnamefont
  {Vernizzi}},\ }\href {\doibase 10.1103/PhysRevLett.119.251302} {\bibfield
  {journal} {\bibinfo  {journal} {Phys. Rev. Lett.}\ }\textbf {\bibinfo
  {volume} {119}},\ \bibinfo {pages} {251302} (\bibinfo {year}
  {2017})}\BibitemShut {NoStop}%
\bibitem [{\citenamefont {Ezquiaga}\ and\ \citenamefont
  {Zumalac\'arregui}(2017)}]{Ezquiaga2017}%
  \BibitemOpen
  \bibfield  {author} {\bibinfo {author} {\bibfnamefont {J.~M.}\ \bibnamefont
  {Ezquiaga}}\ and\ \bibinfo {author} {\bibfnamefont {M.}~\bibnamefont
  {Zumalac\'arregui}},\ }\href {\doibase 10.1103/PhysRevLett.119.251304}
  {\bibfield  {journal} {\bibinfo  {journal} {Phys. Rev. Lett.}\ }\textbf
  {\bibinfo {volume} {119}},\ \bibinfo {pages} {251304} (\bibinfo {year}
  {2017})}\BibitemShut {NoStop}%
\bibitem [{\citenamefont {Sakstein}\ and\ \citenamefont
  {Jain}(2017)}]{Sakstein2017}%
  \BibitemOpen
  \bibfield  {author} {\bibinfo {author} {\bibfnamefont {J.}~\bibnamefont
  {Sakstein}}\ and\ \bibinfo {author} {\bibfnamefont {B.}~\bibnamefont
  {Jain}},\ }\href {\doibase 10.1103/PhysRevLett.119.251303} {\bibfield
  {journal} {\bibinfo  {journal} {Phys. Rev. Lett.}\ }\textbf {\bibinfo
  {volume} {119}},\ \bibinfo {pages} {251303} (\bibinfo {year}
  {2017})}\BibitemShut {NoStop}%
\bibitem [{\citenamefont {Akrami}\ \emph {et~al.}(2018)\citenamefont {Akrami},
  \citenamefont {Brax}, \citenamefont {Davis},\ and\ \citenamefont
  {Vardanyan}}]{Akrami2018}%
  \BibitemOpen
  \bibfield  {author} {\bibinfo {author} {\bibfnamefont {Y.}~\bibnamefont
  {Akrami}}, \bibinfo {author} {\bibfnamefont {P.}~\bibnamefont {Brax}},
  \bibinfo {author} {\bibfnamefont {A.-C.}\ \bibnamefont {Davis}}, \ and\
  \bibinfo {author} {\bibfnamefont {V.}~\bibnamefont {Vardanyan}},\ }\href
  {\doibase 10.1103/PhysRevD.97.124010} {\bibfield  {journal} {\bibinfo
  {journal} {Phys. Rev. D}\ }\textbf {\bibinfo {volume} {97}},\ \bibinfo
  {pages} {124010} (\bibinfo {year} {2018})}\BibitemShut {NoStop}%
\bibitem [{\citenamefont {Amendola}\ \emph {et~al.}(2018)\citenamefont
  {Amendola}, \citenamefont {Kunz}, \citenamefont {Saltas},\ and\ \citenamefont
  {Sawicki}}]{Amendola2018}%
  \BibitemOpen
  \bibfield  {author} {\bibinfo {author} {\bibfnamefont {L.}~\bibnamefont
  {Amendola}}, \bibinfo {author} {\bibfnamefont {M.}~\bibnamefont {Kunz}},
  \bibinfo {author} {\bibfnamefont {I.~D.}\ \bibnamefont {Saltas}}, \ and\
  \bibinfo {author} {\bibfnamefont {I.}~\bibnamefont {Sawicki}},\ }\href
  {\doibase 10.1103/PhysRevLett.120.131101} {\bibfield  {journal} {\bibinfo
  {journal} {Phys. Rev. Lett.}\ }\textbf {\bibinfo {volume} {120}},\ \bibinfo
  {pages} {131101} (\bibinfo {year} {2018})}\BibitemShut {NoStop}%
\bibitem [{\citenamefont {Arai}\ and\ \citenamefont
  {Nishizawa}(2018)}]{Arai2018}%
  \BibitemOpen
  \bibfield  {author} {\bibinfo {author} {\bibfnamefont {S.}~\bibnamefont
  {Arai}}\ and\ \bibinfo {author} {\bibfnamefont {A.}~\bibnamefont
  {Nishizawa}},\ }\href {\doibase 10.1103/PhysRevD.97.104038} {\bibfield
  {journal} {\bibinfo  {journal} {Phys. Rev. D}\ }\textbf {\bibinfo {volume}
  {97}},\ \bibinfo {pages} {104038} (\bibinfo {year} {2018})}\BibitemShut
  {NoStop}%
\bibitem [{\citenamefont {Boran}\ \emph {et~al.}(2018)\citenamefont {Boran},
  \citenamefont {Desai}, \citenamefont {Kahya},\ and\ \citenamefont
  {Woodard}}]{Boran2018}%
  \BibitemOpen
  \bibfield  {author} {\bibinfo {author} {\bibfnamefont {S.}~\bibnamefont
  {Boran}}, \bibinfo {author} {\bibfnamefont {S.}~\bibnamefont {Desai}},
  \bibinfo {author} {\bibfnamefont {E.~O.}\ \bibnamefont {Kahya}}, \ and\
  \bibinfo {author} {\bibfnamefont {R.~P.}\ \bibnamefont {Woodard}},\ }\href
  {\doibase 10.1103/PhysRevD.97.041501} {\bibfield  {journal} {\bibinfo
  {journal} {Phys. Rev. D}\ }\textbf {\bibinfo {volume} {97}},\ \bibinfo
  {pages} {041501} (\bibinfo {year} {2018})}\BibitemShut {NoStop}%
\bibitem [{\citenamefont {Cai}\ \emph {et~al.}(2018)\citenamefont {Cai},
  \citenamefont {Li}, \citenamefont {Saridakis},\ and\ \citenamefont
  {Xue}}]{Cai2018}%
  \BibitemOpen
  \bibfield  {author} {\bibinfo {author} {\bibfnamefont {Y.-F.}\ \bibnamefont
  {Cai}}, \bibinfo {author} {\bibfnamefont {C.}~\bibnamefont {Li}}, \bibinfo
  {author} {\bibfnamefont {E.~N.}\ \bibnamefont {Saridakis}}, \ and\ \bibinfo
  {author} {\bibfnamefont {L.-Q.}\ \bibnamefont {Xue}},\ }\href {\doibase
  10.1103/PhysRevD.97.103513} {\bibfield  {journal} {\bibinfo  {journal} {Phys.
  Rev. D}\ }\textbf {\bibinfo {volume} {97}},\ \bibinfo {pages} {103513}
  (\bibinfo {year} {2018})}\BibitemShut {NoStop}%
\bibitem [{\citenamefont {Copeland}\ \emph {et~al.}(2019)\citenamefont
  {Copeland}, \citenamefont {Kopp}, \citenamefont {Padilla}, \citenamefont
  {Saffin},\ and\ \citenamefont {Skordis}}]{Copeland2019}%
  \BibitemOpen
  \bibfield  {author} {\bibinfo {author} {\bibfnamefont {E.~J.}\ \bibnamefont
  {Copeland}}, \bibinfo {author} {\bibfnamefont {M.}~\bibnamefont {Kopp}},
  \bibinfo {author} {\bibfnamefont {A.}~\bibnamefont {Padilla}}, \bibinfo
  {author} {\bibfnamefont {P.~M.}\ \bibnamefont {Saffin}}, \ and\ \bibinfo
  {author} {\bibfnamefont {C.}~\bibnamefont {Skordis}},\ }\href {\doibase
  10.1103/PhysRevLett.122.061301} {\bibfield  {journal} {\bibinfo  {journal}
  {Phys. Rev. Lett.}\ }\textbf {\bibinfo {volume} {122}},\ \bibinfo {pages}
  {061301} (\bibinfo {year} {2019})}\BibitemShut {NoStop}%
\bibitem [{\citenamefont {Crisostomi}\ and\ \citenamefont
  {Koyama}(2018)}]{Crisostomi2018}%
  \BibitemOpen
  \bibfield  {author} {\bibinfo {author} {\bibfnamefont {M.}~\bibnamefont
  {Crisostomi}}\ and\ \bibinfo {author} {\bibfnamefont {K.}~\bibnamefont
  {Koyama}},\ }\href {\doibase 10.1103/PhysRevD.97.021301} {\bibfield
  {journal} {\bibinfo  {journal} {Phys. Rev. D}\ }\textbf {\bibinfo {volume}
  {97}},\ \bibinfo {pages} {021301} (\bibinfo {year} {2018})}\BibitemShut
  {NoStop}%
\bibitem [{\citenamefont {Emir G\"umr\"uk\ifmmode \mbox{\c{c}}\else
  \c{c}\fi{}\"uo\ifmmode~\breve{g}\else \u{g}\fi{}lu}\ \emph
  {et~al.}(2018)\citenamefont {Emir G\"umr\"uk\ifmmode \mbox{\c{c}}\else
  \c{c}\fi{}\"uo\ifmmode~\breve{g}\else \u{g}\fi{}lu}, \citenamefont
  {Saravani},\ and\ \citenamefont {Sotiriou}}]{Gumrukcuoglu2018}%
  \BibitemOpen
  \bibfield  {author} {\bibinfo {author} {\bibfnamefont {A.}~\bibnamefont {Emir
  G\"umr\"uk\ifmmode \mbox{\c{c}}\else \c{c}\fi{}\"uo\ifmmode~\breve{g}\else
  \u{g}\fi{}lu}}, \bibinfo {author} {\bibfnamefont {M.}~\bibnamefont
  {Saravani}}, \ and\ \bibinfo {author} {\bibfnamefont {T.~P.}\ \bibnamefont
  {Sotiriou}},\ }\href {\doibase 10.1103/PhysRevD.97.024032} {\bibfield
  {journal} {\bibinfo  {journal} {Phys. Rev. D}\ }\textbf {\bibinfo {volume}
  {97}},\ \bibinfo {pages} {024032} (\bibinfo {year} {2018})}\BibitemShut
  {NoStop}%
\bibitem [{\citenamefont {Kreisch}\ and\ \citenamefont
  {Komatsu}(2018)}]{Kreisch2018}%
  \BibitemOpen
  \bibfield  {author} {\bibinfo {author} {\bibfnamefont {C.~D.}\ \bibnamefont
  {Kreisch}}\ and\ \bibinfo {author} {\bibfnamefont {E.}~\bibnamefont
  {Komatsu}},\ }\href {\doibase 10.1088/1475-7516/2018/12/030} {\bibfield
  {journal} {\bibinfo  {journal} {\jcap}\ }12 (\bibinfo {year} {2018})\
  \bibinfo {pages} {030}}\BibitemShut {NoStop}%
\bibitem [{\citenamefont {Nojiri}\ and\ \citenamefont
  {Odintsov}(2018)}]{Nojiri2018-170817}%
  \BibitemOpen
  \bibfield  {author} {\bibinfo {author} {\bibfnamefont {S.}~\bibnamefont
  {Nojiri}}\ and\ \bibinfo {author} {\bibfnamefont {S.~D.}\ \bibnamefont
  {Odintsov}},\ }\href {\doibase 10.1016/j.physletb.2018.01.078} {\bibfield
  {journal} {\bibinfo  {journal} {\plb}\ }\textbf {\bibinfo {volume} {779}},\
  \bibinfo {pages} {425 } (\bibinfo {year} {2018})}\BibitemShut {NoStop}%
\bibitem [{\citenamefont {Oost}\ \emph {et~al.}(2018)\citenamefont {Oost},
  \citenamefont {Mukohyama},\ and\ \citenamefont {Wang}}]{Oost2018}%
  \BibitemOpen
  \bibfield  {author} {\bibinfo {author} {\bibfnamefont {J.}~\bibnamefont
  {Oost}}, \bibinfo {author} {\bibfnamefont {S.}~\bibnamefont {Mukohyama}}, \
  and\ \bibinfo {author} {\bibfnamefont {A.}~\bibnamefont {Wang}},\ }\href
  {\doibase 10.1103/PhysRevD.97.124023} {\bibfield  {journal} {\bibinfo
  {journal} {Phys. Rev. D}\ }\textbf {\bibinfo {volume} {97}},\ \bibinfo
  {pages} {124023} (\bibinfo {year} {2018})}\BibitemShut {NoStop}%
\bibitem [{\citenamefont {Pardo}\ \emph {et~al.}(2018)\citenamefont {Pardo},
  \citenamefont {Fishbach}, \citenamefont {Holz},\ and\ \citenamefont
  {Spergel}}]{Pardo2018}%
  \BibitemOpen
  \bibfield  {author} {\bibinfo {author} {\bibfnamefont {K.}~\bibnamefont
  {Pardo}}, \bibinfo {author} {\bibfnamefont {M.}~\bibnamefont {Fishbach}},
  \bibinfo {author} {\bibfnamefont {D.~E.}\ \bibnamefont {Holz}}, \ and\
  \bibinfo {author} {\bibfnamefont {D.~N.}\ \bibnamefont {Spergel}},\ }\href
  {\doibase 10.1088/1475-7516/2018/07/048} {\bibfield  {journal} {\bibinfo
  {journal} {\jcap}\ }07 (\bibinfo {year} {2018})\ \bibinfo {pages}
  {048}}\BibitemShut {NoStop}%
\bibitem [{\citenamefont {Visinelli}\ \emph {et~al.}(2018)\citenamefont
  {Visinelli}, \citenamefont {Bolis},\ and\ \citenamefont
  {Vagnozzi}}]{Visinelli2018}%
  \BibitemOpen
  \bibfield  {author} {\bibinfo {author} {\bibfnamefont {L.}~\bibnamefont
  {Visinelli}}, \bibinfo {author} {\bibfnamefont {N.}~\bibnamefont {Bolis}}, \
  and\ \bibinfo {author} {\bibfnamefont {S.}~\bibnamefont {Vagnozzi}},\ }\href
  {\doibase 10.1103/PhysRevD.97.064039} {\bibfield  {journal} {\bibinfo
  {journal} {Phys. Rev. D}\ }\textbf {\bibinfo {volume} {97}},\ \bibinfo
  {pages} {064039} (\bibinfo {year} {2018})}\BibitemShut {NoStop}%
\bibitem [{\citenamefont {Casalino}\ \emph {et~al.}(2018)\citenamefont
  {Casalino}, \citenamefont {Rinaldi}, \citenamefont {Sebastiani},\ and\
  \citenamefont {Vagnozzi}}]{Casalino2018}%
  \BibitemOpen
  \bibfield  {author} {\bibinfo {author} {\bibfnamefont {A.}~\bibnamefont
  {Casalino}}, \bibinfo {author} {\bibfnamefont {M.}~\bibnamefont {Rinaldi}},
  \bibinfo {author} {\bibfnamefont {L.}~\bibnamefont {Sebastiani}}, \ and\
  \bibinfo {author} {\bibfnamefont {S.}~\bibnamefont {Vagnozzi}},\ }\href
  {\doibase 10.1088/1361-6382/aaf1fd} {\bibfield  {journal} {\bibinfo
  {journal} {\cqg}\ }\textbf {\bibinfo {volume} {36}},\ \bibinfo {pages}
  {017001} (\bibinfo {year} {2018})}\BibitemShut {NoStop}%
\bibitem [{\citenamefont {Ganz}\ \emph {et~al.}(2019)\citenamefont {Ganz},
  \citenamefont {Bartolo}, \citenamefont {Karmakar},\ and\ \citenamefont
  {Matarrese}}]{Ganz2019}%
  \BibitemOpen
  \bibfield  {author} {\bibinfo {author} {\bibfnamefont {A.}~\bibnamefont
  {Ganz}}, \bibinfo {author} {\bibfnamefont {N.}~\bibnamefont {Bartolo}},
  \bibinfo {author} {\bibfnamefont {P.}~\bibnamefont {Karmakar}}, \ and\
  \bibinfo {author} {\bibfnamefont {S.}~\bibnamefont {Matarrese}},\ }\href
  {\doibase 10.1088/1475-7516/2019/01/056} {\bibfield  {journal} {\bibinfo
  {journal} {\jcap}\ }01 (\bibinfo {year} {2019})\ \bibinfo {pages}
  {056}}\BibitemShut {NoStop}%
\bibitem [{\citenamefont {Jana}\ and\ \citenamefont
  {Mohanty}(2019)}]{Jana2019}%
  \BibitemOpen
  \bibfield  {author} {\bibinfo {author} {\bibfnamefont {S.}~\bibnamefont
  {Jana}}\ and\ \bibinfo {author} {\bibfnamefont {S.}~\bibnamefont {Mohanty}},\
  }\href {\doibase 10.1103/PhysRevD.99.044056} {\bibfield  {journal} {\bibinfo
  {journal} {Phys. Rev. D}\ }\textbf {\bibinfo {volume} {99}},\ \bibinfo
  {pages} {044056} (\bibinfo {year} {2019})}\BibitemShut {NoStop}%
\bibitem [{\citenamefont {Ramos}\ and\ \citenamefont
  {Barausse}(2019)}]{Ramos2019}%
  \BibitemOpen
  \bibfield  {author} {\bibinfo {author} {\bibfnamefont {O.}~\bibnamefont
  {Ramos}}\ and\ \bibinfo {author} {\bibfnamefont {E.}~\bibnamefont
  {Barausse}},\ }\href {\doibase 10.1103/PhysRevD.99.024034} {\bibfield
  {journal} {\bibinfo  {journal} {Phys. Rev. D}\ }\textbf {\bibinfo {volume}
  {99}},\ \bibinfo {pages} {024034} (\bibinfo {year} {2019})}\BibitemShut
  {NoStop}%
\bibitem [{\citenamefont {Moffat}\ and\ \citenamefont {Toth}()}]{Moffat2019}%
  \BibitemOpen
  \bibfield  {author} {\bibinfo {author} {\bibfnamefont {J.~W.}\ \bibnamefont
  {Moffat}}\ and\ \bibinfo {author} {\bibfnamefont {V.~T.}\ \bibnamefont
  {Toth}},\ }\href@noop {} {}\Eprint {http://arxiv.org/abs/1904.04142}
  {arXiv:1904.04142} \BibitemShut {NoStop}%
\bibitem [{\citenamefont {Davoudiasl}\ and\ \citenamefont
  {Denton}(2019)}]{Davoudiasl2019}%
  \BibitemOpen
  \bibfield  {author} {\bibinfo {author} {\bibfnamefont {H.}~\bibnamefont
  {Davoudiasl}}\ and\ \bibinfo {author} {\bibfnamefont {P.~B.}\ \bibnamefont
  {Denton}},\ }\href {\doibase 10.1103/PhysRevLett.123.021102} {\bibfield
  {journal} {\bibinfo  {journal} {Phys. Rev. Lett.}\ }\textbf {\bibinfo
  {volume} {123}},\ \bibinfo {pages} {021102} (\bibinfo {year}
  {2019})}\BibitemShut {NoStop}%
\bibitem [{\citenamefont {Bar}\ \emph {et~al.}(2019)\citenamefont {Bar},
  \citenamefont {Blum}, \citenamefont {Lacroix},\ and\ \citenamefont
  {Panci}}]{Bar2019}%
  \BibitemOpen
  \bibfield  {author} {\bibinfo {author} {\bibfnamefont {N.}~\bibnamefont
  {Bar}}, \bibinfo {author} {\bibfnamefont {K.}~\bibnamefont {Blum}}, \bibinfo
  {author} {\bibfnamefont {T.}~\bibnamefont {Lacroix}}, \ and\ \bibinfo
  {author} {\bibfnamefont {P.}~\bibnamefont {Panci}},\ }\href {\doibase
  10.1088/1475-7516/2019/07/045} {\bibfield  {journal} {\bibinfo  {journal}
  {\jcap}\ }07 (\bibinfo {year} {2019})\ \bibinfo {pages} {045}}\BibitemShut {NoStop}%
\bibitem [{\citenamefont {Abbott~\etal}(2016)}]{Abbott2016_GW150914testGR}%
  \BibitemOpen
  \bibfield  {author} {\bibinfo {author} {\bibfnamefont {B.~P.}\ \bibnamefont
  {Abbott~\etal}},\ }\href {\doibase 10.1103/PhysRevLett.116.221101} {\bibfield
   {journal} {\bibinfo  {journal} {Phys. Rev. Lett.}\ }\textbf {\bibinfo
  {volume} {116}},\ \bibinfo {pages} {221101} (\bibinfo {year}
  {2016})}\BibitemShut {NoStop}%
\bibitem [{\citenamefont {Callister}\ \emph {et~al.}(2017)\citenamefont
  {Callister}, \citenamefont {Biscoveanu}, \citenamefont {Christensen},
  \citenamefont {Isi}, \citenamefont {Matas}, \citenamefont {Minazzoli},
  \citenamefont {Regimbau}, \citenamefont {Sakellariadou}, \citenamefont
  {Tasson},\ and\ \citenamefont {Thrane}}]{Callister2017}%
  \BibitemOpen
  \bibfield  {author} {\bibinfo {author} {\bibfnamefont {T.}~\bibnamefont
  {Callister}}, \bibinfo {author} {\bibfnamefont {A.~S.}\ \bibnamefont
  {Biscoveanu}}, \bibinfo {author} {\bibfnamefont {N.}~\bibnamefont
  {Christensen}}, \bibinfo {author} {\bibfnamefont {M.}~\bibnamefont {Isi}},
  \bibinfo {author} {\bibfnamefont {A.}~\bibnamefont {Matas}}, \bibinfo
  {author} {\bibfnamefont {O.}~\bibnamefont {Minazzoli}}, \bibinfo {author}
  {\bibfnamefont {T.}~\bibnamefont {Regimbau}}, \bibinfo {author}
  {\bibfnamefont {M.}~\bibnamefont {Sakellariadou}}, \bibinfo {author}
  {\bibfnamefont {J.}~\bibnamefont {Tasson}}, \ and\ \bibinfo {author}
  {\bibfnamefont {E.}~\bibnamefont {Thrane}},\ }\href {\doibase
  10.1103/PhysRevX.7.041058} {\bibfield  {journal} {\bibinfo  {journal} {Phys.
  Rev. X}\ }\textbf {\bibinfo {volume} {7}},\ \bibinfo {pages} {041058}
  (\bibinfo {year} {2017})}\BibitemShut {NoStop}%
\bibitem [{\citenamefont {Shao}\ \emph {et~al.}(2017)\citenamefont {Shao},
  \citenamefont {Sennett}, \citenamefont {Buonanno}, \citenamefont {Kramer},\
  and\ \citenamefont {Wex}}]{Shao2017}%
  \BibitemOpen
  \bibfield  {author} {\bibinfo {author} {\bibfnamefont {L.}~\bibnamefont
  {Shao}}, \bibinfo {author} {\bibfnamefont {N.}~\bibnamefont {Sennett}},
  \bibinfo {author} {\bibfnamefont {A.}~\bibnamefont {Buonanno}}, \bibinfo
  {author} {\bibfnamefont {M.}~\bibnamefont {Kramer}}, \ and\ \bibinfo {author}
  {\bibfnamefont {N.}~\bibnamefont {Wex}},\ }\href {\doibase
  10.1103/PhysRevX.7.041025} {\bibfield  {journal} {\bibinfo  {journal} {Phys.
  Rev. X}\ }\textbf {\bibinfo {volume} {7}},\ \bibinfo {pages} {041025}
  (\bibinfo {year} {2017})}\BibitemShut {NoStop}%
\bibitem [{\citenamefont {Abbott~\etal}(2019)}]{Abbott2019_GW170817testGR}%
  \BibitemOpen
  \bibfield  {author} {\bibinfo {author} {\bibfnamefont {B.~P.}\ \bibnamefont
  {Abbott~\etal}},\ }\href {\doibase 10.1103/PhysRevLett.123.011102} {\bibfield
   {journal} {\bibinfo  {journal} {Phys. Rev. Lett.}\ }\textbf {\bibinfo
  {volume} {123}},\ \bibinfo {pages} {011102} (\bibinfo {year}
  {2019})}\BibitemShut {NoStop}%
\bibitem [{\citenamefont {Bustillo}\ \emph {et~al.}()\citenamefont {Bustillo},
  \citenamefont {Evans}, \citenamefont {Clark}, \citenamefont {Kim},
  \citenamefont {Laguna},\ and\ \citenamefont {Shoemaker}}]{Bustillo2019}%
  \BibitemOpen
  \bibfield  {author} {\bibinfo {author} {\bibfnamefont {J.~C.}\ \bibnamefont
  {Bustillo}}, \bibinfo {author} {\bibfnamefont {C.}~\bibnamefont {Evans}},
  \bibinfo {author} {\bibfnamefont {J.~A.}\ \bibnamefont {Clark}}, \bibinfo
  {author} {\bibfnamefont {G.}~\bibnamefont {Kim}}, \bibinfo {author}
  {\bibfnamefont {P.}~\bibnamefont {Laguna}}, \ and\ \bibinfo {author}
  {\bibfnamefont {D.}~\bibnamefont {Shoemaker}},\ }\href@noop {} {}\Eprint
  {http://arxiv.org/abs/1906.01153} {arXiv:1906.01153} \BibitemShut {NoStop}%
\bibitem [{\citenamefont {Giddings}\ \emph {et~al.}(2019)\citenamefont
  {Giddings}, \citenamefont {Koren},\ and\ \citenamefont
  {Trevi\~no}}]{Giddings2019}%
  \BibitemOpen
  \bibfield  {author} {\bibinfo {author} {\bibfnamefont {S.~B.}\ \bibnamefont
  {Giddings}}, \bibinfo {author} {\bibfnamefont {S.}~\bibnamefont {Koren}}, \
  and\ \bibinfo {author} {\bibfnamefont {G.}~\bibnamefont {Trevi\~no}},\ }\href
  {\doibase 10.1103/PhysRevD.100.044005} {\bibfield  {journal} {\bibinfo
  {journal} {Phys. Rev. D}\ }\textbf {\bibinfo {volume} {100}},\ \bibinfo
  {pages} {044005} (\bibinfo {year} {2019})}\BibitemShut {NoStop}%
\bibitem [{\citenamefont {Hughes}\ \emph {et~al.}()\citenamefont {Hughes},
  \citenamefont {Apte}, \citenamefont {Khanna},\ and\ \citenamefont
  {Lim}}]{Hughes2019}%
  \BibitemOpen
  \bibfield  {author} {\bibinfo {author} {\bibfnamefont {S.~A.}\ \bibnamefont
  {Hughes}}, \bibinfo {author} {\bibfnamefont {A.}~\bibnamefont {Apte}},
  \bibinfo {author} {\bibfnamefont {G.}~\bibnamefont {Khanna}}, \ and\ \bibinfo
  {author} {\bibfnamefont {H.}~\bibnamefont {Lim}},\ }\href@noop {} {}\Eprint
  {http://arxiv.org/abs/1901.05900} {arXiv:1901.05900} \BibitemShut {NoStop}%
\bibitem [{\citenamefont {Isi}\ \emph {et~al.}({\natexlab{a}})\citenamefont
  {Isi}, \citenamefont {Giesler}, \citenamefont {Farr}, \citenamefont
  {Scheel},\ and\ \citenamefont {Teukolsky}}]{Isi2019a}%
  \BibitemOpen
  \bibfield  {author} {\bibinfo {author} {\bibfnamefont {M.}~\bibnamefont
  {Isi}}, \bibinfo {author} {\bibfnamefont {M.}~\bibnamefont {Giesler}},
  \bibinfo {author} {\bibfnamefont {W.~M.}\ \bibnamefont {Farr}}, \bibinfo
  {author} {\bibfnamefont {M.~A.}\ \bibnamefont {Scheel}}, \ and\ \bibinfo
  {author} {\bibfnamefont {S.~A.}\ \bibnamefont {Teukolsky}},\ }\href@noop {}
  {}\ \Eprint {http://arxiv.org/abs/1905.00869}
  {arXiv:1905.00869} \BibitemShut {NoStop}%
\bibitem [{\citenamefont {Isi}\ \emph {et~al.}({\natexlab{b}})\citenamefont
  {Isi}, \citenamefont {Chatziioannou},\ and\ \citenamefont {Farr}}]{Isi2019b}%
  \BibitemOpen
  \bibfield  {author} {\bibinfo {author} {\bibfnamefont {M.}~\bibnamefont
  {Isi}}, \bibinfo {author} {\bibfnamefont {K.}~\bibnamefont {Chatziioannou}},
  \ and\ \bibinfo {author} {\bibfnamefont {W.~M.}\ \bibnamefont {Farr}},\
  }\href@noop {} {}\ \Eprint
  {http://arxiv.org/abs/1904.08011} {arXiv:1904.08011} \BibitemShut {NoStop}%
\bibitem [{\citenamefont {Silva}\ and\ \citenamefont {Yunes}()}]{Silva2019}%
  \BibitemOpen
  \bibfield  {author} {\bibinfo {author} {\bibfnamefont {H.~O.}\ \bibnamefont
  {Silva}}\ and\ \bibinfo {author} {\bibfnamefont {N.}~\bibnamefont {Yunes}},\
  }\href@noop {} {}\Eprint {http://arxiv.org/abs/1906.00485} {arXiv:1906.00485}
  \BibitemShut {NoStop}%
\end{thebibliography}
%

\end{document}